\documentclass[smallextended]{svjour3}

\usepackage[most]{tcolorbox}  
\usepackage[utf8]{inputenc}  
\usepackage[framemethod=TikZ]{mdframed}
\usepackage{orcidlink}
\usepackage{hyperref}
\usepackage{url}
\usepackage{array}
\usepackage{multirow}
\usepackage{graphicx}
\usepackage{balance}
\usepackage{esvect}
\usepackage{booktabs}
\usepackage{tabularx}
\usepackage{anyfontsize}
\usepackage[all]{nowidow}
\usepackage[normalem]{ulem}
\usepackage{xcolor}
\usepackage{colortbl}
\usepackage{enumitem}
\usepackage{makecell}
\usepackage{caption}
\usepackage{subcaption}
\usepackage[switch]{lineno}
\usepackage{framed}
\usepackage{wrapfig}
\usepackage{amsmath,amssymb}

\usepackage{newtxmath} 
\usepackage{natbib}
\usepackage{makecell}
\usepackage{algorithm}
\usepackage{algpseudocode}
\usepackage{subcaption}
\usepackage{nicematrix}
\usepackage{rotating}

\newcolumntype{Y}{>{\centering\arraybackslash}X}

\definecolor{commentGray}{RGB}{120,120,120}
\definecolor{darkgreen}{RGB}{1,128,30}

\renewcommand{\algorithmiccomment}[1]{\bgroup\color{commentGray}{//#1}\egroup}

\usepackage{pifont}% http://ctan.org/pkg/pifont
\newcommand{\cmark}{\ding{51}}%
\newcommand{\xmark}{\ding{55}}%
\usepackage{listings}

\usepackage{framed}
\usepackage[most]{tcolorbox}
\usepackage{tikz}
\definecolor{light-gray}{gray}{0.9}
\usepackage[framemethod=TikZ]{mdframed}
\mdfsetup{skipabove=5pt,skipbelow=3pt}
\usepackage{lipsum}
\mdfdefinestyle{RQFrame}{%
	linecolor=black,
	outerlinewidth=0.15pt,
	roundcorner=3pt,
	innertopmargin=2pt,
	innerbottommargin=2pt,
	innerrightmargin=4pt,
	innerleftmargin=4pt,
	backgroundcolor=light-gray}

\newcommand{\sectopic}[1]{\vspace*{0.1em}\par\noindent{\textit{\bfseries #1}}}

\journalname{Empirical Software Engineering}

\newcommand{\approach}[0]{LADEX}

\newcommand{\approachAlignmentLLMStrucutLLM}[0]{\approach-LLM-LLM}
\newcommand{\approachAlignmentLLMStrucutFormal}[0]{\approach-Alg-LLM}
\newcommand{\approachStrucutLLMOnly}[0]{\approach-LLM-NA}
\newcommand{\approachStrucutFormalOnly}[0]{\approach-Alg-NA}

\definecolor{javagreen}{rgb}{0.25,0.5,0.35} % comments

\lstdefinestyle{Alg}{
  basicstyle=\ttfamily\footnotesize,
  breaklines=true,
  tabsize=2,
  mathescape,
  numbers=left,
  xleftmargin=2.5em,
  xrightmargin=0.5em,
  frame=tb,
  framexleftmargin=2em,
  emph={Algorithm,Input,Output,for,each,do,if,else,Function,while,let,be,repeat,until,return,times,and,or,break,in,then,},
  emphstyle={\textbf},
  escapechar=?,
  morecomment=[l][\color{javagreen}]{//},
  columns=flexible,
}

\newcommand\Algphase[1]{%
  \vspace*{-.7\baselineskip}%
  \Statex\hspace*{\dimexpr-\algorithmicindent-2pt\relax}\rule{\columnwidth}{0.4pt}%
  \Statex\hspace*{-\algorithmicindent}\textbf{#1}%
  \vspace*{-.7\baselineskip}%
  \Statex\hspace*{\dimexpr-\algorithmicindent-2pt\relax}\rule{\columnwidth}{0.4pt}%
}

\begin{document}

\title{The Impact of Critique on LLM-Based\\Model Generation from Natural Language:\\The Case of Activity Diagrams}
\titlerunning{The Impact of Critique on LLM-Based Model Generation}

\author{%
  Parham Khamsepour\textsuperscript{1,2}\orcidlink{0009-0002-9187-3845}, 
  Mark Cole\textsuperscript{2}, 
  Ish Ashraf\textsuperscript{2}, 
  DaYuan Tan\textsuperscript{2}, 
  Sandeep Puri\textsuperscript{2}, 
  Mehrdad Sabetzadeh\textsuperscript{1}\orcidlink{0000-0002-4711-8319}, and 
  Shiva Nejati\textsuperscript{1}\orcidlink{0000-0002-0281-8231}%
}

\authorrunning{P. Khamsepour et al.}

\institute{%
  Parham Khamsepour \at \email{parham.khamsepour@uottawa.ca}
  \and
  Mark Cole \at \email{mcole@ciena.com}
  \and
  Ish Ashraf \at \email{iashraf@ciena.com}
  \and
  DaYuan Tan \at \email{datan@ciena.com}
  \and
  Sandeep Puri \at \email{spuri@ciena.com}
  \and
  Mehrdad Sabetzadeh \at \email{m.sabetzadeh@uottawa.ca}
  \and
  Shiva Nejati \at \email{snejati@uottawa.ca} 
  \\
  \textsuperscript{1} University of Ottawa \quad
  \textsuperscript{2} Ciena Corp
}
\date{Received: date / Accepted: date}

\maketitle

\begin{abstract}
Large Language Models (LLMs) show strong potential for automating model generation from natural-language descriptions. A common approach begins with an initial model generation, followed by an iterative critique-refine loop in which the model is evaluated for issues and refined based on those issues.  This process needs to address: (1) structural correctness  -- compliance with well-formedness rules  --  and (2) semantic alignment -- accurate reflection of the intended meaning in the source text.  We present \approach\ (LLM-based Activity Diagram Extractor), a pipeline for deriving activity diagrams from natural-language process descriptions using an LLM-driven critique-refine process. Structural checks in \approach\ can be performed either algorithmically or by an LLM, while alignment checks are performed by an LLM. We design five ablated variants of \approach\ to study: (i)~the impact of the critique-refine loop itself, (ii)~the role of LLM-based semantic checks, and (iii)~the comparative effectiveness of algorithmic versus LLM-based structural checks.

To evaluate \approach, we compare the generated activity diagrams with expert-created ground truths using both trace-based behavioural matcher   and an LLM-based activity-diagram matcher. This enables automated measurement of correctness (whether the generated activity diagram includes the ground-truth nodes) and completeness (how many of the ground-truth nodes the generated activity diagram covers). Experiments on two datasets -- a public-domain dataset and an industry dataset from our collaborator, Ciena -- indicate that: (1)~Both the behavioural and LLM matchers yield similar completeness and correctness comparisons across the \approach\ variants.
(2)~The critique–refine loop improves structural validity, correctness, and completeness compared to single-pass generation. 
(3)~Activity diagrams refined based on algorithmic structural checks achieve structural consistency, whereas those refined based on LLM-based checks often still show structural inconsistencies. (4)~Combining algorithmic structural checks with LLM-based semantic checks (using O4 Mini) delivers the strongest results -- up to 86\% correctness and 92\% completeness -- while requiring fewer than five LLM calls on average. In contrast, using only algorithmic structural checks reaches similar correctness (86\%) and slightly lower completeness (90\%) with just over one LLM call, making it the preferred low-cost option.

\end{abstract}


\section{Introduction}
\label{sec:intro}
Organizations commonly rely on extensive textual documentation to convey complex procedures -- such as system setup, application development processes, and maintenance or troubleshooting tasks -- to stakeholders including developers, product managers, QA engineers, and technical staff~\citep{DBLP:conf/models/AroraSBZ16,DBLP:journals/tosem/AroraSNB19}. Due to its complexity and the inherent limitations of natural language, such documentation is often difficult to interpret or verify, and is generally not amenable to effective monitoring, improvement, or integration into computer-assisted automation. Behavioural workflow models, such as activity diagrams~\citep{Cook2017}, flowcharts~\citep{Cook2017}, and business process models~\citep{omg2011bpmn}, provide intuitive yet precise visual representations of complex procedures. Although these models often include large amounts of text in the form of transition and node labels, they follow standardized notations that support systematic quality validation~\citep{NejatiS0BM16} and enable automation for tasks such as code generation and system monitoring~\citep{6507223}. 

Despite the compelling advantages of models, text has to date remained the primary means of capturing complex procedures in industry. The persistence of text, despite its inherent limitations, stems from its flexibility, ease of writing, and familiarity -- whereas in most real-world settings, building and maintaining models by hand has been prohibitively expensive.

Research has explored ways to automatically generate models from textual descriptions, preserving the simplicity of using natural language while reducing the cost of developing accurate and up-to-date models~\citep{DBLP:conf/models/AroraSBZ16,DBLP:journals/tosem/AroraSNB19,DBLP:conf/re/FerrariAA24, DBLP:conf/refsq/Herwanto24, DBLP:conf/models/JahanHGRRRS24, DBLP:conf/models/YangCCMV24, abscon}. Early approaches to automating model generation from text use traditional Natural Language Processing (NLP) pipelines, such as part-of-speech tagging, rule-based extraction, and syntactic parsing~\citep{DBLP:conf/models/AroraSBZ16,DBLP:journals/tosem/AroraSNB19}. More recently, Large Language Models (LLMs) have been used for this purpose, drawing on their ability to capture semantic context, reason across extended passages, and generate coherent, structured outputs directly from natural-language prompts.  Recent attempts at model generation have investigated prompting strategies ranging from zero-shot and few-shot to more sophisticated multi-step approaches~\citep{DBLP:conf/re/FerrariAA24, DBLP:conf/refsq/Herwanto24, DBLP:conf/models/JahanHGRRRS24, DBLP:conf/models/YangCCMV24,abscon}.

Recent prompting paradigms begin with an initial model generation, followed by an iterative critique-refine loop in which the LLM assesses this initial draft, identifies issues, and revises it accordingly~\citep{DBLP:conf/models/YangCCMV24}. For model generation, the critique-refine loop must address two main types of issues: (1) the model may be structurally (syntactically) incorrect, violating well-formedness rules; and (2) the model may fail to semantically align with the textual description, meaning it does not accurately capture the intended content.

While the idea of a critique-refine loop for model improvement is conceptually simple, its practical implementation and ultimate effectiveness remain open to investigation. Our work in this article is broadly motivated by the following question: \emph{How can we effectively develop an LLM-based critique-refine loop that can identify and resolve both structural and alignment issues in models?} We pursue this question in the context of \emph{behavioural workflow models}, considering these two angles:

\textbf{(1)} Structural correctness can, due to its formal nature, be captured as machine-verifiable rules and can be enforced by algorithmic checks derived from the Unified Modeling Language (UML) specification. In LLM-based model generation pipelines, however, models are often produced in partially formalized textual formats for which no dedicated checker exists, making it attractive to reuse the LLM itself as a structural critic. This leads to an empirical design question: \emph{Should the critique step use algorithms or LLMs to check the structural correctness of the generated models?}

\textbf{(2)} Semantic alignment requires understanding the context, concepts, and relationships described in text and verifying that these are faithfully reflected in the generated models. Bridging heterogeneous modalities (text and model representations) involves interpretive nuances that resist purely algorithmic checks. While human oversight remains the most reliable means of ensuring alignment, it is often resource-intensive and impractical for large-scale or iterative workflows. Therefore, LLMs are explored as scalable proxies for human judgment. This raises the question: \emph{How effective are LLMs in evaluating semantic alignment between models and their textual descriptions?}

In this article, while the questions we raise apply widely to behavioural workflow models -- and, more generally, to models at large -- we concentrate on the specific task of generating \emph{\textbf{activity diagrams}} from natural-language process descriptions. Activity diagrams are commonly used in software engineering, business process management, workflow automation, and project management for their precise modelling of dynamic processes and intuitive, flowchart-like notation~\citep{Cook2017}. We propose an LLM-based pipeline, which we refer to as \approach, \textbf{L}LM-based \textbf{A}ctivity \textbf{D}iagram \textbf{EX}tractor, for generating activity diagrams from text. \approach\ employs prompts that incorporate both (i)~a list of structural constraints, specifying what constitutes a well-formed activity diagram, and (ii)~a list of alignment constraints, outlining how the generated activity diagram should reflect the textual description. The LLM is first instructed to generate an activity diagram that satisfies these constraints. Through an iterative critique-refine loop, \approach\ then evaluates the generated activity diagram against both the structural and alignment constraints. If any constraint violations are detected, the LLM is instructed to update the activity diagram accordingly. The process ends when no issues remain or a user-defined iteration limit is reached.

We evaluate five ablated variants of \approach\ to examine the impact of the critique-refine loop, the role of LLMs in checking alignment constraints in activity diagrams, and the comparative effectiveness of LLMs versus algorithmic methods for checking structural constraints. To isolate the impact of the critique-refine loop, one \approach\ variant removes the loop entirely. The remaining four variants retain the loop but vary in how the constraints are checked.
To study the impact of LLM-based alignment checking, we create two groups of variants: one with LLM-based alignment checking and one without, as alignment constraints can only be critiqued using an LLM. To compare LLM-based and algorithmic methods for structural constraint checking, we create two variants within each group: one using algorithmic structural constraint checking and the other using LLM-based checks.

To evaluate activity diagrams generated by the different \approach\ variants, we compare them against their ground-truth counterparts using two complementary methods: (1) a \emph{behavioural matching algorithm} based on the trace-based operational semantics of activity diagrams~\citep{maoz,NejatiSCEZ07,NejatiSCEZ12}, and (2)~an \emph{LLM-based matcher} instructed to compare activity diagrams based on their textual, behavioural, and structural content. Both the behavioural and the LLM-based matcher compute a mapping between the nodes of an LLM-generated activity diagram and those of the ground truth. Two nodes are considered matched when they exhibit a similar label, behaviour and structure.  We then measure the \textit{correctness} of an LLM-generated activity diagram $A$ by assessing whether the nodes of $A$ are present in the ground truth, and the \textit{completeness} of $A$ by assessing whether the nodes in the ground truth are present in $A$.

We conduct an extensive empirical evaluation of the \approach\ variants, examining \textit{structural consistency}, \textit{semantic correctness}, and \textit{completeness} of the generated activity diagrams, as well as the number of LLM calls required. The evaluation is performed on two datasets: (1)~a public-domain collection of textual process descriptions paired with ground-truth activity diagrams~\citep{du-etal-2024-paged}, and (2)~an industry dataset provided by our partner, Ciena. For the public-domain dataset, we experiment with GPT-4.1 Mini~\citep{openai2024gpt4ocard}, O4 Mini~\citep{jaech2024openai}, and DeepSeek-R1-Distill-Llama-70B~\citep{guo2025deepseek} as the underlying LLMs for the \approach\ variants. For the industry dataset, we evaluate only GPT-4.1 Mini and O4 Mini due to confidentiality and privacy constraints permitting partner-approved LLMs only.

\textbf{Contributions.} Our contributions are as follows: 

\textbf{(1)} We present the first study on designing and evaluating an LLM-based critique–refine loop for behavioural workflow models, specifically activity diagrams. The study examines how LLMs and deterministic algorithms detect and resolve issues of structural correctness and semantic alignment.

\textbf{(2)} Our evaluation results -- based on two evaluation methods, derived from two datasets (one public and one industrial), and tested across three different LLMs -- show that:

\begin{itemize}
\item  Both evaluation methods -- the LLM-based matcher and the behavioural matching algorithm -- produce consistent conclusions regarding the \textit{completeness} and \textit{correctness} of the \approach\ variants. Specifically, the two evaluation methods produce highly correlated \textit{correctness} and \textit{completeness} scores (Spearman's rank correlation in the range $0.8\text{-}1.0$) and never yield conflicting statistical conclusions. Taken together, the results from both evaluation methods mutually reinforce the conclusions, thus improving the overall credibility of our empirical findings.

\item The critique-refine loop produces activity diagrams that are more likely to be structurally and semantically correct and complete than those generated without it.
\item~Iteratively refining activity diagrams using algorithmic structural checks tends to eliminate structural inconsistencies, whereas activity diagrams refined with LLM-based structural checks often still show inconsistencies after multiple iterations. Further, on average, algorithmic structural checking improves \textit{correctness} by 16.95\% and \textit{completeness} by 15.12\% compared to LLM-based structural checking.
\item Compared to using only algorithmic structural checking, adding alignment checking significantly improves correctness in our public dataset without making a significant impact on completeness. However, it introduces a trade-off in our industry dataset: it improves \textit{completeness} by 1.75\% but degrades \textit{correctness} by 4.24\%.
\end{itemize}

\textbf{(3)} A main finding of our work is that the critique approach that yields the best results combines algorithmic structural checking with LLM-based alignment checking and uses the reasoning-based LLM O4 Mini~\citep{jaech2024openai}. This approach produces structurally sound activity diagrams with an average \textit{semantic correctness} of 86\% and an average \textit{completeness} of 92\% across our two datasets and both evaluation methods, while requiring an average of 4.91 LLM calls. However, since LLM-based alignment checking provides only marginal benefits on our industry dataset,
applying only algorithmic structural checking -- without LLM-based alignment checking -- is a competitive alternative. It produces structurally sound activity diagrams with an average \textit{correctness} of 86\% and an average \textit{completeness} of 90\%, while reducing LLM calls to an average of 1.08. This alternative is preferable when activity diagrams are large and complex or when minimizing LLM costs is a priority.

Our full \textbf{replication package} is available online~\citep{replicationPackage}.

\section{Structural and Alignment Constraints}
\label{sec:formalization}
This section defines the activity-diagram notation and presents the structural and alignment constraints used to design \approach.  Our formalization matches prior definitions in the literature and the UML standard for activity-diagram syntax~\citep{Cook2017,NejatiS0BM16,maoz}.

\begin{definition}[\textbf{Activity Diagram~\citep{NejatiS0BM16}}]
\label{def:activitydiagram}
An activity diagram $\mathit{ad}$ is  a tuple $\langle \mathit{NL}, \mathit{TL}, \mathit{N}, \mathit{T} \rangle$, where $\mathit{NL}$ is a set of node labels; $\mathit{TL}$ is a set of transition labels; 
 $\mathit{N} = \{n_1, n_2, \dots, n_k\}$ is a set of nodes partitioned into four subsets:  action nodes ($\mathit{N}^a$), decision nodes ($\mathit{N}^d$), initial nodes ($\mathit{N}^i$), and  end nodes ($\mathit{N}^e$); and $\mathit{T}$ is a set of transitions that may be labelled or unlabelled: $\mathit{T} \subseteq  (\mathit{N}\times  \mathit{N}) \cup  (\mathit{N} \times TL\times  \mathit{N})$. 
Each node $\mathit{n} \in \mathit{N}$ is associated with a node label, denoted by $\mathit{label(n)} \in \mathit{NL}$. 
Labels of transitions originating from decision nodes represent guard conditions.  
\end{definition}

 \begin{figure}
	\centering
	\includegraphics[width=1.02\linewidth]{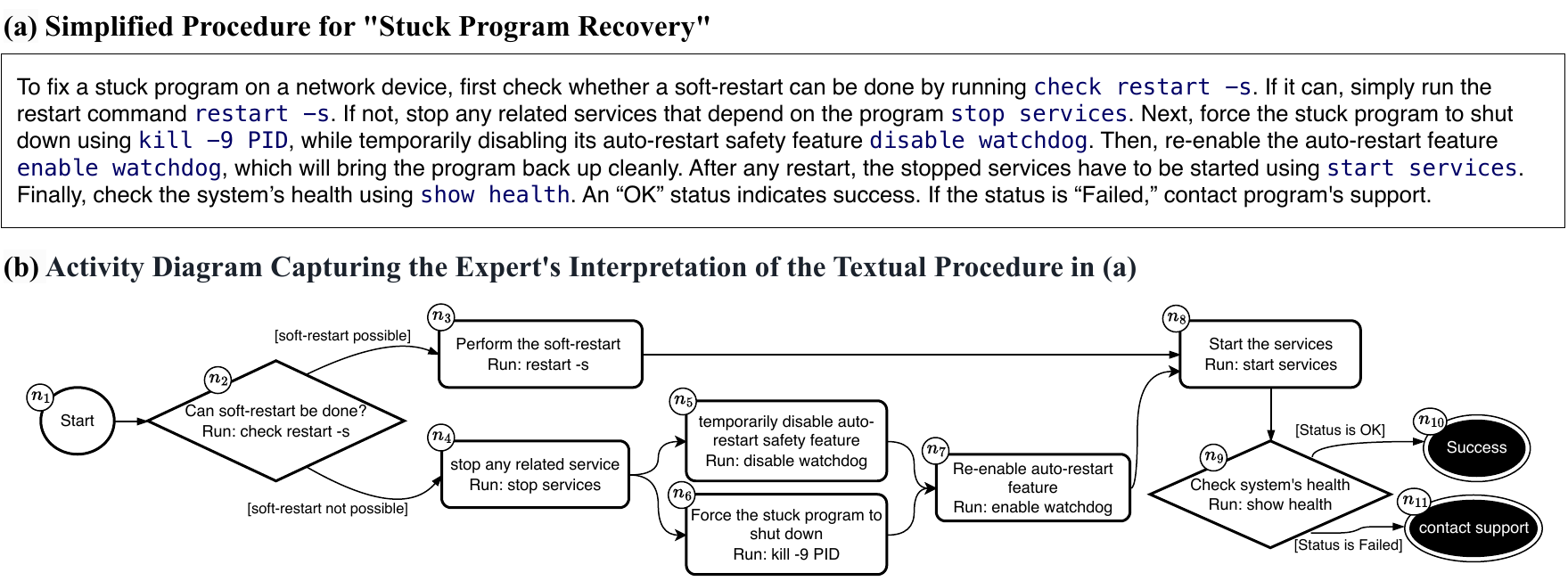} 
        \vspace*{-.05cm}
        \caption{Motivating example}
	\label{figure:example}
    \vspace*{-.45cm}
\end{figure}

For example, Figure~\ref{figure:example}(a) illustrates a simplified, step-by-step procedure -- adapted from Ciena’s configuration documents -- for recovering a stuck program in a network system. Figure~\ref{figure:example}(b) shows an activity diagram created by an expert, based on their interpretation of the procedure in Figure~\ref{figure:example}(a). The process description and the corresponding activity diagram are adapted, anonymized, and simplified from Ciena's dataset. The activity  diagram in Figure~\ref{figure:example}(b) includes one initial node ($n_1$), two decision nodes ($n_2$ and $n_9$), two end nodes ($n_{10}$ and $n_{11}$) and six action nodes ($n_3, n_4,\cdots, n_8)$. Nodes $n_5$ and $n_6$ are parallel nodes. The transitions from decision nodes are labelled with guard conditions.

\begin{figure}
	\centering
	\includegraphics[width=\linewidth]{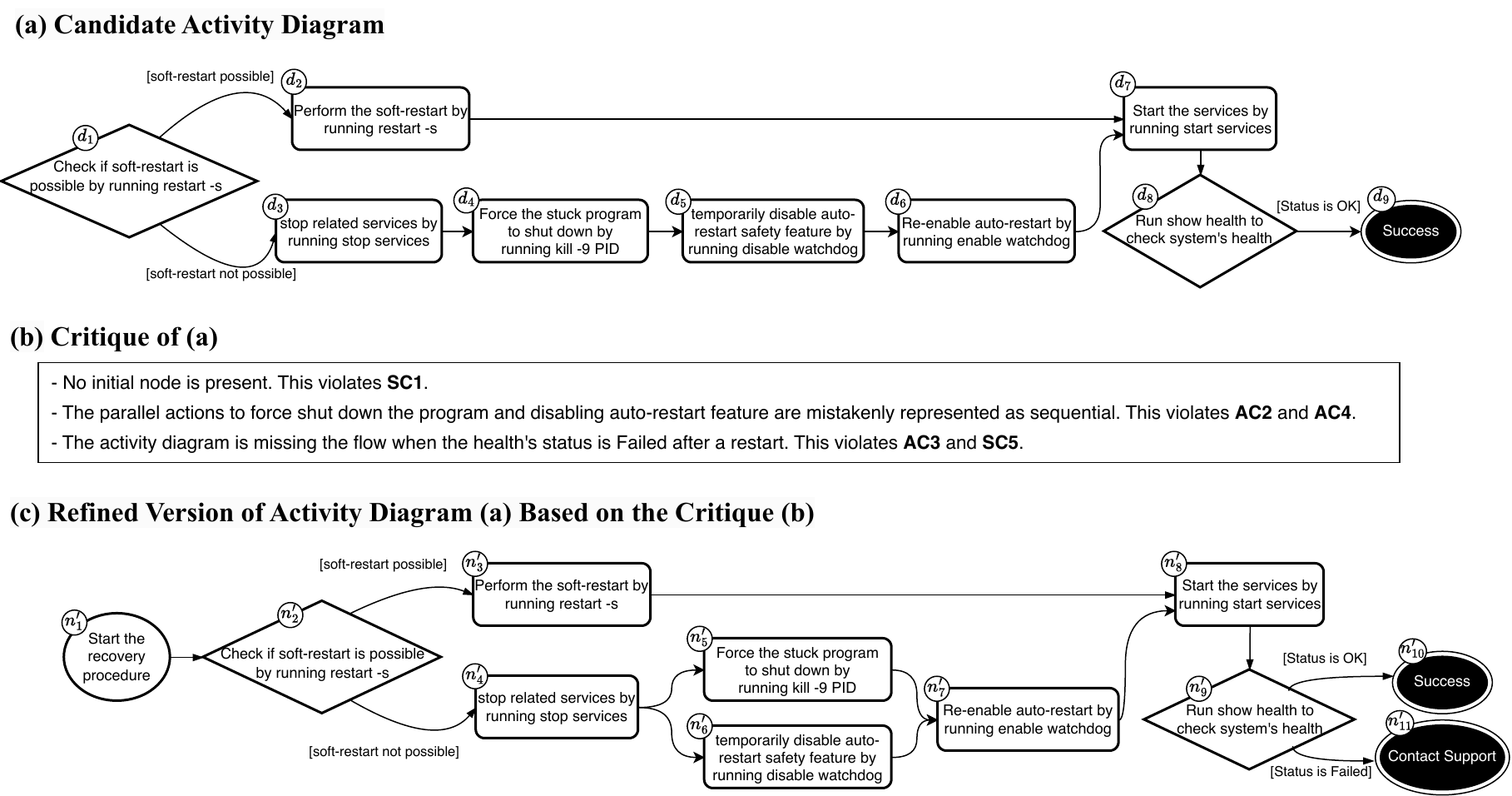} 
    \caption{Example of one execution iteration of \approach: (a) a candidate activity diagram generated from the process description in Figure~\ref{figure:example}(a); (b) a critique of (a), evaluating the candidate against the structural and alignment constraints in Table~\ref{table:prompt_constraint} and identifying any violations; (c) a refined version of (a), revised based on the critique in (b).}
    \label{figure:examp} 
    \vspace*{-.4cm}
\end{figure}

We use two sets of constraints for extracting activity diagrams from text, shown in Table~\ref{table:prompt_constraint}.
The first set, which we refer to as \emph{structural constraints}, ensures the syntactic correctness of the generated activity diagrams. 
Specifically, the structural constraints SC1 to SC6 require that an activity diagram contain exactly one initial node, at least one end node, and a valid number of incoming and outgoing transitions for both the initial and end nodes. These constraints also state that decision nodes must have multiple outgoing transitions, each labelled with a guard condition, and that every node must be reachable from the initial node to ensure full connectivity.
Table~\ref{table:prompt_constraint} presents textual descriptions of the structural constraints, along with formalizations of these constraints using the notation introduced in Definition~\ref{def:activitydiagram}. All elements and notation used in Table~\ref{table:prompt_constraint} are based on Definition~\ref{def:activitydiagram}, except for the predicate $\mathit{path}(x, y)$, defined for a pair of nodes $x$ and $y$. This predicate holds when there is a path from node $x$ to node $y$. Its formal definition, based on the notation in Definition~\ref{def:activitydiagram}, is provided in Appendix~\ref{ap:path}.
For example, the activity diagram in Figure~\ref{figure:examp}(a) has two structural flaws: (1)~it lacks an initial node, violating SC1, and (2)~the decision node ``Run show health to check system's health'' has only one outgoing transition, violating SC5.

\begin{table}[t]
\caption{Structural and alignment constraints used in \approach's prompts. The structural constraints are derived from the UML 2.5.1 specification~\citep{Cook2017} to ensure syntactic correctness of the generated activity diagrams. The alignment constraints are derived from an analysis of UML semantics for sequential flows, decision nodes, and parallelism in activity diagrams~\citep{Cook2017}, and from a study of how existing LLM-based techniques address semantic alignment\citep{DBLP:conf/refsq/Herwanto24, DBLP:conf/models/YangCCMV24}.}  \label{table:prompt_constraint}
  \vspace*{-.3cm}
\begin{center}
\scalebox{0.85}{\begin{tabular}{p{\linewidth}}
\hline
\multicolumn{1}{c}{\textbf{Structural Constraints}} \\ \hline \\
{\bf $\mathbf{SC1}$.} An activity diagram must have exactly one initial node. \\
$|\mathit{N}^i| = 1$
\\ \vspace*{3pt}
{\bf $\mathbf{SC2}$.} An activity diagram must have at least one end node. \\
$|\mathit{N}^e| \geq 1$.
\\  \vspace*{3pt}
{\bf $\mathbf{SC3}$.} The initial node must have no incoming transitions.  \\ $\forall n \in \mathit{N}^i : \nexists x \in \mathit{N} : ( (x, n) \in \mathit{T} \lor \exists l \in \mathit{TL} : (x, l, n) \in \mathit{T} )$ \\  \vspace*{3pt}
{\bf $\mathbf{SC4}$.} End nodes must have no outgoing transitions. \\ $\forall n \in \mathit{N}^e : \nexists y \in \mathit{N} : ( (n, y) \in \mathit{T} \lor \exists l \in \mathit{TL} : (n, l, y) \in \mathit{T} )$ \\  \vspace*{3pt}
{\bf $\mathbf{SC5}$.} Each decision node must have at least two outgoing transitions, each labelled by a guard condition. \\ $\forall d \in \mathit{N}^d : |\{ y \in \mathit{N} \mid \exists l \in \mathit{TL} : (d, l, y) \in \mathit{T} \}| \geq 2 \land \nexists y \in \mathit{N} : (d, y) \in \mathit{T}$ \\  \vspace*{3pt}
{\bf $\mathbf{SC6}$.} An activity diagram must be fully connected so that every node is reachable from the initial node. \\ $\forall n \in \mathit{N} \setminus \mathit{N}^i : \exists y \in \mathit{N}^i : \mathit{path}(y, n)$ \\ \\ \hline
\multicolumn{1}{c}{\textbf{Alignment Constraints}} \\ \hline \\
{\bf $\mathbf{AC1}$.} Action and transition labels in the activity diagram must be consistent with and accurately reflect the process description. \\  \vspace*{3pt}
{\bf $\mathbf{AC2}$.} The sequence of actions and transitions must accurately represent the order of actions and their triggers described in the process description. \\  \vspace*{3pt}
{\bf $\mathbf{AC3}$.} All possible action flows described in the process description must be represented in the activity diagram. The activity diagram must not introduce any actions or transitions that are not present in the process description. \\  \vspace*{3pt}
{\bf $\mathbf{AC4}$.} Concurrency occurs when actions happen simultaneously and is modelled using multiple parallel flows originating from a single action node. The parallel flows may synchronize into a single flow after some steps. \\  \vspace*{3pt}
{\bf $\mathbf{AC5}$.} Only procedural steps from the process description should be incorporated into the activity diagram. Examples, explanatory text, and commentary should be excluded. \\ \\ \hline
\end{tabular}%
}
\end{center}
\vspace*{-.6cm}
\end{table}

We formulated the structural constraints by first inspecting the UML 2.5.1 specification~\citep{Cook2017} and identifying every normative statement related to the well-formedness of activity diagrams. We then refined the constraints by cross-checking them against the OMG standard for UML's formal execution semantics~\citep{Cook2017}. This process yielded 17 constraints. Of these, ten constraints concern syntactic variations or visual notations in UML activity diagrams that do not require explicit enforcement in our work since our formalism unifies different UML syntactic elements; for example, we treat forks and merges uniformly as action nodes (Definition \ref{def:activitydiagram}). One rule -- namely, the exclusion of dangling transitions -- is intrinsically enforced in our generation process, as our encoding (see Section~\ref{sec:approach}) does not admit transitions without both source and target nodes. The remaining six constraints are listed in Table~\ref{table:prompt_constraint}. The full list of 17 constraints is available in Appendix~\ref{ap:cons}.

The second set, which we refer to as \emph{alignment constraints}, ensures that the activity diagrams accurately reflect their textual descriptions. To identify these constraints, we followed a process similar to that for structural constraints, reviewing UML specifications~\citep{Cook2017} on activity-diagram semantics to capture the meaning of the three main control-flow concepts in activity diagrams: sequential flows, decision nodes, and parallelism. We then assessed how the prompts used in existing LLM-based software model generation approaches~\citep{DBLP:conf/refsq/Herwanto24, DBLP:conf/models/YangCCMV24} guide the mapping of text to these three main control-flow constructs. This process resulted in three alignment constraints -- AC1, AC2, and AC3 -- listed in Table~\ref{table:prompt_constraint}.  Specifically, AC1 states that the generated action and transition labels must be consistent with the input text.  AC2 is concerned with the correctness of the ordering of actions and transitions in the generated activity diagrams compared to the input text.  AC3 relates to the completeness and correctness of the generated activity diagrams, ensuring they include all action flows described in the input text without introducing any actions or transitions not present in the process description.
While we did not find prompts related to parallelism in the literature~\citep{DBLP:conf/refsq/Herwanto24, DBLP:conf/models/YangCCMV24}, due to the importance of parallelism and its presence in our datasets, we introduced AC4 to explicitly define concurrency and to guide how actions that occur simultaneously, according to the input text, should be represented in the generated activity diagrams. 
Finally, specification documents often contain additional content beyond strict requirements and procedures, including explanatory notes, clarifications, background information, and examples. This supplementary material should be excluded during activity diagram generation. Prior requirements engineering research on specification-content classification has addressed the identification and removal of such supplementary content, typically using custom learning techniques~\citep{10.1007/978-3-031-29786-1_8,DBLP:journals/ese/AbualhaijaASBT20}. LLMs simplify this process considerably, as simple prompts can now instruct an LLM to omit such supplementary material. Accordingly, we introduced AC5 to direct the LLM to exclude such additional content, if present.

For example, suppose the activity diagram in Figure~\ref{figure:examp}(a) is generated by an LLM prompted to capture the textual description in Figure~\ref{figure:example}(a). In this case, since LLMs can sometimes misinterpret the semantic nuances of the source text, this activity diagram contains three alignment flaws: (1)~the order of actions in Figure~\ref{figure:examp}(a) is inconsistent with the process description, as two action nodes -- ``Force the stuck program to shut down by running kill -9 PID'' and ``temporarily disable auto-restart safety feature by running disable watchdog'' -- are connected sequentially even though the description specifies that they should be performed in parallel, thereby violating AC2; (2)~the representation of these actions also violates AC4, due to the incorrect specification of concurrency; and (3)~the flow related to when the health status is ``Failed'' at the end of the procedure is not captured in the activity diagram, thus violating AC3.

\section{Activity Diagram Generation and Refinement}
\label{sec:approach}
\begin{figure}[t]
\centering
    \includegraphics[width=0.65\linewidth]{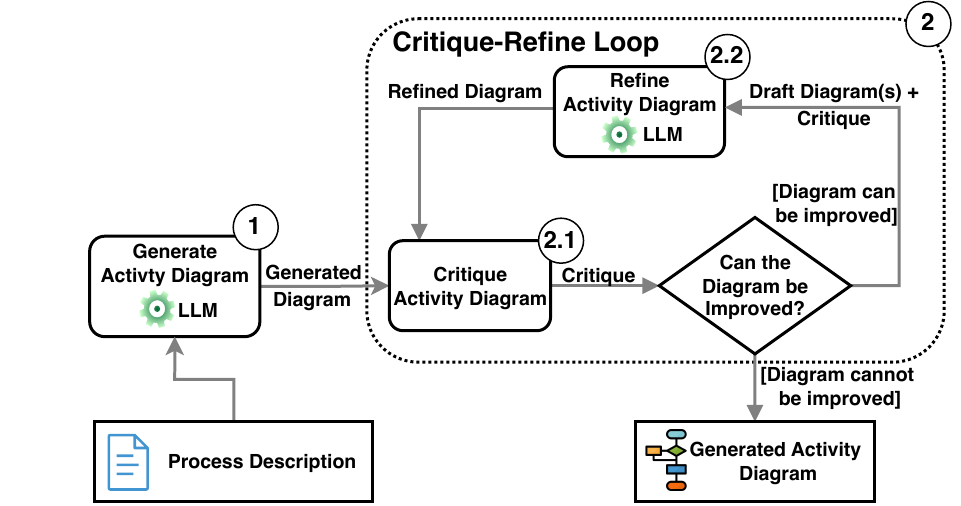} 
    \caption{An overview of the architecture of \approach.}
    \label{figure:gen_pipeline}
    \vspace*{-.2cm}
\end{figure}

Figure~\ref{figure:gen_pipeline} provides an overview of \approach, which takes a textual process description as input and transforms it into an activity diagram. Briefly, the pipeline starts by generating a candidate activity diagram in the generation step (Step~1). The candidate activity diagram is then assessed by the critique step (Step~2.1) and refined in the refinement step (Step~2.2) based on the critique. The structural and alignment constraints from Table~\ref{table:prompt_constraint} are included into the prompts for \emph{both} the generation and refinement steps, (Steps~1 and~2.2 in Figure~\ref{figure:gen_pipeline}). In addition, the critique step (Step~2.1) checks the compliance of the candidate activity diagram against the structural and alignment constraints from Table~\ref{table:prompt_constraint}. If no structural or alignment issues are identified in the critique, the loop terminates.

\begin{table}
    \captionof{table}{Elements included in the prompts at each step of the \approach\ pipeline shown in Figure~\ref{figure:gen_pipeline}.}
    \label{table:prompt_outline}
    \centering
    \scalebox{0.75}{\begin{tabular}{@{}llccc@{}}
    \toprule
    & & \textbf{\begin{tabular}[c]{@{}c@{}}Generate \\ Activity Diagram\end{tabular}} & \textbf{\begin{tabular}[c]{@{}c@{}}Critique \\ Activity Diagram\end{tabular}} & \textbf{\begin{tabular}[c]{@{}c@{}}Refine \\ Activity Diagram\end{tabular}} \\ 
    \midrule
    \textbf{(I)}   & Role Definition                   & \cmark & \cmark & \cmark \\
    \textbf{(II)}  & Constraints from Table~\ref{table:prompt_constraint} & \cmark & \cmark & \cmark \\
    \textbf{(III)} & Process Description               & \cmark & \cmark & \cmark \\
    \textbf{(IV)}  & Output Format Definition          & \cmark &        & \cmark \\
    \textbf{(V)}   & One-shot Example                  & \cmark &        & \cmark \\
    \textbf{(VI)}  & Generated Candidate Activity Diagram           &        & \cmark & \cmark \\
    \textbf{(VII)} & History of Candidate Activity Diagrams         &        &        & \cmark \\
    \textbf{(VIII)}& Critique from Step 2.1 of Figure~\ref{figure:gen_pipeline}            &        &        & \cmark \\
    \bottomrule
    \end{tabular}}
\end{table}

As discussed in Section~\ref{sec:intro}, the generation step (Step~1) and refinement step (Step~2.2) of \approach\ rely exclusively on LLMs. The critique step (Step~2.1), nevertheless,  while still requiring LLMs to check alignment constraints, can use either LLMs or deterministic algorithms for structural constraints. Depending on how the critique step is implemented and whether the critique-refine loop (Step~2 in Figure~\ref{figure:gen_pipeline}) is included, we develop five variants of \approach.  Below, we first outline \approach's prompts and then present the \approach\ variants.

\textbf{Prompts.} Table~\ref{table:prompt_outline} summarizes the elements used in the prompts for the generation and refinement steps (Step~1 and Step~2.2), as well as for the critique step (Step~2.1) when critique is LLM-based. The outline for these prompts is available in Appendix~\ref{ap:outlines}. The first three elements are common among the prompts for all the three steps. The \emph{role definition} element indicates the role that the LLM takes in each step, e.g., a generator in Step 1 and a refiner in Step~2.2. The \emph{constraints} element lists the constraints from Table~\ref{table:prompt_constraint}. For Step~1 and Step~2.2, both structural constraints and alignment constraints are always included. For the critique step (Step~2.1), the prompt includes only the constraints that are assessed by an LLM -- whether structural, alignment, or both. Constraints that are checked algorithmically, or not checked at all, are not included in the critique prompt.

The \emph{output format definition} element specifies the desired format of the generated activity diagram. In our work, we represent activity diagrams in an intermediate Comma Separated Values (CSV) format. 
This CSV representation is aligned with Definition~\ref{def:activitydiagram} and preserves the abstract syntax of an activity diagram -- its nodes, node types, and transitions -- as well as the control-flow semantics induced by transitions.  We deliberately adopt this lightweight intermediate representation instead of requiring the LLM to directly generate UML extensible markup language metadata interchange (XMI)~\citep{Cook2017,fUML}. This choice is motivated by current limitations of LLMs: complex, highly constrained serialization formats such as XMI are more susceptible to syntactic hallucination and malformed output~\citep{DBLP:journals/ijgi/LiZWJZ24}. By contrast, the CSV representation provides a compact and less error-prone representation format while retaining all the information required for transformation into standard Model-Driven Engineering (MDE)-compatible formats; these transformations are described in Section~\ref{sec:tools}.
In our intermediate CSV encoding, each node in the activity diagram is represented by at least one row in the CSV if it has one or no predecessors. Nodes with multiple predecessors are represented by multiple rows -- one for each incoming transition from a distinct predecessor. Specifically, each node $n$ with a predecessor $s$ such that $s \xrightarrow{a} n$ is represented by the following row in the CSV output: ``\texttt{ID of $n$, type of $n$, ID of $s$, transition label $a$;}''. That is, the row corresponding to $s \xrightarrow{a} n$ includes, respectively, the ID of $n$, its type -- either ``action'', ``decision'', ``initial'', or ``end'' -- the ID of the predecessor of $n$, and the transition label between $s$ and $n$. The row for the initial node uses empty strings for the predecessor node ID and the transition label fields. Similarly, the transition label field is empty in the rows related to unlabelled transitions.  For example, node $n_4$ in Figure~\ref{figure:example}(b) is represented in the CSV file as: ``$n_4, \text{action}, n_2, \text{[soft-restart not possible]}$;'', whereas node $n_7$ is represented by two rows:``$n_7, \text{action}, n_5, \epsilon$;'' and ``$n_7, \text{action}, n_6, \epsilon$;''.

The \emph{one-shot example} provides a complete example of a valid mapping from a process description to its corresponding activity diagram encoded in the CSV format described above. The \emph{output format definition} and \emph{one-shot example} elements appear only in the prompts for Step~1 and Step~2.2, as these two steps output a generated activity diagram. The \emph{generated candidate activity diagram} element indicates the current activity diagram version and appears only in the prompts for Step~2.1 and Step~2.2. 
The \emph{history of candidate activity diagrams} is provided only to Step~2.2 and contains the previous candidate activity diagrams that were rejected by the critique step. Finally, the \emph{critique} is also provided only to Step~2.2 and includes the critique from  Step~2.1 on the latest candidate.

Figure~\ref{figure:examp} shows the step-by-step outputs of one iteration of \approach\ applied to the process description in Figure~\ref{figure:example}(a). First, Step~1 generates the  activity diagram shown in Figure~\ref{figure:examp}(a). As discussed in Section~\ref{sec:formalization}, this activity diagram violates two structural constraints and three alignment constraints. In Step~2.1, the critique step assesses the candidate activity diagram and generates the critique shown in Figure~\ref{figure:examp}(b), which summarizes the violated structural and alignment constraints. This feedback is then included in the prompt for Step~2.2, leading to the refined activity diagram in Figure~\ref{figure:examp}(c).

\textbf{\approach\ variants.} We develop five variants of the \approach\ pipeline (Table~\ref{table:approach_variants}). Each variant uses an LLM to generate the initial candidate activity diagram. The refinement and alignment-checking steps, when present, also use an LLM, while structural constraint checking in the critique step can be done either by an LLM or algorithmically.

\begin{table}[t]
\centering
\caption{\approach\ variants.  
Each variant, other than Baseline, is named using the format \approach-[\emph{Structure}]-[\emph{Alignment}], where  \emph{[Structure]} and \emph{[Alignment]} indicate the methods used to critique structural and alignment constraints, respectively. Possible values include \emph{LLM} (for LLM-based methods), and \emph{NA} (when the alignment constraints are not applied) for alignment constraint checking, and \emph{LLM} and \emph{Alg} (for algorithmic methods) for structural constraints checking. 
We assess these variants in Section~\ref{sec:res} based on different instances of instruction-following and reasoning-based LLMs} 
\label{table:approach_variants}
\scalebox{.9}{\begin{tabular}{lcll}
\toprule
\textbf{Variant} & \textbf{Refinement} & \textbf{Structural Checks} & \textbf{Alignment Checks} \\ 
&  & \textbf{in Critique} & \textbf{in Critique}\\ 
\midrule
Baseline & \xmark & \xmark & \xmark \\
\approachAlignmentLLMStrucutLLM & \cmark (LLM) & \cmark (LLM) & \cmark (LLM) \\
\approachAlignmentLLMStrucutFormal & \cmark (LLM) &  \cmark  (Algorithmically) & \cmark (LLM) \\
\approachStrucutLLMOnly & \cmark  (LLM)  &  \cmark  (LLM) & \xmark \\
\approachStrucutFormalOnly & \cmark (LLM)  & \cmark (Algorithmically) & \xmark \\
\bottomrule
\end{tabular}}
\vspace*{-.2cm}
\end{table}

We call the variant that excludes the critique-refine loop in Figure~\ref{figure:gen_pipeline} and only includes the generation step (Step~1 in Figure~\ref{figure:gen_pipeline}) the \emph{Baseline}. The Baseline disentangles the impact of the critique-refine loop. We choose to refer to this variant as Baseline because it follows the prevailing state-of-the-art approaches, which employ LLMs to generate complete behavioural models directly from a given textual description~\citep{DBLP:conf/re/FerrariAA24, DBLP:conf/refsq/Herwanto24, DBLP:conf/models/JahanHGRRRS24}. 

The other four variants that include the critique-refine loop in Figure~\ref{figure:gen_pipeline} are named using the format \approach-[Structure]-[Alignment], where [Structure] and [Alignment] indicate the methods used to check structural and alignment constraints, respectively. Specifically, [Structure] is replaced by LLM or Alg, denoting that structural checks are performed using an LLM or algorithmically, respectively. Similarly, [Alignment] is replaced with LLM or NA, indicating that alignment checks are LLM-based or not performed, respectively. 

More precisely, these four variants are  as follows: 
(1)~\emph{\approachAlignmentLLMStrucutLLM}: the critique uses an LLM to check both structural and alignment constraints. 
(2)~\emph{\approachAlignmentLLMStrucutFormal}:  the critique uses an LLM for alignment constraints, but structural constraints are checked algorithmically. 
(3)~\emph{\approachStrucutLLMOnly}: the critique uses an LLM for structural constraints only (no alignment checking). 
(4)~\emph{\approachStrucutFormalOnly}:  the critique checks structural constraints algorithmically (no alignment checking).

We use the four variants above for the following comparisons:
(1) We compare all four variants with Baseline to assess the impact of the critique-refine loop;
(2) We compare \approachAlignmentLLMStrucutFormal\ and \approachStrucutFormalOnly\ with \approachAlignmentLLMStrucutLLM\ and \approachStrucutLLMOnly\ to determine the impact of checking structural constraints algorithmically versus using an LLM; and
(3) We compare \approachAlignmentLLMStrucutFormal\ with \approachStrucutFormalOnly\  to assess the impact of alignment checking in the critique.

\section{Activity Diagram Matching}
\label{sec:matchings}
To evaluate LLM-generated activity diagrams, we must ensure that: (1) they are structurally sound, i.e.,  they satisfy the structural constraints in Table~\ref{table:prompt_constraint}; and (2) their behaviour aligns with the given textual process description. As described in Section~\ref{sec:intro}, we assess the latter by comparing the generated activity diagrams with ground-truth activity diagrams  using two alternative methods: (1) behavioural matching of activity diagrams, and (2) an LLM-based activity diagram matcher. We present both methods below. 

\subsection{Behavioural Matching of Activity Diagrams (B-Match)}
\label{sec:compare}
The behaviour of activity diagrams is formally characterized as the set of traces they produce~\citep{maoz}. A \emph{trace} is a sequence of nodes and transitions that can be executed from the initial node according to the flow of control. Prior research on comparing behavioural models based on their trace-based semantics relies on the process-algebraic notion of a \emph{simulation} preorder from a source model to a target model~\citep{10.1007/11691372_28}. In a simulation preorder, every step that the source model can perform must also be reproducible by the target model, which ensures that every execution trace of the source can be replicated by the target.

Traditional simulation relations require two traces to correspond only when their sequences of node labels and transition labels match exactly at every step. However, exact trace matching is too restrictive, as it fails to capture partial similarities or overlapping behaviours that arise naturally when models are generated independently (e.g., using LLMs versus manual model generation). To address this, we propose Algorithm~\ref{alg:alg}, the Activity Diagram Behavioural Matching (B-Match) algorithm, which builds on our earlier behavioural matching approach~\citep{NejatiSCEZ07,NejatiSCEZ12}. Our earlier approach relies on quantitative notions of simulation and heuristics that combine textual features, i.e.,  node and transition labels, and trace-based behaviours, enabling model matching even when the sets of traces are not exact matches. B-Match adapts this earlier behavioural matching method originally defined for state machines, to activity diagrams.

\begin{figure}
	\centering
	\includegraphics[width=\linewidth]{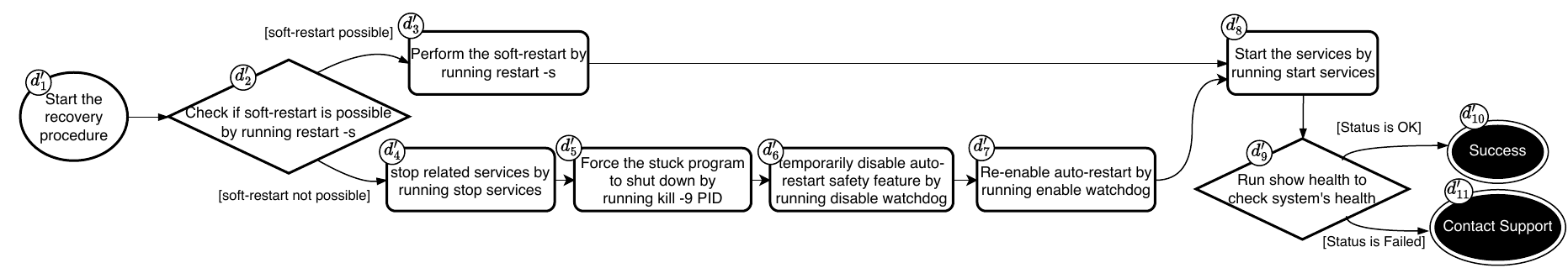} 
    \vspace*{-.3cm}
    \caption{A candidate activity diagram generated from the process description in Figure\ref{figure:example}(a) that is structurally sound based on the constraints outlined in Table\ref{table:prompt_constraint}, but is semantically misaligned with the textual process description.}
\label{figure:examp-struc-sound}
\vspace*{-.2cm}
\end{figure}

Before presenting the B-Match algorithm, we first illustrate why it is useful to be able to assess the correctness of LLM-generated activity diagrams based on their trace-based semantics.
Figure~\ref{figure:examp}(c) and Figure~\ref{figure:examp-struc-sound} show two LLM-generated activity diagrams. Both are structurally sound and generated from the process description in Figure~\ref{figure:example}(a). All nodes in Figure~\ref{figure:examp}(c) and Figure~\ref{figure:examp-struc-sound} have labels similar to those in the ground-truth activity diagram in Figure~\ref{figure:example}(b). Therefore, when evaluated only based on node matching against the ground truth, both activity diagrams are considered equally matching.

However, according to a trace-based similarity measure, the ground truth is behaviourally closer to the activity diagram in Figure~\ref{figure:examp}(c) than to the one in Figure~\ref{figure:examp-struc-sound}. 
That is, a larger number of nodes in Figure~\ref{figure:examp}(c) are correctly matched to nodes in the ground truth, with higher similarity scores, compared to those in Figure~\ref{figure:examp-struc-sound}.
In particular, the parallel nodes $n'_5$ and $n'_6$ in Figure~\ref{figure:examp}(c) are, respectively, behaviourally close to the parallel nodes $n_6$ and $n_5$ in the ground-truth activity diagram. 
We define B-Match in a way that ensures this behavioural similarity is taken into consideration. B-Match therefore identifies high-similarity matches for $(n_5, n'_6)$, $(n_6, n'_5)$, and subsequently matches the successors of $n_6$ and $n_5$ with those of $n'_5$ and $n'_6$  as being highly similar. In contrast, B-Match does not recognize pairs such as $(n_5, d'_6)$ and $(n_7, d'_7)$ due to the differing trace-based behaviour of $d'_6$ and $d'_7$ in Figure~\ref{figure:examp-struc-sound} compared to $n_5$ and $n_7$ in the ground truth.  Therefore, the activity diagram in Figure~\ref{figure:examp-struc-sound} yields fewer node matches with the ground truth than the activity diagram in Figure~\ref{figure:examp}(c).

\small
\begin{algorithm}[t]
\caption{Activity Diagram Behavioural Matching (B-Match)}
\label{alg:alg}
{\scriptsize
\begin{flushleft}
\textbf{Input} $\mathit{ad}$, $\mathit{ad'}$: Input activity diagrams: $\mathit{ad}$ is the source activity diagram, and  $\mathit{ad'}$ is the target activity diagram. \\
\textbf{Output} $\rho$ : A matching relating the nodes of  $\mathit{ad}$ to those of $\mathit{ad'}$\\

\end{flushleft}
\begin{algorithmic}[1]
\Algphase{Phase 0 - Initialization of Auxiliary Variables}

\State $\mathit{queue} \gets \emptyset$; \Comment{\ Queue of candidate node pairs $(\mathit{n}, \mathit{n'})$ to be checked}
\State $\rho \gets \emptyset$; \Comment{\ Set of matched node pairs $(\mathit{n}, \mathit{n'})$ to track visited pairs and prevent revisiting pairs and entering cycles}
\Algphase{Phase 1 - Root Matching}
    \State $\mathit{i} \gets \mathit{ad}.N^i$; $\mathit{i'} \gets \mathit{ad'}.N^i$ 
    \Comment{ Get the initial node of $\mathit{ad}$ and $\mathit{ad}$'}
    \State Enqueue $(\mathit{i}, \mathit{i'})$ to $\mathit{queue}$ \Comment{Add the pair $(\mathit{i}, \mathit{i'})$ to the queue, populating the queue for BFS traversal}

\Algphase{Phase 2 - Breadth First Search (BFS) Traversal and Node Matching}
\While{$\mathit{queue}$ is not empty}
    \State $(\mathit{n}, \mathit{n'}) \gets$ dequeue from $\mathit{queue}$
    \If{$(\mathit{n}, \mathit{n'}) \in \rho$} \Comment{ If nodes $n$ and $n'$ are already matched there is no need to process this pair further}
        \State \textbf{continue}
    \EndIf
    \State Add $(\mathit{n}, \mathit{n'})$ to $\rho$ \Comment{ Match $n$ to $n'$}

    \For{every node $s$ such that $\mathit{n} \xrightarrow{\,a\,} \mathit{s}$} 
        \State $(\mathit{bestMatch}, \mathit{bestScore}) \gets (\mathit{null}, 0.0)$
        \For{every node $s'$ such that $\mathit{n'} \xrightarrow{\,b\,} \mathit{s'}$}
            \State $\mathit{score} \gets \operatorname{\texttt{simStep}}(\mathit{n} \xrightarrow{\,\mathit{a}\,} \mathit{s}, \mathit{n'} \xrightarrow{\,\mathit{b}\,} \mathit{s'})$ 
            \If{$\mathit{score} \geq \mathit{bestScore}$} 
                \State $(\mathit{bestMatch}, \mathit{bestScore}) \gets (\mathit{s'}, \mathit{score})$
            \EndIf
        \EndFor
        \If{$\mathit{bestMatch} \neq null$}
            \State Enqueue $(\mathit{s}, \mathit{bestMatch})$ to $\mathit{queue}$ \Comment{ Map successor $s$ of $n$ to its best-matching successor of $n'$.}
        \EndIf
    \EndFor
\EndWhile
\State \Return $\rho$

\end{algorithmic}}
\end{algorithm}
\normalsize

B-Match takes as input two structurally sound activity diagrams, $\mathit{ad}$ and $\mathit{ad}'$, and computes a matching $\rho \subseteq N \times N'$, where $N$ and $N'$ are the sets of nodes in $\mathit{ad}$ and $\mathit{ad}'$, respectively. 
B-Match identifies all pairs of nodes from $\mathit{ad}$ and $\mathit{ad}'$ that are matched based on the trace-based semantic similarity of the activity diagrams. Each tuple $(n, n') \in \rho$ indicates that node $n$ in $\mathit{ad}$ is matched to node $n'$ in $\mathit{ad}'$.

Algorithm \ref{alg:alg} relies on two similarity functions. The first, $\texttt{simLabel}$, measures similarity between the textual elements of activity diagrams (node and transition labels). The second, $\texttt{simStep}$, evaluates similarity between a step in $\mathit{ad}$, i.e., a node and its successor, with a corresponding step in $\mathit{ad'}$. We first describe these functions and then present the algorithm.

-- \textbf{Label similarity {\fontfamily{lmtt}\fontseries{b}\selectfont simLabel()}:} Let $a$ and $b$ be two node or transition labels. Traditional approaches rely on exact matching, scoring one for strict string equivalence between $a$ and $b$, and zero otherwise~\citep{maoz}. To accommodate the natural-language variations in LLM-generated content, we compute $\operatorname{\texttt{simLabel}}$ as the semantic similarity between the strings $a$ and $b$, using state-of-the-art string comparison methods and embedding models. Details of the embedding approach are provided in Section~\ref{sec:implementation}.

-- \textbf{Step similarity {\fontfamily{lmtt}\fontseries{b}\selectfont simStep()}:} Let $n$ and $n'$ be a node of $\mathit{ad}$ and $\mathit{ad'}$, respectively, such that $n$ is matched to $n'$. For any successor $s$ of $n$ such that $n \xrightarrow{\,a\,} s$, and for any successor $s'$ of $n'$ such that $n' \xrightarrow{\,b\,} s'$, we denote the similarity degree between these two pairs of transitions by  $\operatorname{\texttt{simStep}}(n \xrightarrow{\,a\,} s, n' \xrightarrow{\,b\,} s')$ and compute it as follows: 

\[
\scalebox{0.65}{$
\displaystyle
\operatorname{\texttt{simStep}}(n \xrightarrow{\,a\,} s, n' \xrightarrow{\,b\,} s') = 
\begin{cases} 
\operatorname{\texttt{simLabel}}(label(s), label(s')) & \text{if } a = b = \epsilon; \\[10pt]
\dfrac{\operatorname{\texttt{simLabel}}(label(s), label(s')) + \operatorname{\texttt{simLabel}}(a, b)}{2} & \text{otherwise}.
\end{cases}
$}
\]

Specifically, if both transition labels are absent, i.e., $a = b = \epsilon$, the similarity degree of the nodes is determined only by their labels. Otherwise, we take the average of the similarity between the successor node labels and the similarity between the transition labels. This ensures that, when matching the step $n \xrightarrow{\,a\,} s$ to the step $n' \xrightarrow{\,b\,} s'$, we give equal weight to the similarity of the transition labels and the successor-node labels. This equal weighting is consistent with established model-matching practices~\citep{NejatiSCEZ07,NejatiSCEZ12}. In activity diagrams, transition labels are often absent; when they are absent,  $\texttt{simStep}$  relies only on node-label similarity. When transition labels are present, they either capture guard conditions on outgoing edges from decision nodes, or, less commonly, additional clarifying information about the flow. Such information is semantically important because it helps determine under what conditions a particular action or branch is reached. For this reason, when both node and transition labels are available, we treat them as equally important components of the step’s meaning and assign them equal weight.
Note that, in the trace-based semantics of activity diagrams, a step $n \xrightarrow{\,a\,} s$ is encoded as \texttt{Label(n).a.Label(s)}. 
The \texttt{simStep} function is invoked only when the parent nodes $n$ and $n'$ have already been matched. Given the pairing of the parent nodes, it evaluates the similarity of the corresponding successor steps by considering the similarity between transition labels and the similarity between successor-node labels.

Having defined the two functions, we now introduce B-Match, which applies a Breadth-First Traversal (BFS) to the source activity diagram, evaluates node similarities with the target activity diagram, and identifies the pairs with the highest similarity scores.  B-Match begins with the initial nodes and iteratively matches each pair of nodes before examining their successors. This traversal strategy inherently guarantees that parent nodes are always matched before \texttt{simStep} evaluates the similarity of their successor steps. 
Note that BFS intrinsically avoids cycles by tracking visited node pairs, ensuring that each pair is explored only once. This is desirable in our work, since we want to avoid revisiting already visited node pairs.

Algorithm \ref{alg:alg} begins by pairing the initial nodes of the two activity diagrams. The resulting pair is then stored in a \emph{queue} (lines 3--4). Next, each pair $(n, n')$ of nodes in \emph{queue} is stored in $\rho$ if the pair $(n, n')$ is not already matched (lines 5--10). For any matched pair $(n, n')$, Algorithm~\ref{alg:alg} iteratively matches any successor $s$ of $n$ to the most similar successor $s'$ of $n'$ (lines 11--18). Newly matched nodes are then added to \emph{queue} for processing in the next iteration (lines 19--21). Once all pairs in \emph{queue} have been processed, the algorithm returns the matches in $\rho$ (line 24).

\subsection{LLM-Based Matching of Activity Diagrams (L-Match)}
\label{sec:calibration}
In this section, we describe our LLM-based matcher (L-Match for short) for comparing automatically generated activity diagrams with their ground truths. Similar to  B-Match (Section~\ref{sec:compare}),  L-Match takes two structurally sound activity diagrams as input and returns a node-level matching, where mapped nodes are judged by the LLM to have similar behaviour.

Table~\ref{table:intent} shows the criteria provided to L-Match. These criteria are developed based on the alignment constraints in Table~\ref{table:prompt_constraint}. 
As prior work has shown that the order of information presented to an LLM is important~\citep{DBLP:conf/naacl/PezeshkpourH24}, we provide L-Match with the criteria in Table~\ref{table:intent} in the same order as the table, so that textual correspondence is given priority over behavioural correspondence, which is in turn considered more important than structural correspondence.
The criteria in Table~\ref{table:intent}, along with a one-shot example and the expected output format, are provided as prompts to L-Match.  
The complete set of prompts is available in our replication package~\citep{replicationPackage}.

\begin{table}[t]
\centering
\caption{Textual, behavioural, and structural criteria provided to the LLM activity diagram matcher (L-Match).}
\label{table:intent}
\resizebox{\textwidth}{!}{%
\begin{tabular}{@{}lp{0.8\textwidth}@{}}
\toprule
\textbf{Criteria} & \textbf{Description} \\ \midrule
\textit{Textual Correspondence} &
Matched nodes should have labels with similar meanings, and their incoming and outgoing transitions should also have labels with similar meanings.\vspace*{5pt}\\
\textit{Behavioural Correspondence} &
Matched nodes should exhibit similar behaviour, meaning that some of their predecessors and successors should also be matched. The numbers of predecessors and successors do not need to be identical, as the input activity diagrams may have been specified at different levels of granularity.  \vspace*{5pt} \\ 
\textit{Structural Correspondence} &
Matched nodes should have the same type (e.g., action, decision, initial, or end node) to preserve the consistency of the activity diagram’s structure. \\ \bottomrule
\end{tabular}%
}
\end{table}

To evaluate the reliability of L-Match, we selected five activity diagrams from our public-domain dataset and from our partner Ciena’s dataset -- both datasets are introduced in Section~\ref{sec:datsets}. We asked domain experts -- two from Ciena for their dataset and two graduate students experienced in UML for the public-domain dataset -- to annotate node-level correspondences between the LLM-generated and ground-truth activity diagrams. 
To measure inter-rater agreement, we considered node-level matches identified by both experts as agreements, and matches found by only one expert as disagreements. 
The two annotators achieved Cohen's kappa values ($\kappa$) of 0.916 and 0.888 for the Ciena and public-domain datasets, respectively, indicating almost perfect agreement~\citep{cohen1960coefficient}.
Disagreements were subsequently discussed and resolved by both experts to establish a final consensus.
We then computed standard precision-recall metrics by comparing the expert-consensus and LLM-produced matchings as follows: a pair of nodes matched by both is a \emph{true positive (TP)}; a pair matched only by the LLM is a \emph{false positive (FP)}; and a pair matched only by the expert is a \emph{false negative (FN)}. Table~\ref{tab:LLM-expert} shows the precision, recall, and F1-score comparing the LLM-produced and expert-produced matchings for both Ciena and public-domain datasets. 
The high recall and precision values indicate that L-Match is highly correlated with expert judgment when comparing a generated activity diagram with its ground-truth counterpart.

In Section~\ref{sec:metrics}, we define metrics for evaluating the \textit{semantic correctness} and \textit{completeness} of LLM-generated activity diagrams, based on node matchings with ground-truth activity diagrams, which can be obtained using the L-Match or using our B-Match algorithm in Section~\ref{sec:compare}. 

\begin{table}[t]
\caption{Comparing the L-Match and expert-consensus matchings.}\label{tab:LLM-expert}
\centering
\scalebox{1}{
\begin{tabular}{lccc}
\hline
\textbf{Dataset} & \textbf{Precision} & \textbf{Recall} & \textbf{F1-score} \\
\hline
\textbf{Ciena's dataset (Industry*)} & 96.59\% & 95.60\% & 96.03\% \\
\textbf{Public-domain dataset (PAGED*)} & 90.36\% & 94.03\% & 91.40\% \\
\hline
\end{tabular}
}
\caption*{\textsuperscript{*} \footnotesize The Industry and \textsc{Paged} datasets are introduced in detail  in Section~\ref{sec:datsets}.}
\end{table}

\section{Evaluation Setup}
\label{sec:eval}
This section presents our research questions and provides a detailed description of our evaluation setup, including the datasets, metrics, and our experiment procedure.
We address four research questions, RQ1--RQ4, in our evaluation; each plays a distinct but complementary role. RQ1 concerns the evaluation methodology: it investigates whether LLMs can be used as reliable evaluators for generated activity diagrams by comparing LLM-based evaluation with a behavioural matching method. Specifically, RQ1 examines whether the two methods produce consistent conclusions when comparing two diagrams.
RQ2 to RQ4 then use the metrics from the evaluation methods whose consistency is established in RQ1 to evaluate different components of \approach. RQ2 studies the impact of the critique-refine loop on the quality of the generated activity diagrams. RQ3 examines the effect of different structural-checking mechanisms. And, RQ4 studies the role of semantic alignment checking. The presentation and analysis of the results for our research questions are provided in Section~\ref{sec:res}.
Throughout this section and Section~\ref{sec:res}, every mention of the \emph{refinement loop} should be understood as referring to the critique-refine loop of \approach\ shown in Figure~\ref{figure:gen_pipeline}.

Having outlined the role of each research question above, we now state them precisely.

\textbf{RQ1 (Evaluation Consistency)} \emph{How consistent are the evaluation results obtained based on our behavioural matching algorithm versus an LLM-based activity diagram matcher?}  This research question examines the consistency and agreement between two evaluation methods for activity diagrams: B-Match (Section~\ref{sec:compare}), and L-Match (Section~\ref{sec:calibration}).

\textbf{RQ2 (Impact of the Refinement Loop)} \emph{How does including the refinement loop versus excluding it affect the quality and cost of the generated activity diagrams?} We compare the Baseline variant versus the other four variants  in Table~\ref{table:approach_variants}  to assess the impact of the refinement loop.

\textbf{RQ3 (LLM-based  vs. Algorithmic Structural Checking)} \emph{How does using LLM-based structural checking versus algorithmic structural checking within the refinement loop  of \approach\ affect the quality and  cost of the generated activity diagrams?} We compare the variants that perform structural checking with an LLM to the ones that perform structural checking algorithmically.

\textbf{RQ4 (Impact of Alignment Checking)} \emph{How does including the alignment checking versus excluding it affect the quality and cost of the generated activity  diagrams?} With this research question, we investigate the impact of alignment checking, which is performed by LLMs, on the generated activity diagrams.

\subsection{Datasets}
\label{sec:datsets}
Our evaluation is based on two datasets: (1)~our industrial dataset from Ciena, referred to as Industry, which consists of procedural documents specifying configuration steps for hardware and software product families, along with ground-truth activity diagrams created by domain experts; and (2)~a publicly available dataset, PAGED~\citep{du-etal-2024-paged}, which contains textual process descriptions and their corresponding ground-truth activity diagrams, constructed from a collection of procedural graphs. The original PAGED dataset contains 3394 pairs of activity diagrams and process descriptions. To keep the cost, time, and resources required for our empirical evaluation manageable, we randomly selected 200 entries from it and refer to this subset as the \textsc{Paged} dataset in our evaluation.

Table~\ref{table:dataset} provides summary statistics for our two datasets. The Industry dataset consists of 20 process description documents paired with ground-truth activity diagrams containing an average of 29.15 nodes and 31.35 transitions, with process descriptions averaging 5187.8 characters and 1411.5 tokens, respectively. The \textsc{Paged} dataset includes 200 pairs of activity diagrams and their corresponding process descriptions, whose ground-truth activity diagrams contain on average 10.5 nodes and 10.9 transitions, with process descriptions averaging 641.9 characters and 132.9 tokens.
On average, the process descriptions in the Industry dataset are considerably longer than those in \textsc{Paged}, containing 8.1 times more characters and 10.6 times more tokens.

\begin{table}[t]
\centering
\caption{Summary statistics for our two datasets.}
\label{table:dataset}
\scalebox{0.8}{\begin{tabular}{@{}lcc@{}}
\toprule
~ & \textbf{Industry} & \textbf{\textsc{Paged}} \\
\midrule
Number of Documents/Diagrams & 20 & 200 \\
Average number of characters in the textual process descriptions  &  5187.8 & 641.9 \\
Average number of tokens in the textual process descriptions  &  1411.5 & 132.9 \\
Average nodes per ground-truth activity diagram & 29.15 & 10.5 \\
Average transitions per ground-truth activity diagram & 31.35 & 10.9 \\
\bottomrule
\end{tabular}}

\end{table}

\subsection{Evaluation Metrics} 
\label{sec:metrics}
We use four metrics in our evaluation: \textit{structural consistency}, \textit{semantic correctness}, \textit{semantic completeness}, and \textit{cost}. The first three assess the \emph{quality} of generated activity diagrams. \textit{Structural consistency} is evaluated by constraint violations aggregated across all activity diagrams, whereas \textit{semantic correctness} and \textit{semantic completeness} are evaluated per activity diagram against its ground truth. The fourth metric, \textit{cost}, complements these by measuring the efficiency of generation. The metrics are defined below: 

\textbf{Structural consistency} is the number of activity diagrams generated by each variant that violate at least one structural constraint from Table~\ref{table:prompt_constraint}. 

\textbf{Semantic Correctness and Completeness.} \textit{Semantic correctness} and \textit{completeness} are computed based on a matching between the similar nodes of the LLM-generated and ground-truth activity diagrams, obtained using either B-Match (Section~\ref{sec:compare}) or L-Match (Section~\ref{sec:calibration}). The \textit{semantic correctness} and \textit{completeness} metrics are then defined based on the node matching as follows:

\emph{Semantic correctness.} We consider an LLM-generated activity diagram correct if every one of its nodes is matched to some node in the ground-truth activity diagram. In a matching between $\mathit{ad}_{\mathit{llm}}$ and $\mathit{ad}_g$, a matched node pair \((n, n')\), indicates that \(n \in \mathit{ad}_{\mathit{llm}}.N\) and \(n' \in \mathit{ad}_g.N\). 
Given a matching \(\rho\), \textit{correctness} is defined as the proportion of nodes in the generated activity diagram that are matched: $\mathit{cor} = \frac{|A|}{|\mathit{ad}_{\mathit{llm}}.N|}$,
where \(A = \{ n \in \mathit{ad}_{\mathit{llm}}.N \mid \exists n' \in \mathit{ad}_g.N: (n,n') \in \rho \}\).

\emph{Semantic completeness.} We consider an LLM-generated activity diagram complete if all the nodes in the ground-truth activity diagram are matched to some node in the LLM-generated activity diagram. Here, in a matching between nodes of \(\mathit{ad}_g\) and \(\mathit{ad}_{\mathit{llm}}\), a matched node pair \((n, n')\), indicates \(n \in \mathit{ad}_{\mathit{g}}.N\) and \(n' \in \mathit{ad}_{\mathit{llm}}.N\). 
Given a matching \(\rho\), \textit{semantic completeness} is defined as the proportion of ground-truth nodes that are matched:$\mathit{com} = \frac{|B|}{|\mathit{ad}_g.N|}$, where \(B = \{ n \in \mathit{ad}_g.N \mid \exists n' \in \mathit{ad}_{\mathit{llm}}.N: (n,n') \in \rho\}\).

In addition to B-Match and L-Match, we computed \textit{semantic correctness} and \textit{completeness} using a traditional \emph{exact matching algorithm} discussed as part of the \texttt{simLabel} function definition in Section~\ref{sec:compare}. The results indicate that exact matching is too brittle for comparing LLM-generated activity diagrams with ground truths, as small textual differences in node and transition labels (e.g., ``start the process'' versus ``begin the process'') can lead to the diagrams being considered entirely dissimilar. Specifically, on the Industry dataset, the average \textit{semantic correctness} and \textit{completeness} scores obtained with exact matching are zero across all LLMs and \approach\ variants, while on the \textsc{Paged} dataset, the corresponding averages are below $1\%$. For transparency, detailed exact-matching results are provided in Appendix~\ref{app:exact_match}.

\textbf{Cost.} To assess cost, we report: (1)~the number of LLM calls made for activity-diagram generation, as well as for critiquing and refining generated activity diagrams; and (2)~the average number of tokens required per LLM call for activity-diagram generation, including the number of input tokens  (the average number of tokens in the prompts provided to the LLM), the number of output tokens (the average number of tokens generated by the LLM), and the number of reasoning tokens (the average number of tokens consumed by reasoning-based LLMs during their internal chain-of-thought reasoning).

\subsection{Implementation}
\label{sec:implementation}
Our implementation supports all variants of \approach, as presented in Table~\ref{table:approach_variants}, and is applied to both the Industry dataset from Ciena and the public-domain \textsc{Paged} dataset. To ensure confidentiality and respect privacy requirements at Ciena, and because the data cannot leave partner systems, we restrict the Industry dataset experiments to LLMs approved for use under the company's business subscriptions. In contrast, experiments on the public-domain \textsc{Paged} dataset are not subject to these company policies and can therefore be conducted with a broader range of LLMs.

We use OpenAI's GPT-4.1 Mini~\citep{openai2024gpt4ocard} and O4 Mini~\citep{jaech2024openai} via API access for both datasets. For the \textsc{Paged} dataset, we also apply the open-source DeepSeek-R1-Distill-Llama-70B~\citep{guo2025deepseek}, hosted locally using Ollama~\citep{ollama2023}. GPT-4.1 Mini is an instruction-following LLM that generates outputs based on explicit instruction provided in the prompt, while O4 Mini and DeepSeek-R1-Distill-Llama-70B use an internal chain-of-thought reasoning step to generate more accurate and human-aligned responses~\citep{jaech2024openai,guo2025deepseek}.
For L-Match (Section~\ref{sec:calibration}), we use OpenAI's O4 Mini~\citep{jaech2024openai} via API access for both datasets.

We use embedding-based string comparison in B-Match (Section \ref{sec:compare}) by applying the Sentence Transformers library (v4.1.0) \citep{reimers-2019-sentence-bert} with the Alibaba-NLP/gte-base-en-v1.5 embedding model, which provides high-quality and efficient semantic representations.  These embeddings are used to project strings into a shared vector space, where cosine similarity is computed between the resulting vectors~\citep{zhang2024mgte, li2023gte}. Our full implementation is available online~\citep{replicationPackage}.

\subsection{Experimental Procedure}
\label{subsec:expproc}
We applied all variants of \approach\ from Table~\ref{table:approach_variants} to both datasets. OpenAI-based experiments were executed via API on OpenAI's infrastructure, while the experiments involving DeepSeek were run on a machine equipped with two Intel Xeon Gold 6338 CPUs, 512GB of RAM, and one NVIDIA A40 GPU (46GB memory). To mitigate randomness, each experiment was repeated \emph{five times}.
Across both datasets, we submitted a total of 16000 process descriptions and evaluated the generated activity diagrams using the metrics in Section~\ref{sec:metrics}. 

All runs were executed sequentially, with dedicated GPU usage and minimal CPU interference for the local LLM.

For each \approach\ variant with a refinement loop, we cap the number of refinement iterations at five. If the loop does not converge within these five iterations, i.e., if the critique step continues to identify issues, we discard the activity diagram and restart the variant from its generation step. In other words, an activity diagram produced by a given \approach\ variant is accepted as that variant's output only if it passes the critique check. Across the 12,800 diagram generation tasks that utilized a refinement loop, this five-iteration cap was reached in only 703 cases (approximately 5.5\% of all tasks). Overall, the refinement process required an average of only 1.14 iterations per generated diagram. Further, for the 703 cases that were restarted, we encountered 14 cases (approximately 1.99\% of the tasks that were restarted, and 0.1\% of all tasks) in which the generated activity diagrams had to be discarded more than once. For all process descriptions in our datasets, we ultimately obtained an activity diagram that passed the critique step.

\section{Evaluation Results}
\label{sec:res}
\begin{table}[t]
\centering
\caption{\textit{Structural consistency} results report how many activity diagrams violate at least one structural constraint (Table \ref{table:prompt_constraint}). For the Industry dataset, 100 activity diagrams were generated in total. For the \textsc{Paged} dataset, 1000 activity diagrams were generated in total. Highlighted cells indicate cases with zero violations.}
\label{table:structural_constraint_result}
\scalebox{0.8}{
\begin{tabular}{@{}llcc@{}}
\toprule
& & \textbf{Industry (out of 100)} & \textbf{\textsc{Paged} (out of 1000)} \\
\midrule
\textbf{Baseline} 
& GPT-4.1 Mini & 49 (49.00\%) & 235 (23.50\%) \\
& O4 Mini & 12 (12.0\%) & 12 (1.20\%) \\
& Deepseek-R1 & -- & 233 (23.30\%) \\
\midrule

\textbf{\approachAlignmentLLMStrucutLLM}
& GPT-4.1 Mini & 42 (42.0\%) & 201 (20.10\%) \\
& O4 Mini & 13 (13.0\%) & 7 (0.70\%) \\
& Deepseek-R1 & -- & 216 (21.60\%) \\
\midrule

\textbf{\approachAlignmentLLMStrucutFormal}
& GPT-4.1 Mini & {\cellcolor{yellow!25}0 (0.00\%)} & {\cellcolor{yellow!25}0 (0.00\%)} \\
& O4 Mini & {\cellcolor{yellow!25}0 (0.00\%)} & {\cellcolor{yellow!25}0 (0.00\%)} \\
& Deepseek-R1 & -- & {\cellcolor{yellow!25}0 (0.00\%)} \\
\midrule

\textbf{\approachStrucutLLMOnly}
& GPT-4.1 Mini & 58 (58.0\%) & 129 (12.90\%) \\
& O4 Mini & 3 (3.0\%) & 2 (0.20\%) \\
& Deepseek-R1 & -- & 176 (17.60\%) \\
\midrule

\textbf{\approachStrucutFormalOnly}
& GPT-4.1 Mini & {\cellcolor{yellow!25}0 (0.00\%)} & {\cellcolor{yellow!25}0 (0.00\%)} \\
& O4 Mini & {\cellcolor{yellow!25}0 (0.00\%)} & {\cellcolor{yellow!25}0 (0.00\%)} \\
& Deepseek-R1 & -- & {\cellcolor{yellow!25}0 (0.00\%)} \\
\bottomrule
\end{tabular}}
\end{table}

\begin{table}[t]
\centering
\caption{\textit{Semantic correctness} and \textit{completeness} results for the Industry and \textsc{Paged} datasets using the L-Match and B-Match methods. The table shows mean (\%) and standard deviation across runs for each variant and LLM. Highlighted cells indicate the best-performing variant for each metric–LLM pair within each dataset and method.}
\label{table:semantic_results_combined}
\begin{subtable}[t]{\textwidth}
    \centering
    \caption{L-Match evaluation method}
    \label{table:semantic_results_a}
    \scalebox{0.55}{
    \begin{tabular}{@{}ll l cc cc cc cc cc@{}}
        \toprule
        \multirow{2}{*}{\textbf{Dataset}} & \multirow{2}{*}{\textbf{LLM}} & \multirow{2}{*}{\textbf{Metric}} 
        & \multicolumn{2}{c}{\textbf{Baseline}} 
        & \multicolumn{2}{c}{\textbf{\approachAlignmentLLMStrucutLLM}} 
        & \multicolumn{2}{c}{\textbf{\approachAlignmentLLMStrucutFormal}} 
        & \multicolumn{2}{c}{\textbf{\approachStrucutLLMOnly}} 
        & \multicolumn{2}{c}{\textbf{\approachStrucutFormalOnly}} \\
        \cmidrule(lr){4-5} \cmidrule(lr){6-7} \cmidrule(lr){8-9} \cmidrule(lr){10-11} \cmidrule(lr){12-13}
         & & & \textit{\textbf{Avg (\%)}} & \textit{\textbf{SD}} & \textit{\textbf{Avg (\%)}} & \textit{\textbf{SD}} 
         & \textit{\textbf{Avg (\%)}} & \textit{\textbf{SD}} & \textit{\textbf{Avg (\%)}} & \textit{\textbf{SD}} & \textit{\textbf{Avg (\%)}} & \textit{\textbf{SD}} \\
        \midrule
        
        % ---------------- INDUSTRY ----------------
        \multirow{4}{*}{\textit{\textbf{Industry}}} & \multirow{2}{*}{\textit{\textbf{O4 Mini}}} & \textit{\textbf{Completeness}} & 79.66 & 6.71 &  80.62 & 6.17 &\cellcolor{yellow!25} 93.56 & 0.96 & 90.74 & 2.54 & 91.81 & 1.37 \\
         &  & \textit{\textbf{Correctness}} & 78.43 & 6.53 & 75.99 & 7.28 &86.51 & 2.59 & 86.01 & 2.69 & \cellcolor{yellow!25} 90.75 & 0.87 \\
        \cmidrule(l){2-13}
         & \multirow{2}{*}{\textit{\textbf{GPT-4.1 Mini}}} & \textit{\textbf{Completeness}} & 45.39 & 8.47 & 51.56 & 8.52 &87.07 & 1.25 & 37.23 & 12.33 & \cellcolor{yellow!25} 87.26 & 1.24 \\
         &  & \textit{\textbf{Correctness}} & 41.23 & 8.12 & 46.05 & 9.52 &77.9 & 1.67 & 33.37 & 10.26 & \cellcolor{yellow!25} 78.16 & 2.91 \\
        \midrule
        
        % ---------------- PAGED ----------------
        \multirow{6}{*}{\textit{\textbf{\textsc{Paged}}}} & \multirow{2}{*}{\textit{\textbf{O4 Mini}}} & \textit{\textbf{Completeness}} & 96.13 & 0.82 & 96.84 & 0.56 &97.31 & 0.38 & 96.98 & 0.31 & \cellcolor{yellow!25} 97.44 & 0.13 \\
         &  & \textit{\textbf{Correctness}} & 84.51 & 0.45 & 85.91 & 0.57 &\cellcolor{yellow!25}86.81 & 0.64 & 86.27 & 0.6 & 85.89 & 1.01 \\
        \cmidrule(l){2-13}
         & \multirow{2}{*}{\textit{\textbf{GPT-4.1 Mini}}} & \textit{\textbf{Completeness}} &74.72 & 2.01 & 78.1 & 2.02 &97.2 & 0.14 & 85.07 & 2.26 & \cellcolor{yellow!25}97.37 & 0.23 \\
         &  & \textit{\textbf{Correctness}} &65.7 & 1.39 & 67.88 & 1.05 &82.46 & 0.55 & 72.56 & 1.93 &\cellcolor{yellow!25} 82.64 & 0.6 \\
        \cmidrule(l){2-13}
         & \multirow{2}{*}{\textit{\textbf{Deepseek-R1}}} & \textit{\textbf{Completeness}} &73.36 & 2.96 &  75.76 & 2.06 &\cellcolor{yellow!25}95.77 & 0.59 &78.84 & 1.71 & 94.93 & 0.46 \\
         &  & \textit{\textbf{Correctness}} & 67.86 & 2.66 & 69.57 & 2.71 &\cellcolor{yellow!25}87.28 & 0.3 & 71.4 & 1.53 & 86.68 & 1.0 \\
        \bottomrule
    \end{tabular}
    }
\end{subtable}

\hfill 

\begin{subtable}[t]{\textwidth}
    \centering
    \caption{B-Match evaluation method}
    \label{table:semantic_results_b}
    \scalebox{0.55}{
    \begin{tabular}{@{}ll l cc cc cc cc cc@{}}
        \toprule
        \multirow{2}{*}{\textbf{Dataset}} & \multirow{2}{*}{\textbf{LLM}} & \multirow{2}{*}{\textbf{Metric}} 
        & \multicolumn{2}{c}{\textbf{Baseline}} 
        & \multicolumn{2}{c}{\textbf{\approachAlignmentLLMStrucutLLM}} 
        & \multicolumn{2}{c}{\textbf{\approachAlignmentLLMStrucutFormal}} 
        & \multicolumn{2}{c}{\textbf{\approachStrucutLLMOnly}} 
        & \multicolumn{2}{c}{\textbf{\approachStrucutFormalOnly}} \\
        \cmidrule(lr){4-5} \cmidrule(lr){6-7} \cmidrule(lr){8-9} \cmidrule(lr){10-11} \cmidrule(lr){12-13}
         & & & \textit{\textbf{Avg (\%)}} & \textit{\textbf{SD}} & \textit{\textbf{Avg (\%)}} & \textit{\textbf{SD}} 
         & \textit{\textbf{Avg (\%)}} & \textit{\textbf{SD}} & \textit{\textbf{Avg (\%)}} & \textit{\textbf{SD}} & \textit{\textbf{Avg (\%)}} & \textit{\textbf{SD}} \\
        \midrule

        % ---------------- INDUSTRY ----------------
        \multirow{4}{*}{\textit{\textbf{Industry}}} 
         & \multirow{2}{*}{\textit{\textbf{O4 Mini}}} 
         & \textit{\textbf{Completeness}} & 68.59 & 3.35 & 70.19 & 11.48 & \cellcolor{yellow!25}82.56 & 2.84 & 74.92 & 2.58 & 78.47 & 4.51 \\
         & & \textit{\textbf{Correctness}} & 75.40 & 2.96 & 69.93 & 6.85 & \cellcolor{yellow!25}86.41 & 5.12 & 82.15 & 3.59 & 85.26 & 5.50 \\
        \cmidrule(l){2-13}
         & \multirow{2}{*}{\textit{\textbf{GPT-4.1 Mini}}} 
         & \textit{\textbf{Completeness}} & 39.39 & 9.48 & 47.55 & 6.49 & 68.68 & 1.21 & 32.36 & 7.57 & \cellcolor{yellow!25}74.61 & 2.46 \\
         & & \textit{\textbf{Correctness}} & 39.21 & 8.52 & 46.41 & 6.66 & 76.37 & 3.29 & 32.97 & 12.06 & \cellcolor{yellow!25}80.34 & 3.57 \\

        \midrule

        % ---------------- PAGED ----------------
        \multirow{6}{*}{\textit{\textbf{\textsc{Paged}}}} 
         & \multirow{2}{*}{\textit{\textbf{O4 Mini}}} 
         & \textit{\textbf{Completeness}} & 93.80 & 0.64 & 94.06 & 1.03 & 94.57 & 0.49 & \cellcolor{yellow!25}95.04 & 0.42 & 94.51 & 0.42 \\
         & & \textit{\textbf{Correctness}} & 83.04 & 0.43 & 85.15 & 0.89 & \cellcolor{yellow!25}86.33 & 0.60 & 84.85 & 0.67 & 84.81 & 0.88 \\
        \cmidrule(l){2-13}
         & \multirow{2}{*}{\textit{\textbf{GPT-4.1 Mini}}} 
         & \textit{\textbf{Completeness}} & 71.92 & 1.56 & 75.40 & 2.17 & \cellcolor{yellow!25}92.33 & 0.42 & 82.23 & 2.06 & 92.11 & 0.38 \\
         & & \textit{\textbf{Correctness}} & 63.79 & 1.25 & 66.33 & 1.52 & \cellcolor{yellow!25}81.15 & 1.25 & 70.33 & 1.95 & 79.94 & 0.88 \\
        \cmidrule(l){2-13}
         & \multirow{2}{*}{\textit{\textbf{Deepseek-R1}}} 
         & \textit{\textbf{Completeness}} & 69.84 & 3.16 & 72.57 & 1.67 & \cellcolor{yellow!25}90.28 & 1.01 & 73.51 & 2.33 & 89.20 & 0.56 \\
         & & \textit{\textbf{Correctness}} & 68.09 & 2.44 & 69.56 & 3.31 & \cellcolor{yellow!25}86.49 & 0.55 & 65.04 & 9.06 & 85.70 & 1.19 \\

        \bottomrule
    \end{tabular}
    }
\end{subtable}

\end{table}

\begin{table}[t]
\centering
\caption{Pairwise statistical comparisons of \approach\ variants for \emph{correctness} and \emph{completeness}, computed separately for each dataset, LLM, evaluation method, and metric using the per-LLM results in Table~\ref{table:semantic_results_combined}. Each entry is reported as A vs. B and indicates whether variant A significantly outperforms variant B. Yellow cells indicate statistically significant improvements of A over B, while blue cells indicate statistically significant improvements of B over A.}
\label{table:all_stat_results}
% ---------------- Top-level subtable 1: L-Match ----------------
\begin{subtable}[t]{\textwidth}
\centering
\caption{L-Match evaluation method}

% --- First 5 comparisons ---
\begin{subtable}[t]{\textwidth}
\centering
\scalebox{0.47}{
\begin{tabular}{@{}ll l cc cc cc cc cc@{}}
\toprule
\multirow{2}{*}{\textbf{Dataset}} & \multirow{2}{*}{\textbf{LLM}} & \multirow{2}{*}{\textbf{Metric}} 
& \multicolumn{2}{c}{\textbf{\begin{tabular}[c]{@{}c@{}}\approachAlignmentLLMStrucutLLM(A) \\ vs Baseline(B)\end{tabular}}} 
& \multicolumn{2}{c}{\textbf{\begin{tabular}[c]{@{}c@{}}\approachAlignmentLLMStrucutFormal(A) \\ vs Baseline(B)\end{tabular}}} 
& \multicolumn{2}{c}{\textbf{\begin{tabular}[c]{@{}c@{}}\approachStrucutLLMOnly(A) \\ vs Baseline(B)\end{tabular}}} 
& \multicolumn{2}{c}{\textbf{\begin{tabular}[c]{@{}c@{}}\approachStrucutFormalOnly(A) \\ vs Baseline(B)\end{tabular}}} 
& \multicolumn{2}{c}{\textbf{\begin{tabular}[c]{@{}c@{}}\approachStrucutFormalOnly(A) \\ vs\ \approachStrucutLLMOnly(B) \end{tabular}}} \\
\cmidrule(lr){4-5} \cmidrule(lr){6-7} \cmidrule(lr){8-9} \cmidrule(lr){10-11} \cmidrule(lr){12-13}
 & & & \textit{\textbf{p-value}} & \textit{\textbf{$\mathbf{\hat{A}_{12}}$}}  & \textit{\textbf{p-value}} & \textit{\textbf{$\mathbf{\hat{A}_{12}}$}}  & \textit{\textbf{p-value}} & \textit{\textbf{$\mathbf{\hat{A}_{12}}$}}  & \textit{\textbf{p-value}} & \textit{\textbf{$\mathbf{\hat{A}_{12}}$}}  & \textit{\textbf{p-value}} & \textit{\textbf{$\mathbf{\hat{A}_{12}}$}}  \\
\midrule

\multirow{4}{*}{\textit{\textbf{Industry}}} & \multirow{2}{*}{\textit{\textbf{O4 Mini}}} & \textit{\textbf{Completeness}} & 0.23 & 0.54 & \cellcolor{yellow!25}8.24E-06 & 0.64 (S) & \cellcolor{yellow!25}2.13E-05 & 0.62 (S) & \cellcolor{yellow!25}6.48E-04 & 0.58 (S) & 0.22 & 0.46 \\
& & \textit{\textbf{Correctness}}  & 0.26 & 0.43 & 0.74 & 0.47 & 0.40 & 0.51 & \cellcolor{yellow!25}0.02 & 0.59 (S) & 0.07 & 0.57 \\
  \cmidrule(l){2-13}
 & \multirow{2}{*}{\textit{\textbf{GPT-4.1 Mini}}} & \textit{\textbf{Completeness}} & 0.42 & 0.53 & \cellcolor{yellow!25}4.53E-10 & 0.72 (L) & 0.19 & 0.45 & \cellcolor{yellow!25}4.02E-10 & 0.72 (L) & \cellcolor{yellow!25}2.14E-12 & 0.78 (L) \\
& & \textit{\textbf{Correctness}}  & 0.48 & 0.52 & \cellcolor{yellow!25}1.45E-09 & 0.71 (L) & 0.18 & 0.45 & \cellcolor{yellow!25}3.58E-09 & 0.71 (L) & \cellcolor{yellow!25}2.64E-13 & 0.78 (L) \\

\midrule

\multirow{6}{*}{\textit{\textbf{\textsc{Paged}}}} & \multirow{2}{*}{\textit{\textbf{O4 Mini}}} & \textit{\textbf{Completeness}} & 0.10 & 0.51 & 0.19 & 0.51 & 0.44 & 0.50 & \cellcolor{yellow!25}0.05 & 0.51 & 0.19 & 0.51 \\
& & \textit{\textbf{Correctness}}  & \cellcolor{yellow!25}0.02 & 0.52 & \cellcolor{yellow!25}1.11E-04 & 0.53 & \cellcolor{yellow!25}5.14E-03 & 0.52 & \cellcolor{yellow!25}0.04 & 0.51 & 0.19 & 0.49 \\
  \cmidrule(l){2-13}
 & \multirow{2}{*}{\textit{\textbf{GPT-4.1 Mini}}} & \textit{\textbf{Completeness}} & \cellcolor{yellow!25}0.03 & 0.52 & \cellcolor{yellow!25}9.53E-47 & 0.60 (S) & \cellcolor{yellow!25}1.11E-12 & 0.55 & \cellcolor{yellow!25}9.06E-48 & 0.60 (S) & \cellcolor{yellow!25}3.96E-24 & 0.55 \\
& & \textit{\textbf{Correctness}}  & 0.22 & 0.51 & \cellcolor{yellow!25}3.68E-31 & 0.57 (S) & \cellcolor{yellow!25}6.69E-07 & 0.53 & \cellcolor{yellow!25}9.89E-36 & 0.58 (S) & \cellcolor{yellow!25}5.93E-16 & 0.56 (S) \\
  \cmidrule(l){2-13}
 & \multirow{2}{*}{\textit{\textbf{Deepseek-R1}}} & \textit{\textbf{Completeness}} & 0.15 & 0.52 & \cellcolor{yellow!25}4.56E-47 & 0.61 (S) & \cellcolor{yellow!25}1.75E-03 & 0.53 & \cellcolor{yellow!25}2.53E-44 & 0.60 (S) & \cellcolor{yellow!25}1.80E-31 & 0.58 (S) \\
& & \textit{\textbf{Correctness}}  & 0.44 & 0.51 & \cellcolor{yellow!25}3.59E-31 & 0.59 (S) & 0.17 & 0.51 & \cellcolor{yellow!25}3.59E-31 & 0.59 (S) & \cellcolor{yellow!25}1.35E-26 & 0.58 (S) \\

\bottomrule
\end{tabular}
}
\end{subtable}

\vspace{0.5em}

% --- Second 5 comparisons ---
\begin{subtable}[t]{\textwidth}
\centering
\scalebox{0.43}{
\begin{tabular}{@{}ll l cc cc cc cc cc@{}}
\toprule
\multirow{2}{*}{\textbf{Dataset}} & \multirow{2}{*}{\textbf{LLM}} & \multirow{2}{*}{\textbf{Metric}} 
& \multicolumn{2}{c}{\textbf{\begin{tabular}[c]{@{}c@{}}\approachAlignmentLLMStrucutFormal(A) \\ vs\ \approachAlignmentLLMStrucutLLM(B)\end{tabular}}} 
& \multicolumn{2}{c}{\textbf{\begin{tabular}[c]{@{}c@{}}\approachStrucutFormalOnly(A) \\ vs\ \approachAlignmentLLMStrucutLLM(B) \end{tabular}}} 
& \multicolumn{2}{c}{\textbf{\begin{tabular}[c]{@{}c@{}}\approachStrucutFormalOnly(A) \\ vs\ \approachAlignmentLLMStrucutFormal(B) \end{tabular}}} 
& \multicolumn{2}{c}{\textbf{\begin{tabular}[c]{@{}c@{}}\approachStrucutLLMOnly(A) \\ vs\ \approachAlignmentLLMStrucutLLM(B) \end{tabular}}} 
& \multicolumn{2}{c}{\textbf{\begin{tabular}[c]{@{}c@{}}\approachStrucutLLMOnly(A) \\ vs\ \approachAlignmentLLMStrucutFormal(B)\end{tabular}}} \\
\cmidrule(lr){4-5} \cmidrule(lr){6-7} \cmidrule(lr){8-9} \cmidrule(lr){10-11} \cmidrule(lr){12-13}
 & && \textit{\textbf{p-value}} & \textit{\textbf{$\mathbf{\hat{A}_{12}}$}}  & \textit{\textbf{p-value}} & \textit{\textbf{$\mathbf{\hat{A}_{12}}$}}  & \textit{\textbf{p-value}} & \textit{\textbf{$\mathbf{\hat{A}_{12}}$}}  & \textit{\textbf{p-value}} & \textit{\textbf{$\mathbf{\hat{A}_{12}}$}}  & \textit{\textbf{p-value}} & \textit{\textbf{$\mathbf{\hat{A}_{12}}$}}  \\
\midrule

\multirow{4}{*}{\textit{\textbf{Industry}}} & \multirow{2}{*}{\textit{\textbf{O4 Mini}}} & \textit{\textbf{Completeness}} & \cellcolor{yellow!25}2.17E-03 & 0.58 (S) & 0.11 & 0.54 & \cellcolor{blue!25}0.03 & 0.45 & \cellcolor{yellow!25}0.01 & 0.58 (S) & 0.58 & 0.49 \\
& & \textit{\textbf{Correctness}}  & 0.42 & 0.54 & \cellcolor{yellow!25}1.30E-05 & 0.66 (M) & \cellcolor{yellow!25}2.33E-03 & 0.63 (S) & \cellcolor{yellow!25}0.04 & 0.59 (S) & 0.18 & 0.55 \\
  \cmidrule(l){2-13}
 & \multirow{2}{*}{\textit{\textbf{GPT-4.1 Mini}}} & \textit{\textbf{Completeness}} & \cellcolor{yellow!25}3.20E-09 & 0.69 (M) & \cellcolor{yellow!25}3.85E-09 & 0.69 (M) & 0.62 & 0.51 & \cellcolor{blue!25}0.04 & 0.42 (S) & \cellcolor{blue!25}4.71E-12 & 0.23 (L) \\
& & \textit{\textbf{Correctness}}  & \cellcolor{yellow!25}9.82E-08 & 0.69 (M) & \cellcolor{yellow!25}4.73E-08 & 0.70 (M) & 0.83 & 0.51 & \cellcolor{blue!25}0.03 & 0.42 (S) & \cellcolor{blue!25}8.18E-13 & 0.22 (L) \\

 \midrule

\multirow{6}{*}{\textit{\textbf{\textsc{Paged}}}} & \multirow{2}{*}{\textit{\textbf{O4 Mini}}} & \textit{\textbf{Completeness}} & 0.54 & 0.50 & 0.85 & 0.50 & 0.58 & 0.50 & 0.19 & 0.49 & 0.62 & 0.49 \\
& & \textit{\textbf{Correctness}}  & 0.23 & 0.51 & 0.44 & 0.49 & \cellcolor{blue!25}0.02 & 0.48 & 0.78 & 0.50 & 0.21 & 0.49 \\
  \cmidrule(l){2-13}
 & \multirow{2}{*}{\textit{\textbf{GPT-4.1 Mini}}} & \textit{\textbf{Completeness}} & \cellcolor{yellow!25}2.94E-38 & 0.58 (S) & \cellcolor{yellow!25}3.82E-40 & 0.58 (S) & 0.38 & 0.50 & \cellcolor{yellow!25}6.68E-06 & 0.54 & \cellcolor{blue!25}2.17E-21 & 0.46 \\
& & \textit{\textbf{Correctness}}  & \cellcolor{yellow!25}1.39E-25 & 0.57 (S) & \cellcolor{yellow!25}2.44E-28 & 0.57 (S) & 0.78 & 0.51 & \cellcolor{yellow!25}1.89E-03 & 0.52 & \cellcolor{blue!25}4.59E-14 & 0.45 \\
  \cmidrule(l){2-13}
 & \multirow{2}{*}{\textit{\textbf{Deepseek-R1}}} & \textit{\textbf{Completeness}} & \cellcolor{yellow!25}3.95E-39 & 0.59 (S) & \cellcolor{yellow!25}1.09E-36 & 0.58 (S) & 0.12 & 0.49 & 0.12 & 0.51 & \cellcolor{blue!25}5.30E-34 & 0.42 (S) \\
& & \textit{\textbf{Correctness}}  & \cellcolor{yellow!25}4.96E-28 & 0.58 (S) & \cellcolor{yellow!25}4.03E-28 & 0.58 (S) & 0.46 & 0.49 & 0.42 & 0.50 & \cellcolor{blue!25}7.46E-28 & 0.41 (S) \\

\bottomrule
\end{tabular}
}
\end{subtable}

\end{subtable}

\vspace{1em}

% ---------------- Top-level subtable 2: B-Match/Formal ----------------
\begin{subtable}[t]{\textwidth}
\centering
\caption{B-Match evaluation method}

% --- First 5 comparisons ---
\begin{subtable}[t]{\textwidth}
\centering
\scalebox{0.47}{
\begin{tabular}{@{}ll l cc cc cc cc cc@{}}
\toprule
\multirow{2}{*}{\textbf{Dataset}} & \multirow{2}{*}{\textbf{LLM}} & \multirow{2}{*}{\textbf{Metric}} 
& \multicolumn{2}{c}{\textbf{\begin{tabular}[c]{@{}c@{}}\approachAlignmentLLMStrucutLLM(A) \\ vs Baseline(B)\end{tabular}}} 
& \multicolumn{2}{c}{\textbf{\begin{tabular}[c]{@{}c@{}}\approachAlignmentLLMStrucutFormal(A) \\ vs Baseline(B)\end{tabular}}} 
& \multicolumn{2}{c}{\textbf{\begin{tabular}[c]{@{}c@{}}\approachStrucutLLMOnly(A) \\ vs Baseline(B)\end{tabular}}} 
& \multicolumn{2}{c}{\textbf{\begin{tabular}[c]{@{}c@{}}\approachStrucutFormalOnly(A) \\ vs Baseline(B)\end{tabular}}} 
& \multicolumn{2}{c}{\textbf{\begin{tabular}[c]{@{}c@{}}\approachStrucutFormalOnly(A) \\ vs\ \approachStrucutLLMOnly(B) \end{tabular}}} \\
\cmidrule(lr){4-5} \cmidrule(lr){6-7} \cmidrule(lr){8-9} \cmidrule(lr){10-11} \cmidrule(lr){12-13}
 & & & \textit{\textbf{p-value}} & \textit{\textbf{$\mathbf{\hat{A}_{12}}$}}  & \textit{\textbf{p-value}} & \textit{\textbf{$\mathbf{\hat{A}_{12}}$}}  & \textit{\textbf{p-value}} & \textit{\textbf{$\mathbf{\hat{A}_{12}}$}}  & \textit{\textbf{p-value}} & \textit{\textbf{$\mathbf{\hat{A}_{12}}$}}  & \textit{\textbf{p-value}} & \textit{\textbf{$\mathbf{\hat{A}_{12}}$}}  \\
\midrule
\multirow{4}{*}{\textit{\textbf{Industry}}} & \multirow{2}{*}{\textit{\textbf{O4 Mini}}} & \textit{\textbf{Completeness}} & 0.66 & 0.53 & \cellcolor{yellow!25}4.76E-04 & 0.62 (S) & 0.18 & 0.55 & \cellcolor{yellow!25}0.01 & 0.57 (S) & 0.69 & 0.51 \\
& & \textit{\textbf{Correctness}}  & 0.26 & 0.44 & \cellcolor{yellow!25}0.04 & 0.56 (S) & 0.22 & 0.54 & \cellcolor{yellow!25}0.04 & 0.57 (S) & 0.48 & 0.53 \\
  \cmidrule(l){2-13}
 & \multirow{2}{*}{\textit{\textbf{GPT-4.1 Mini}}} & \textit{\textbf{Completeness}} & 0.17 & 0.54 & \cellcolor{yellow!25}1.47E-07 & 0.70 (M) & 0.18 & 0.45 & \cellcolor{yellow!25}6.25E-09 & 0.74 (L) & \cellcolor{yellow!25}1.18E-10 & 0.78 (L) \\
& & \textit{\textbf{Correctness}}  & 0.20 & 0.55 & \cellcolor{yellow!25}6.25E-10 & 0.75 (L) & 0.30 & 0.46 & \cellcolor{yellow!25}1.42E-11 & 0.77 (L) & \cellcolor{yellow!25}3.29E-13 & 0.80 (L) \\

\midrule

\multirow{6}{*}{\textit{\textbf{\textsc{Paged}}}} & \multirow{2}{*}{\textit{\textbf{O4 Mini}}} & \textit{\textbf{Completeness}} & 0.92 & 0.50 & 0.76 & 0.50 & \cellcolor{yellow!25}0.04 & 0.51 & 0.61 & 0.50 & 0.06 & 0.49 \\
& & \textit{\textbf{Correctness}}  & \cellcolor{yellow!25}0.01 & 0.52 & \cellcolor{yellow!25}7.50E-05 & 0.52 & \cellcolor{yellow!25}0.04 & 0.51 & \cellcolor{yellow!25}0.02 & 0.51 & 0.73 & 0.50 \\
  \cmidrule(l){2-13}
& \multirow{2}{*}{\textit{\textbf{GPT-4.1 Mini}}} & \textit{\textbf{Completeness}} & \cellcolor{yellow!25}0.02 & 0.52 & \cellcolor{yellow!25}2.82E-44 & 0.60 (S) & \cellcolor{yellow!25}1.11E-12 & 0.56 (S) & \cellcolor{yellow!25}5.11E-44 & 0.60 (S) & \cellcolor{yellow!25}1.72E-18 & 0.54 \\
& & \textit{\textbf{Correctness}}  & 0.09 & 0.51 & \cellcolor{yellow!25}5.64E-37 & 0.60 (S) & \cellcolor{yellow!25}2.63E-07 & 0.53 & \cellcolor{yellow!25}1.39E-32 & 0.59 (S) & \cellcolor{yellow!25}1.96E-14 & 0.55 \\
  \cmidrule(l){2-13}
 & \multirow{2}{*}{\textit{\textbf{Deepseek-R1}}} & \textit{\textbf{Completeness}} & 0.06 & 0.52 & \cellcolor{yellow!25}5.11E-44 & 0.61 (S) & \cellcolor{yellow!25}0.04 & 0.52 & \cellcolor{yellow!25}5.01E-42 & 0.60 (S) & \cellcolor{yellow!25}8.45E-31 & 0.58 (S) \\
& & \textit{\textbf{Correctness}}  & 0.49 & 0.50 & \cellcolor{yellow!25}5.88E-33 & 0.59 (S) & 0.10 & 0.48 & \cellcolor{yellow!25}2.09E-30 & 0.59 (S) & \cellcolor{yellow!25}3.73E-37 & 0.62 (S) \\

\bottomrule
\end{tabular}
}
\end{subtable}

\vspace{0.5em}

% --- Second 5 comparisons ---
\begin{subtable}[t]{\textwidth}
\centering
\scalebox{0.43}{
\begin{tabular}{@{}ll l cc cc cc cc cc@{}}
\toprule
\multirow{2}{*}{\textbf{Dataset}} & \multirow{2}{*}{\textbf{LLM}} & \multirow{2}{*}{\textbf{Metric}} 
& \multicolumn{2}{c}{\textbf{\begin{tabular}[c]{@{}c@{}}\approachAlignmentLLMStrucutFormal(A) \\ vs\ \approachAlignmentLLMStrucutLLM(B)\end{tabular}}} 
& \multicolumn{2}{c}{\textbf{\begin{tabular}[c]{@{}c@{}}\approachStrucutFormalOnly(A) \\ vs\ \approachAlignmentLLMStrucutLLM(B) \end{tabular}}} 
& \multicolumn{2}{c}{\textbf{\begin{tabular}[c]{@{}c@{}}\approachStrucutFormalOnly(A) \\ vs\ \approachAlignmentLLMStrucutFormal(B) \end{tabular}}} 
& \multicolumn{2}{c}{\textbf{\begin{tabular}[c]{@{}c@{}}\approachStrucutLLMOnly(A) \\ vs\ \approachAlignmentLLMStrucutLLM(B) \end{tabular}}} 
& \multicolumn{2}{c}{\textbf{\begin{tabular}[c]{@{}c@{}}\approachStrucutLLMOnly(A) \\ vs\ \approachAlignmentLLMStrucutFormal(B)\end{tabular}}} \\
\cmidrule(lr){4-5} \cmidrule(lr){6-7} \cmidrule(lr){8-9} \cmidrule(lr){10-11} \cmidrule(lr){12-13}
 & && \textit{\textbf{p-value}} & \textit{\textbf{$\mathbf{\hat{A}_{12}}$}}  & \textit{\textbf{p-value}} & \textit{\textbf{$\mathbf{\hat{A}_{12}}$}}  & \textit{\textbf{p-value}} & \textit{\textbf{$\mathbf{\hat{A}_{12}}$}}  & \textit{\textbf{p-value}} & \textit{\textbf{$\mathbf{\hat{A}_{12}}$}}  & \textit{\textbf{p-value}} & \textit{\textbf{$\mathbf{\hat{A}_{12}}$}}  \\
\midrule
\multirow{4}{*}{\textit{\textbf{Industry}}} & \multirow{2}{*}{\textit{\textbf{O4 Mini}}} & \textit{\textbf{Completeness}} & \cellcolor{yellow!25}0.02 & 0.58 (S) & 0.13 & 0.53 & 0.18 & 0.45 & 0.61 & 0.52 & 0.06 & 0.43 \\
& & \textit{\textbf{Correctness}}  & \cellcolor{yellow!25}4.55E-04 & 0.64 (M) & \cellcolor{yellow!25}2.38E-04 & 0.64 (M) & 0.61 & 0.51 & \cellcolor{yellow!25}6.25E-03 & 0.61 (S) & 0.28 & 0.48 \\
  \cmidrule(l){2-13}
 & \multirow{2}{*}{\textit{\textbf{GPT-4.1 Mini}}} & \textit{\textbf{Completeness}} & \cellcolor{yellow!25}1.00E-04 & 0.64 (M) & \cellcolor{yellow!25}1.11E-06 & 0.68 (M) & 0.09 & 0.56 & \cellcolor{blue!25}0.02 & 0.41 (S) & \cellcolor{blue!25}1.67E-09 & 0.25 (L) \\
& & \textit{\textbf{Correctness}}  & \cellcolor{yellow!25}1.67E-07 & 0.69 (M) & \cellcolor{yellow!25}2.35E-09 & 0.71 (L) & 0.30 & 0.53 & \cellcolor{blue!25}0.03 & 0.42 (S) & \cellcolor{blue!25}1.54E-11 & 0.22 (L) \\

 \midrule
 \multirow{6}{*}{\textit{\textbf{\textsc{Paged}}}} & \multirow{2}{*}{\textit{\textbf{O4 Mini}}} & \textit{\textbf{Completeness}} & 1.00 & 0.50 & 0.87 & 0.50 & 0.94 & 0.50 & 0.06 & 0.51 & 0.08 & 0.51 \\
& & \textit{\textbf{Correctness}}  & 0.16 & 0.51 & 0.51 & 0.49 & \cellcolor{blue!25}0.04 & 0.49 & 0.56 & 0.50 & \cellcolor{blue!25}0.03 & 0.49 \\
  \cmidrule(l){2-13}
& \multirow{2}{*}{\textit{\textbf{GPT-4.1 Mini}}} & \textit{\textbf{Completeness}} & \cellcolor{yellow!25}8.27E-37 & 0.58 (S) & \cellcolor{yellow!25}9.18E-36 & 0.58 (S) & 0.61 & 0.50 & \cellcolor{yellow!25}2.63E-06 & 0.54 & \cellcolor{blue!25}1.90E-19 & 0.46 \\
& & \textit{\textbf{Correctness}}  & \cellcolor{yellow!25}7.91E-30 & 0.58 (S) & \cellcolor{yellow!25}2.76E-26 & 0.57 (S) & \cellcolor{blue!25}0.01 & 0.49 & \cellcolor{yellow!25}4.35E-03 & 0.52 & \cellcolor{blue!25}4.95E-21 & 0.43 (S) \\
  \cmidrule(l){2-13}
 & \multirow{2}{*}{\textit{\textbf{Deepseek-R1}}} & \textit{\textbf{Completeness}} & \cellcolor{yellow!25}3.29E-37 & 0.59 (S) & \cellcolor{yellow!25}7.55E-37 & 0.58 (S) & 0.06 & 0.49 & 0.69 & 0.50 & \cellcolor{blue!25}4.74E-35 & 0.41 (S) \\
& & \textit{\textbf{Correctness}}  & \cellcolor{yellow!25}2.79E-30 & 0.59 (S) & \cellcolor{yellow!25}3.36E-28 & 0.59 (S) & 0.66 & 0.50 & \cellcolor{blue!25}0.02 & 0.48 & \cellcolor{blue!25}1.65E-39 & 0.38 (S) \\

\bottomrule
\end{tabular}
}
\end{subtable}

\end{subtable}
\end{table}

\begin{table}[t]
\centering
\caption{\textit{Cost} results. The average number of tokens in  prompts (Input), responses (Output), and reasoning steps (Reasoning) per LLM call, as well as the average number of LLM calls (\# Calls) required by each variant.}
\label{table:token_cost}
\scalebox{0.65}{\begin{tabular}{@{}llcccccccc@{}}
\toprule
& & \multicolumn{4}{c}{\textbf{Industry}} & \multicolumn{4}{c}{\textbf{\textsc{Paged}}} \\
\cmidrule(lr){3-6} \cmidrule(lr){7-10}
& & \multicolumn{3}{c}{\textbf{Avg. Tokens}} & \textbf{\# Calls}
  & \multicolumn{3}{c}{\textbf{Avg. Tokens}} & \textbf{\# Calls} \\
\cmidrule(lr){3-5} \cmidrule(lr){7-9}
\textbf{~} & \textbf{LLM} & \textbf{Input} & \textbf{Output} & \textbf{Reasoning}
           &              & \textbf{Input} & \textbf{Output} & \textbf{Reasoning} & \\
\midrule
\textbf{Baseline} 
& GPT-4.1 Mini & 1927.62 & 763.34 & 0.0 & 1 & 882.42 & 240.74 & 0.0 & 1 \\
& O4 Mini & 1928.02 & 668.86 & 7116.80 & 1 & 881.42 & 194.98 & 3921.60 & 1 \\
& Deepseek-R1 & -- & -- & -- & -- & 873.90 & 197.43 & 985.69 & 1 \\
\midrule

\textbf{\approachAlignmentLLMStrucutLLM}
& GPT-4.1 Mini & 2818.92 & 756.90 & 0.0 & 3.09 & 1532.67 & 519.00 & 0.0 & 3.40 \\
& O4 Mini & 3864.62 & 650.85 & 6474.38 & 6.51 & 1237.45 & 267.89 & 3063.82 & 3.61 \\
& Deepseek-R1 & -- & -- & -- & -- & 1146.34 & 206.83 & 918.10 & 4.24 \\
\midrule

\textbf{\approachAlignmentLLMStrucutFormal}
& GPT-4.1 Mini & 3991.16 & 785.50 & 0.0 & 13.50 & 1833.85 & 548.91 & 0.0 & 6.43 \\
& O4 Mini & 3033.39 & 553.11 & 5682.94 & 5.96 & 1238.70 & 270.16 & 4113.77 & 3.87 \\
& Deepseek-R1 & -- & -- & -- & -- & 1258.91 & 216.58 & 834.52 & 6.47 \\
\midrule

\textbf{\approachStrucutLLMOnly}
& GPT-4.1 Mini & 2038.95 & 657.49 & 0.0 & 2.18 & 1240.35 & 529.49 & 0.0 & 3.52 \\
& O4 Mini & 2816.98 & 546.41 & 5399.49 & 5.14 & 1000.40 & 270.72 & 3818.95 & 3.22 \\
& Deepseek-R1 & -- & -- & -- & -- & 1137.58 & 250.40 & 1137.76 & 9.21 \\
\midrule

\textbf{\approachStrucutFormalOnly}
& GPT-4.1 Mini & 4003.06 & 825.39 & 0.0 & 3.77 & 1662.92 & 362.86 & 0.0 & 1.74 \\
& O4 Mini & 2251.48 & 708.17 & 7073.99 & 1.16 & 890.09 & 194.35 & 4702.67 & 1.01 \\
& Deepseek-R1 & -- & -- & -- & -- & 1218.17 & 240.88 & 804.93 & 1.67 \\
\bottomrule
\end{tabular}}
\end{table}

In this section, we present the results of our evaluation to answer research questions RQ1–RQ4, described in Section~\ref{sec:eval}. Before addressing these questions individually, we first provide an overview of the evaluation results for all metrics described in Section~\ref{sec:metrics}. The subsequent subsections systematically answer the research questions based on these results. Specifically, we validate the consistency of our evaluation methods (RQ1) in Section~\ref{sec:res_validation}, discuss the impact of the refinement loop on the quality of the generated diagrams (RQ2) in Section~\ref{sec:res_main}, and analyze the impact of our structural and alignment checking methods (RQ3 and RQ4) in Section~\ref{sec:res_ablation}.
We evaluate \textit{structural consistency}, \textit{semantic correctness}, \textit{completeness}, and \textit{cost} for all the five \approach\ variants on both the Industry dataset (using two LLMs) and the \textsc{Paged} dataset (using three LLMs). \textit{Structural consistency} results are shown in Table~\ref{table:structural_constraint_result}; \textit{Semantic correctness} and \textit{completeness}, computed based on the node matchings produced by L-Match and B-Match, are presented in Tables~\ref{table:semantic_results_combined}(a) and (b), respectively; and \textit{cost} results are provided in Table~\ref{table:token_cost}.

Table~\ref{table:structural_constraint_result} reports the number of generated activity diagrams that violate at least one of the structural constraints listed in Table~\ref{table:prompt_constraint}, based on 100 activity diagrams generated for the Industry dataset and 1000 activity diagrams generated for the \textsc{Paged} dataset for each variant and each LLM. 
The \textit{semantic correctness} and \textit{completeness} results in Tables~\ref{table:semantic_results_combined}(a) and~(b) report the averages and standard deviations over five runs for each variant, dataset, and LLM. Table \ref{table:semantic_results_combined}(a) reports the results based on L-Match (Section~\ref{sec:calibration}), while Table~\ref{table:semantic_results_combined}(b) shows the results based on B-Match (Section~\ref{sec:compare}).  Each row highlights in yellow the variant with the highest average \textit{correctness} or \textit{completeness}.

To statistically compare the \textit{correctness} and \textit{completeness} results for answering RQ1--RQ4, we use the Wilcoxon Rank-Sum Test~\citep{wilcoxon1992individual} along with the Vargha-Delaney effect size ($\hat{A}_{12}$)~\citep{vargha2000critique}. To compare variant~\emph{A} with variant~\emph{B}, we use a significance level of 5\%. A difference is deemed significant if the 
$p$-value falls below this threshold. When \emph{A} outperforms \emph{B}, the effect size is classified as small, medium, or large for $\hat{A}{12} \geq 0.56$, $0.64$, and $0.71$, respectively. Otherwise, when \emph{B} outperforms \emph{A}, the effect size is classified as small, medium, or large for $\hat{A}{12} \leq 0.44$, $0.36$, and $0.29$, respectively.
The difference between \emph{A} and \emph{B} is negligible when $0.44 < \hat{A}{12} < 0.56$~\citep{vargha2000critique}. 
The results of all pairwise comparisons among the five \approach\ variants, based on statistical tests for \textit{semantic correctness} and \textit{completeness} obtained by L-Match and B-Match evaluation methods are presented in Table~\ref{table:all_stat_results}. All statistical tests are conducted on results aggregated across LLMs from Table~\ref{table:semantic_results_combined} for each dataset, method, and metric.
In addition,  all statistical significance tests  are reported with p-values adjusted using the Benjamini–Hochberg (BH) procedure~\citep{benjamini1995controlling}.
Yellow-highlighted cells indicate a significant improvement of variant~\emph{A} over variant~\emph{B}, while blue-highlighted cells indicate a significant improvement of variant~\emph{B} over variant~\emph{A}. Cells without colour indicate that there is no significant difference between the two variants being compared. We use these statistical test results to answer RQ1, RQ2, RQ3, and RQ4.

For the \textit{cost} metric, Table~\ref{table:token_cost} reports the average number of LLM calls and the average number of tokens required per LLM for inputs, outputs, and reasoning. Overall, our results show that across all variants, datasets and evaluation methods, O4 Mini generates, on average, 17.54\% fewer activity diagrams with structural-constraint violations than GPT-4.1 Mini and 9.49\% fewer than DeepSeek. In addition, O4 Mini improves \textit{semantic correctness} by an average of 19.28\% compared to GPT-4.1 Mini and 7.75\% compared to DeepSeek, and increases \textit{semantic completeness} by 17.51\% and 6.98\%, respectively. Finally, O4 Mini requires an average of 0.71 fewer LLM calls than GPT-4.1 Mini \hbox{and 1.25 fewer than DeepSeek.}

\subsection{RQ1 - Consistency of Evaluation Methods}
\label{sec:res_validation}

\begin{figure}[t]
    \centering
    \begin{subfigure}[b]{0.48\linewidth}
        \includegraphics[width=\linewidth]{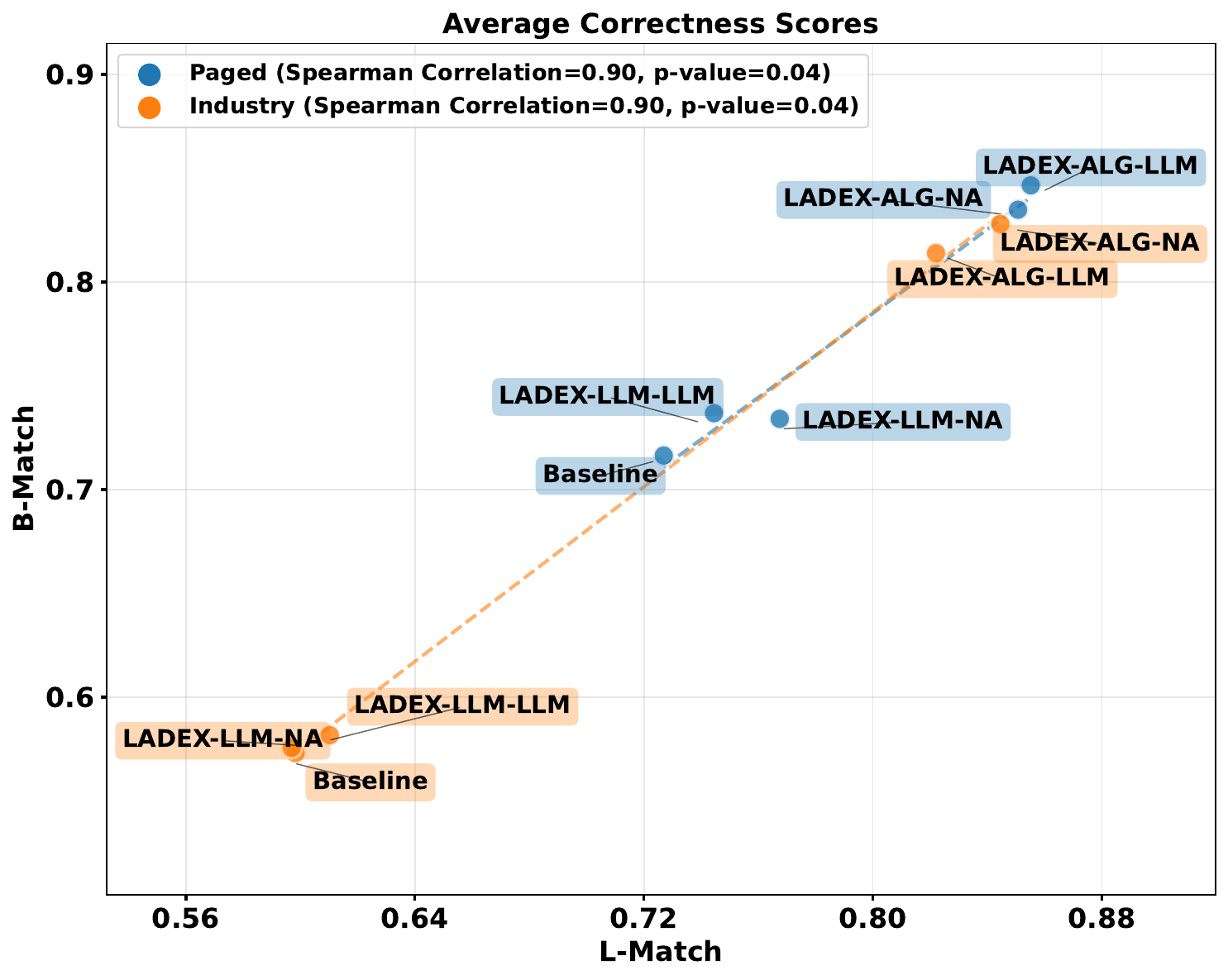}
        \caption{Correctness scores.}
    \end{subfigure}
    \hfill
    \begin{subfigure}[b]{0.48\linewidth}
        \includegraphics[width=\linewidth]{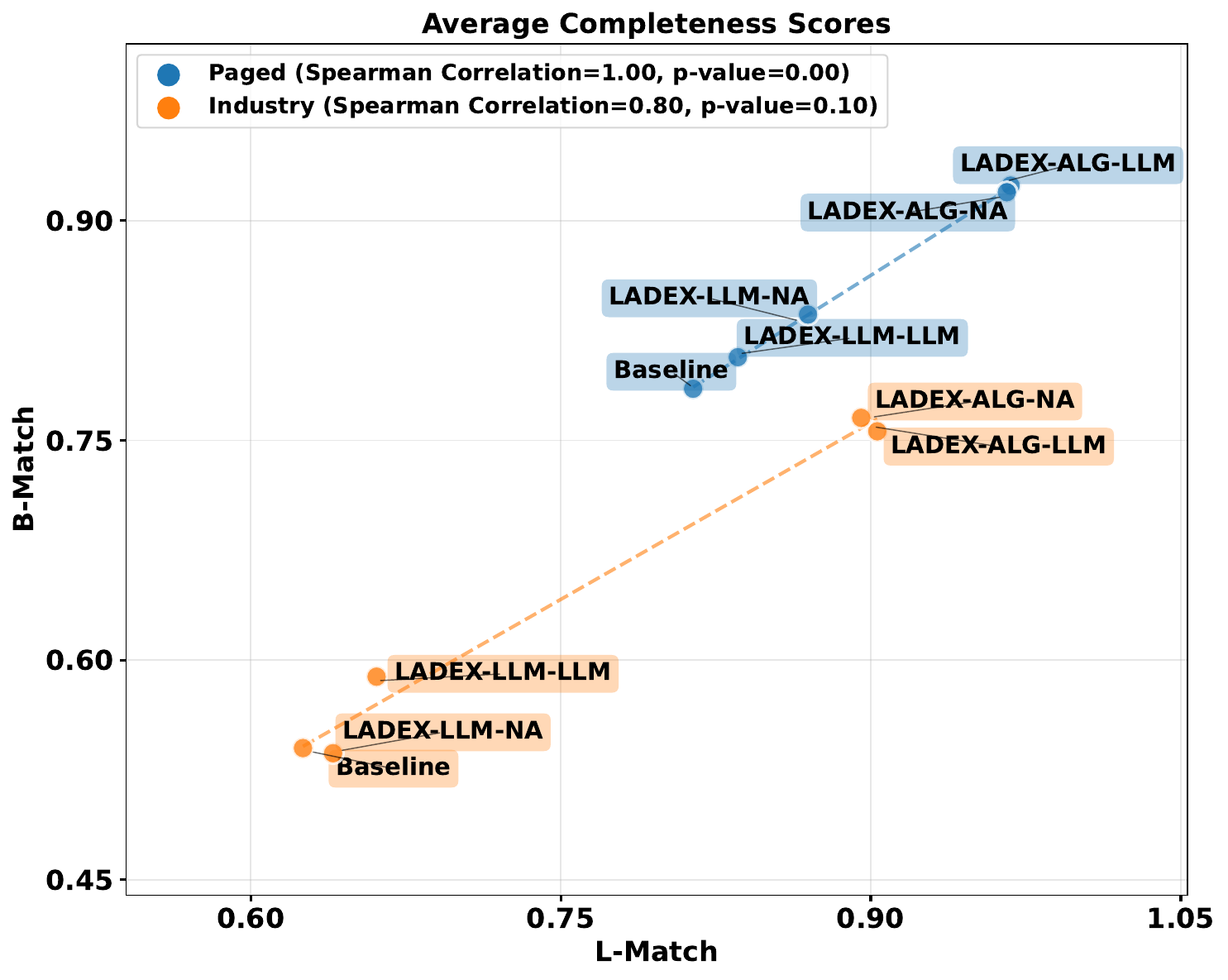}
        \caption{Completeness scores.}
    \end{subfigure}
    \caption{Comparison of average \textit{correctness} and \textit{completeness} scores between L-Match (x-axis) and B-Match (y-axis): Blue and orange points, respectively, represent the average results of each variant across all runs and LLMs for the \textsc{Paged} and Industry datasets.  Dashed lines indicate the best-fit linear trend for each dataset with Spearman correlation coefficients shown in the legend.}
    \label{fig:compare_scores}
    \vspace*{-.3cm}
\end{figure}

To assess agreement between the L-Match and the B-Match evaluation methods, we examine whether the two methods provide consistent comparisons of \approach’s variants. Specifically, when evaluating variants using \textit{correctness} and \textit{completeness} scores -- whether through average values or through statistical tests --  both methods yield consistent conclusions. Below, we assess the agreement between the two methods by first comparing the average \textit{correctness} and \textit{completeness} scores they produce, and then by comparing the statistical tests conducted on those scores:

\emph{(1)~Agreement between L-Match and B-Match based on the average \textit{correctness} and \textit{completeness} scores they produce.} Figure~\ref{fig:compare_scores} shows two plots comparing the L-Match average scores (x-axis) with the B-Match average scores (y-axis) for the five \approach\ variants across our two datasets with  plot (a) showing average \textit{correctness} scores and plot (b) showing average \textit{completeness} scores. Each plot also shows the best-fit linear trend for each dataset. As shown in the figure, across both plots, the relative ordering of the \approach\ variants is consistent between the L-Match and B-Match: whenever a variant ranks higher (or lower) in \textit{correctness} or \textit{completeness} according to the L-Match, B-Match assigns it a correspondingly higher (or lower) position as well. In other words, both  methods agree on the relative performance of the five variants with respect to \textit{correctness} and \textit{completeness}. The Spearman’s rank correlation coefficients~\citep{spearman} for the two plots and the two datasets in Figure~\ref{fig:compare_scores} range from 0.8 to 1. For the \textsc{Paged} dataset, the  correlations for \textit{correctness} and \textit{completeness} are 0.9 and 1.0, respectively. For the Industry dataset, the  correlations for \textit{correctness} and \textit{completeness} are 0.9 and 0.8, respectively. The high coefficient scores confirm the strong monotonic agreement between the two evaluation methods.

\emph{(2)~Agreement between L-Match and B-Match based on the statistical test results.} Table~\ref{table:statistical_comparison} summarizes the outcomes of the statistical tests reported in Table~\ref{table:all_stat_results} for both L-Match and B-Match, across all ten pairwise comparisons of the \approach\ variants on both datasets, three LLMs, and for both \textit{completeness} and \textit{correctness}. As the table shows,  the two evaluation methods never contradict each other: in no case did they identify different variants as significantly better. In every comparison, they either agreed on the better variant, both found no statistically significant difference, or one found a significant difference while the other did not.
\begin{table*}[t]
\centering
\caption{Results of the statistical pairwise comparisons between \approach\ variants (A vs. B) on \textit{completeness} and \textit{correctness} based on the statistical tests reported in Table~\ref{table:all_stat_results}. For each metric and evaluation method, A indicates that variant A performs significantly better, B indicates that variant B performs significantly better, and -- denotes no statistically significant difference.}
\label{table:statistical_comparison}

\begin{minipage}[t]{0.49\textwidth}
\centering
\scalebox{0.42}{
\begin{tabular}{llllcc}
\toprule
\textbf{Comparison (A vs B)} & \textbf{Dataset} & \textbf{LLM} & \textbf{Metric} & \textbf{L-Match} & \textbf{B-Match} \\
\midrule

\multirow{10}{*}{\begin{tabular}[c]{@{}l@{}}\textbf{\approachAlignmentLLMStrucutLLM}\\ \textbf{vs Baseline}\end{tabular}} 
& \multirow{4}{*}{\textbf{\textit{Industry}}} 
& \multirow{2}{*}{\textbf{\textit{O4 Mini}}} & \textbf{\textit{Completeness}} & -- & -- \\
& & & \textbf{\textit{Correctness}} & -- & -- \\
\cline{3-6}
& & \multirow{2}{*}{\textbf{\textit{GPT-4.1 Mini}}} & \textbf{\textit{Completeness}} & -- & -- \\
& & & \textbf{\textit{Correctness}} & -- & -- \\
\cline{2-6}
& \multirow{6}{*}{\textbf{\textit{\textsc{Paged}}}} 
& \multirow{2}{*}{\textbf{\textit{O4 Mini}}} & \textbf{\textit{Completeness}} & -- & -- \\
& & & \textbf{\textit{Correctness}} & A & A \\
\cline{3-6}
& & \multirow{2}{*}{\textbf{\textit{GPT-4.1 Mini}}} & \textbf{\textit{Completeness}} & A & A \\
& & & \textbf{\textit{Correctness}} & -- & -- \\
\cline{3-6}
& & \multirow{2}{*}{\textbf{\textit{DeepSeek-R1}}} & \textbf{\textit{Completeness}} & -- & -- \\
& & & \textbf{\textit{Correctness}} & -- & -- \\

\midrule
\multirow{10}{*}{\begin{tabular}[c]{@{}l@{}}\textbf{\approachAlignmentLLMStrucutFormal}\\ \textbf{vs Baseline}\end{tabular}}
& \multirow{4}{*}{\textbf{\textit{Industry}}} 
& \multirow{2}{*}{\textbf{\textit{O4 Mini}}} & \textbf{\textit{Completeness}} & A & A \\
& & & \textbf{\textit{Correctness}} & -- & A \\
\cline{3-6}
& & \multirow{2}{*}{\textbf{\textit{GPT-4.1 Mini}}} & \textbf{\textit{Completeness}} & A & A \\
& & & \textbf{\textit{Correctness}} & A & A \\
\cline{2-6}
& \multirow{6}{*}{\textbf{\textit{\textsc{Paged}}}} 
& \multirow{2}{*}{\textbf{\textit{O4 Mini}}} & \textbf{\textit{Completeness}} & -- & -- \\
& & & \textbf{\textit{Correctness}} & A & A \\
\cline{3-6}
& & \multirow{2}{*}{\textbf{\textit{GPT-4.1 Mini}}} & \textbf{\textit{Completeness}} & A & A \\
& & & \textbf{\textit{Correctness}} & A & A \\
\cline{3-6}
& & \multirow{2}{*}{\textbf{\textit{DeepSeek-R1}}} & \textbf{\textit{Completeness}} & A & A \\
& & & \textbf{\textit{Correctness}} & A & A \\

\midrule
\multirow{10}{*}{\begin{tabular}[c]{@{}l@{}}\textbf{\approachStrucutLLMOnly}\\ \textbf{vs Baseline}\end{tabular}}
& \multirow{4}{*}{\textbf{\textit{Industry}}} 
& \multirow{2}{*}{\textbf{\textit{O4 Mini}}} & \textbf{\textit{Completeness}} & A & -- \\
& & & \textbf{\textit{Correctness}} & -- & -- \\
\cline{3-6}
& & \multirow{2}{*}{\textbf{\textit{GPT-4.1 Mini}}} & \textbf{\textit{Completeness}} & -- & -- \\
& & & \textbf{\textit{Correctness}} & -- & -- \\
\cline{2-6}
& \multirow{6}{*}{\textbf{\textit{\textsc{Paged}}}} 
& \multirow{2}{*}{\textbf{\textit{O4 Mini}}} & \textbf{\textit{Completeness}} & -- & A \\
& & & \textbf{\textit{Correctness}} & A & A \\
\cline{3-6}
& & \multirow{2}{*}{\textbf{\textit{GPT-4.1 Mini}}} & \textbf{\textit{Completeness}} & A & A \\
& & & \textbf{\textit{Correctness}} & A & A \\
\cline{3-6}
& & \multirow{2}{*}{\textbf{\textit{DeepSeek-R1}}} & \textbf{\textit{Completeness}} & A & A \\
& & & \textbf{\textit{Correctness}} & -- & -- \\

\midrule
\multirow{10}{*}{\begin{tabular}[c]{@{}l@{}}\textbf{\approachStrucutFormalOnly}\\ \textbf{vs Baseline}\end{tabular}}
& \multirow{4}{*}{\textbf{\textit{Industry}}} 
& \multirow{2}{*}{\textbf{\textit{O4 Mini}}} & \textbf{\textit{Completeness}} & A & A \\
& & & \textbf{\textit{Correctness}} & A & A \\
\cline{3-6}
& & \multirow{2}{*}{\textbf{\textit{GPT-4.1 Mini}}} & \textbf{\textit{Completeness}} & A & A \\
& & & \textbf{\textit{Correctness}} & A & A \\
\cline{2-6}
& \multirow{6}{*}{\textbf{\textit{\textsc{Paged}}}} 
& \multirow{2}{*}{\textbf{\textit{O4 Mini}}} & \textbf{\textit{Completeness}} & -- & -- \\
& & & \textbf{\textit{Correctness}} & A & A \\
\cline{3-6}
& & \multirow{2}{*}{\textbf{\textit{GPT-4.1 Mini}}} & \textbf{\textit{Completeness}} & A & A \\
& & & \textbf{\textit{Correctness}} & A & A \\
\cline{3-6}
& & \multirow{2}{*}{\textbf{\textit{DeepSeek-R1}}} & \textbf{\textit{Completeness}} & A & A \\
& & & \textbf{\textit{Correctness}} & A & A \\

\midrule
\multirow{10}{*}{\begin{tabular}[c]{@{}l@{}}\textbf{\approachStrucutFormalOnly}\\ \textbf{vs \approachStrucutLLMOnly}\end{tabular}}
& \multirow{4}{*}{\textbf{\textit{Industry}}} 
& \multirow{2}{*}{\textbf{\textit{O4 Mini}}} & \textbf{\textit{Completeness}} & -- & -- \\
& & & \textbf{\textit{Correctness}} & -- & -- \\
\cline{3-6}
& & \multirow{2}{*}{\textbf{\textit{GPT-4.1 Mini}}} & \textbf{\textit{Completeness}} & A & A \\
& & & \textbf{\textit{Correctness}} & A & A \\
\cline{2-6}
& \multirow{6}{*}{\textbf{\textit{\textsc{Paged}}}} 
& \multirow{2}{*}{\textbf{\textit{O4 Mini}}} & \textbf{\textit{Completeness}} & -- & -- \\
& & & \textbf{\textit{Correctness}} & -- & -- \\
\cline{3-6}
& & \multirow{2}{*}{\textbf{\textit{GPT-4.1 Mini}}} & \textbf{\textit{Completeness}} & A & A \\
& & & \textbf{\textit{Correctness}} & A & A \\
\cline{3-6}
& & \multirow{2}{*}{\textbf{\textit{DeepSeek-R1}}} & \textbf{\textit{Completeness}} & A & A \\
& & & \textbf{\textit{Correctness}} & A & A \\

\bottomrule
\end{tabular}}
\end{minipage}\hfill
%
% SECOND TABLE (SIDE-BY-SIDE)
%
\begin{minipage}[t]{0.49\textwidth}
\centering
\scalebox{0.42}{
\begin{tabular}{llllcc}
\toprule
\textbf{Comparison (A vs B)} & \textbf{Dataset} & \textbf{LLM} & \textbf{Metric} & \textbf{L-Match} & \textbf{B-Match} \\
\midrule

\multirow{10}{*}{\begin{tabular}[c]{@{}l@{}}\textbf{\approachAlignmentLLMStrucutFormal}\\ \textbf{vs \approachAlignmentLLMStrucutLLM}\end{tabular}}
& \multirow{4}{*}{\textbf{\textit{Industry}}} 
& \multirow{2}{*}{\textbf{\textit{O4 Mini}}} & \textbf{\textit{Completeness}} & A & A \\
& & & \textbf{\textit{Correctness}} & -- & A \\
\cline{3-6}
& & \multirow{2}{*}{\textbf{\textit{GPT-4.1 Mini}}} & \textbf{\textit{Completeness}} & A & A \\
& & & \textbf{\textit{Correctness}} & A & A \\
\cline{2-6}
& \multirow{6}{*}{\textbf{\textit{\textsc{Paged}}}} 
& \multirow{2}{*}{\textbf{\textit{O4 Mini}}} & \textbf{\textit{Completeness}} & -- & -- \\
& & & \textbf{\textit{Correctness}} & -- & -- \\
\cline{3-6}
& & \multirow{2}{*}{\textbf{\textit{GPT-4.1 Mini}}} & \textbf{\textit{Completeness}} & A & A \\
& & & \textbf{\textit{Correctness}} & A & A \\
\cline{3-6}
& & \multirow{2}{*}{\textbf{\textit{DeepSeek-R1}}} & \textbf{\textit{Completeness}} & A & A \\
& & & \textbf{\textit{Correctness}} & A & A \\

\midrule
\multirow{10}{*}{\begin{tabular}[c]{@{}l@{}}\textbf{\approachStrucutFormalOnly}\\ \textbf{vs \approachAlignmentLLMStrucutLLM}\end{tabular}}
& \multirow{4}{*}{\textbf{\textit{Industry}}} 
& \multirow{2}{*}{\textbf{\textit{O4 Mini}}} & \textbf{\textit{Completeness}} & -- & -- \\
& & & \textbf{\textit{Correctness}} & A & A \\
\cline{3-6}
& & \multirow{2}{*}{\textbf{\textit{GPT-4.1 Mini}}} & \textbf{\textit{Completeness}} & A & A \\
& & & \textbf{\textit{Correctness}} & A & A \\
\cline{2-6}
& \multirow{6}{*}{\textbf{\textit{\textsc{Paged}}}} 
& \multirow{2}{*}{\textbf{\textit{O4 Mini}}} & \textbf{\textit{Completeness}} & -- & -- \\
& & & \textbf{\textit{Correctness}} & -- & -- \\
\cline{3-6}
& & \multirow{2}{*}{\textbf{\textit{GPT-4.1 Mini}}} & \textbf{\textit{Completeness}} & A & A \\
& & & \textbf{\textit{Correctness}} & A & A \\
\cline{3-6}
& & \multirow{2}{*}{\textbf{\textit{DeepSeek-R1}}} & \textbf{\textit{Completeness}} & A & A \\
& & & \textbf{\textit{Correctness}} & A & A \\

\midrule
\multirow{10}{*}{\begin{tabular}[c]{@{}l@{}}\textbf{\approachStrucutFormalOnly}\\ \textbf{vs \approachAlignmentLLMStrucutFormal}\end{tabular}}
& \multirow{4}{*}{\textbf{\textit{Industry}}} 
& \multirow{2}{*}{\textbf{\textit{O4 Mini}}} & \textbf{\textit{Completeness}} & B & -- \\
& & & \textbf{\textit{Correctness}} & A & -- \\
\cline{3-6}
& & \multirow{2}{*}{\textbf{\textit{GPT-4.1 Mini}}} & \textbf{\textit{Completeness}} & -- & -- \\
& & & \textbf{\textit{Correctness}} & -- & -- \\
\cline{2-6}
& \multirow{6}{*}{\textbf{\textit{\textsc{Paged}}}} 
& \multirow{2}{*}{\textbf{\textit{O4 Mini}}} & \textbf{\textit{Completeness}} & -- & -- \\
& & & \textbf{\textit{Correctness}} & B & B \\
\cline{3-6}
& & \multirow{2}{*}{\textbf{\textit{GPT-4.1 Mini}}} & \textbf{\textit{Completeness}} & -- & -- \\
& & & \textbf{\textit{Correctness}} & -- & B \\
\cline{3-6}
& & \multirow{2}{*}{\textbf{\textit{DeepSeek-R1}}} & \textbf{\textit{Completeness}} & -- & -- \\
& & & \textbf{\textit{Correctness}} & -- & -- \\

\midrule
\multirow{10}{*}{\begin{tabular}[c]{@{}l@{}}\textbf{\approachStrucutLLMOnly}\\ \textbf{vs \approachAlignmentLLMStrucutLLM}\end{tabular}}
& \multirow{4}{*}{\textbf{\textit{Industry}}} 
& \multirow{2}{*}{\textbf{\textit{O4 Mini}}} & \textbf{\textit{Completeness}} & A & -- \\
& & & \textbf{\textit{Correctness}} & A & A \\
\cline{3-6}
& & \multirow{2}{*}{\textbf{\textit{GPT-4.1 Mini}}} & \textbf{\textit{Completeness}} & B & B \\
& & & \textbf{\textit{Correctness}} & B & B \\
\cline{2-6}
& \multirow{6}{*}{\textbf{\textit{\textsc{Paged}}}} 
& \multirow{2}{*}{\textbf{\textit{O4 Mini}}} & \textbf{\textit{Completeness}} & -- & -- \\
& & & \textbf{\textit{Correctness}} & -- & -- \\
\cline{3-6}
& & \multirow{2}{*}{\textbf{\textit{GPT-4.1 Mini}}} & \textbf{\textit{Completeness}} & A & A \\
& & & \textbf{\textit{Correctness}} & A & A \\
\cline{3-6}
& & \multirow{2}{*}{\textbf{\textit{DeepSeek-R1}}} & \textbf{\textit{Completeness}} & -- & -- \\
& & & \textbf{\textit{Correctness}} & -- & B \\

\midrule
\multirow{10}{*}{\begin{tabular}[c]{@{}l@{}}\textbf{\approachStrucutLLMOnly}\\ \textbf{vs \approachAlignmentLLMStrucutFormal}\end{tabular}}
& \multirow{4}{*}{\textbf{\textit{Industry}}} 
& \multirow{2}{*}{\textbf{\textit{O4 Mini}}} & \textbf{\textit{Completeness}} & -- & -- \\
& & & \textbf{\textit{Correctness}} & -- & -- \\
\cline{3-6}
& & \multirow{2}{*}{\textbf{\textit{GPT-4.1 Mini}}} & \textbf{\textit{Completeness}} & B & B \\
& & & \textbf{\textit{Correctness}} & B & B \\
\cline{2-6}
& \multirow{6}{*}{\textbf{\textit{\textsc{Paged}}}} 
& \multirow{2}{*}{\textbf{\textit{O4 Mini}}} & \textbf{\textit{Completeness}} & -- & -- \\
& & & \textbf{\textit{Correctness}} & -- & B \\
\cline{3-6}
& & \multirow{2}{*}{\textbf{\textit{GPT-4.1 Mini}}} & \textbf{\textit{Completeness}} & B & B \\
& & & \textbf{\textit{Correctness}} & B & B \\
\cline{3-6}
& & \multirow{2}{*}{\textbf{\textit{DeepSeek-R1}}} & \textbf{\textit{Completeness}} & B & B \\
& & & \textbf{\textit{Correctness}} & B & B \\

\bottomrule
\end{tabular}}
\end{minipage}
\end{table*}

\begin{tcolorbox}[breakable,colback=gray!10!white,colframe=black!75!black,boxrule=0.5mm,arc=1mm,left=1mm,right=1mm,top=1mm,bottom=1mm,fonttitle=\bfseries]
The answer to \emph{RQ1} is that the comparisons between LLM-generated activity diagrams and their ground truths are highly consistent across the two evaluation methods --  
L-Match and B-Match. The two evaluation methods produce highly correlated \textit{correctness} and \textit{completeness} scores (Spearman’s $\rho = 0.8$--$1.0$) and never yield conflicting statistical conclusions. 
\end{tcolorbox}

\subsection{RQ2 - Impact of the Refinement Loop}
\label{sec:res_main}

We  compare the four \approach\ variants that include the refinement loop against Baseline using the  metrics in Section~\ref{sec:metrics}:

\emph{Structural consistency.} On average, 21.80\% of the activity diagrams generated by Baseline are structurally unsound, whereas the variants with a refinement loop and algorithmic structural checks (i.e., \approachAlignmentLLMStrucutFormal\ and \approachStrucutFormalOnly) produce no structurally unsound activity diagrams. 

\begin{figure}[t]
    \centering
    \begin{subfigure}[b]{0.48\linewidth}
        \includegraphics[width=\linewidth]{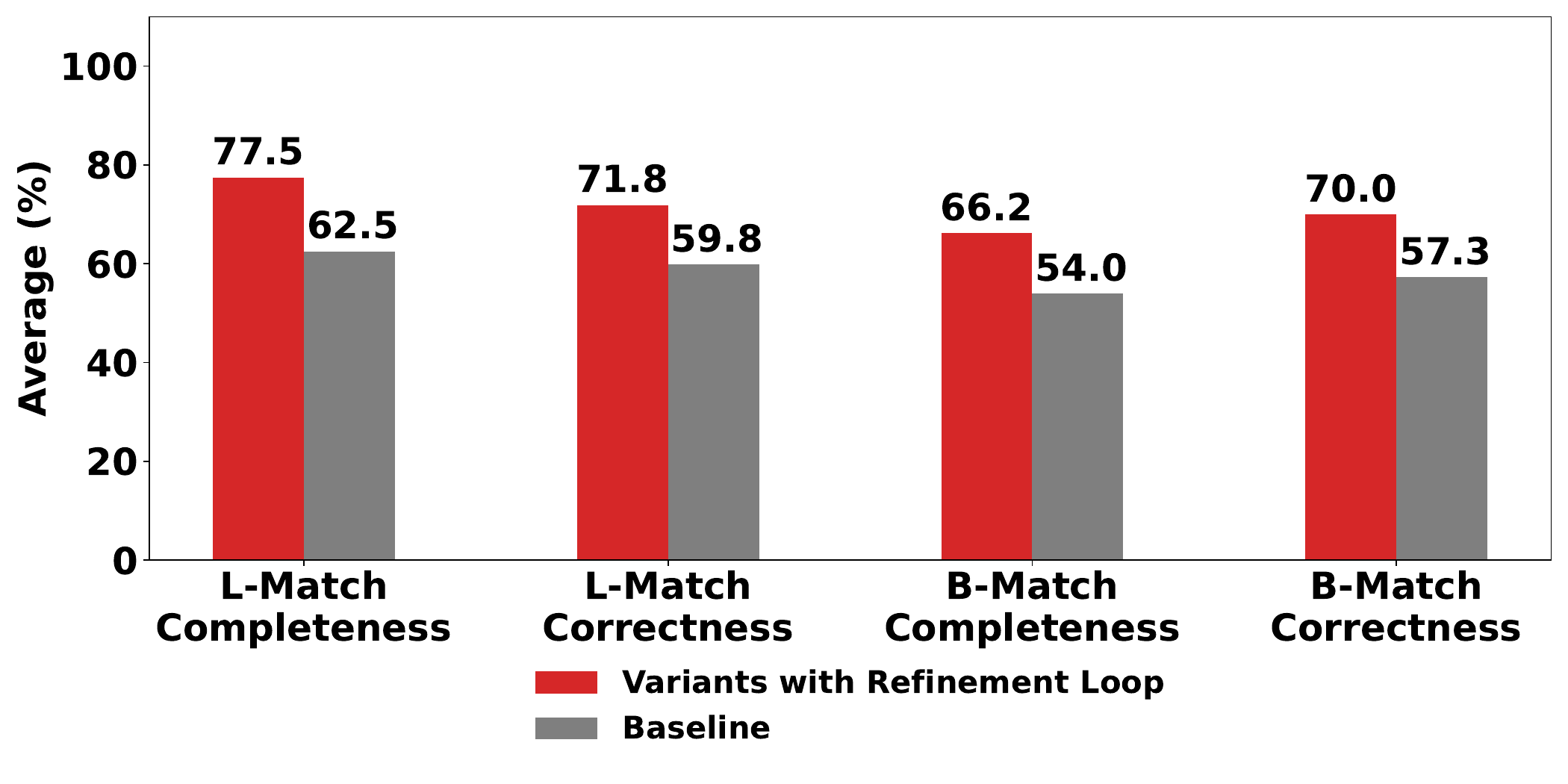}
        \caption{Industry Dataset.}
    \end{subfigure}
    \hfill
    \begin{subfigure}[b]{0.48\linewidth}
        \includegraphics[width=\linewidth]{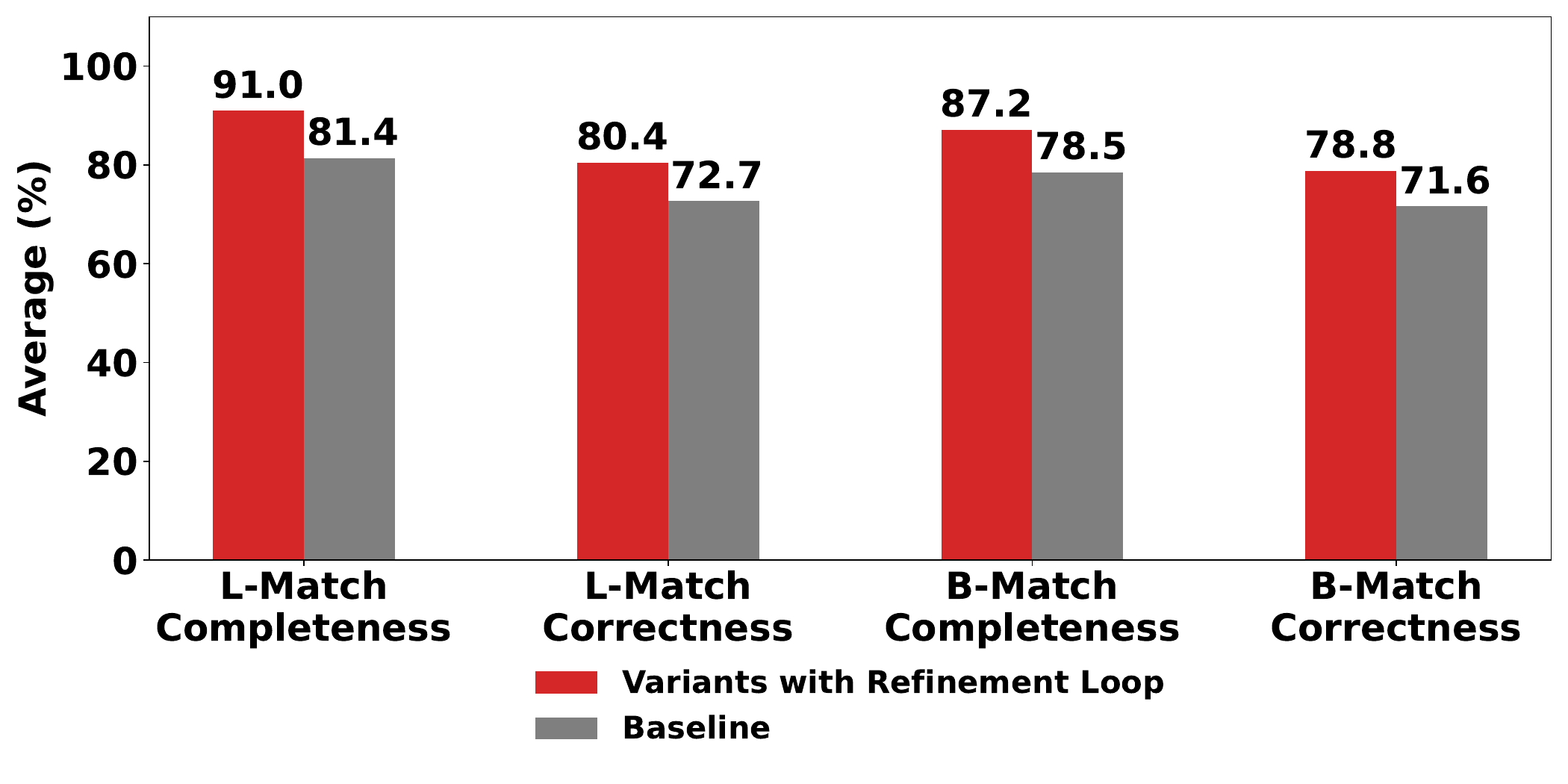}
        \caption{\textsc{Paged} Dataset.}
    \end{subfigure}
    \caption{Comparison of the \approach\ variants with a refinement loop and Baseline in terms of average \emph{correctness} and \emph{completeness}, obtained using  L-Match and B-Match  on the \textsc{Paged} and Industry datasets. All values shown in the bar charts are averaged across all evaluated LLMs. The red bars represent the \approach\ variants with a refinement loop, and the grey bars represent Baseline. }
    \label{fig:rq2-compare}
    \vspace*{-.3cm}
\end{figure}

\emph{Correctness and completeness.} Across the 40 comparisons (4 variant pairs $\times$ 5 datasets-LLM pairs $\times$ 2 evaluation methods), the variants of \approach\ that include a refinement loop significantly outperform Baseline in terms of \textit{correctness} in 25 comparisons and in terms of \textit{completeness} in 24 comparisons with large to negligible effect sizes. In contrast, Baseline never outperforms any \approach\ variants that include a refinement loop. In addition, as shown in Figure~\ref{fig:rq2-compare}, the average \textit{correctness} and \textit{completeness} scores for the variants with a refinement loop are always higher than those for Baseline across both datasets and under both evaluation methods, i.e., L-Match and B-Match.

\emph{Cost.} As expected, the variants with a refinement loop require more LLM calls than Baseline. Specifically, \approachAlignmentLLMStrucutFormal, \approachStrucutFormalOnly, \approachAlignmentLLMStrucutLLM, and \approachStrucutLLMOnly\ require, on average, 6.25, 0.87, 3.17, and 3.65 additional LLM calls, respectively, compared to Baseline. Regarding token usage, including the refinement loop increases input token usage -- an expected outcome given the additional LLM calls that may include prior activity diagram history and the critique, hence increasing average input sizes. However, the number of output and reasoning tokens remains consistent across all variants and LLMs.

\begin{tcolorbox}[breakable,colback=gray!10!white,colframe=black!75!black,boxrule=0.5mm,arc=1mm,left=1mm,right=1mm,top=1mm,bottom=1mm,fonttitle=\bfseries]
The answer to \emph{RQ2} is that  the inclusion of a refinement loop results in activity diagrams that are more likely to be structurally correct, as well as more semantically correct and complete, compared with those generated without a refinement loop. 
Depending on whether and how structural and alignment constraints are checked, the refinement loop leads to an average of 0.87 to 6.25 more LLM calls compared with not having a refinement loop. 
\end{tcolorbox}

\subsection{RQ3 and RQ4 - Impact of Structural and Alignment Checking}
\label{sec:res_ablation}
\textbf{RQ3 (LLM-based  vs. Algorithmic Structural Checking).} We compare the variants of \approach\ that employ algorithmic structural checking  with those that rely on LLM-based structural checking  using the  metrics in Section~\ref{sec:metrics}. Specifically, we compare 
\approachAlignmentLLMStrucutFormal\ with \approachAlignmentLLMStrucutLLM, and  \approachStrucutFormalOnly\  with \approachStrucutLLMOnly.

\emph{Structural consistency.} The variants that perform structural checking algorithmically always generate structurally sound activity diagrams. In contrast, on average, 19.48\% and 18.34\% of the activity diagrams generated by \approachAlignmentLLMStrucutLLM\ and \approachStrucutLLMOnly\ are structurally unsound, respectively.

\begin{figure}[t]
    \centering
    \begin{subfigure}[b]{0.48\linewidth}
        \includegraphics[width=\linewidth]{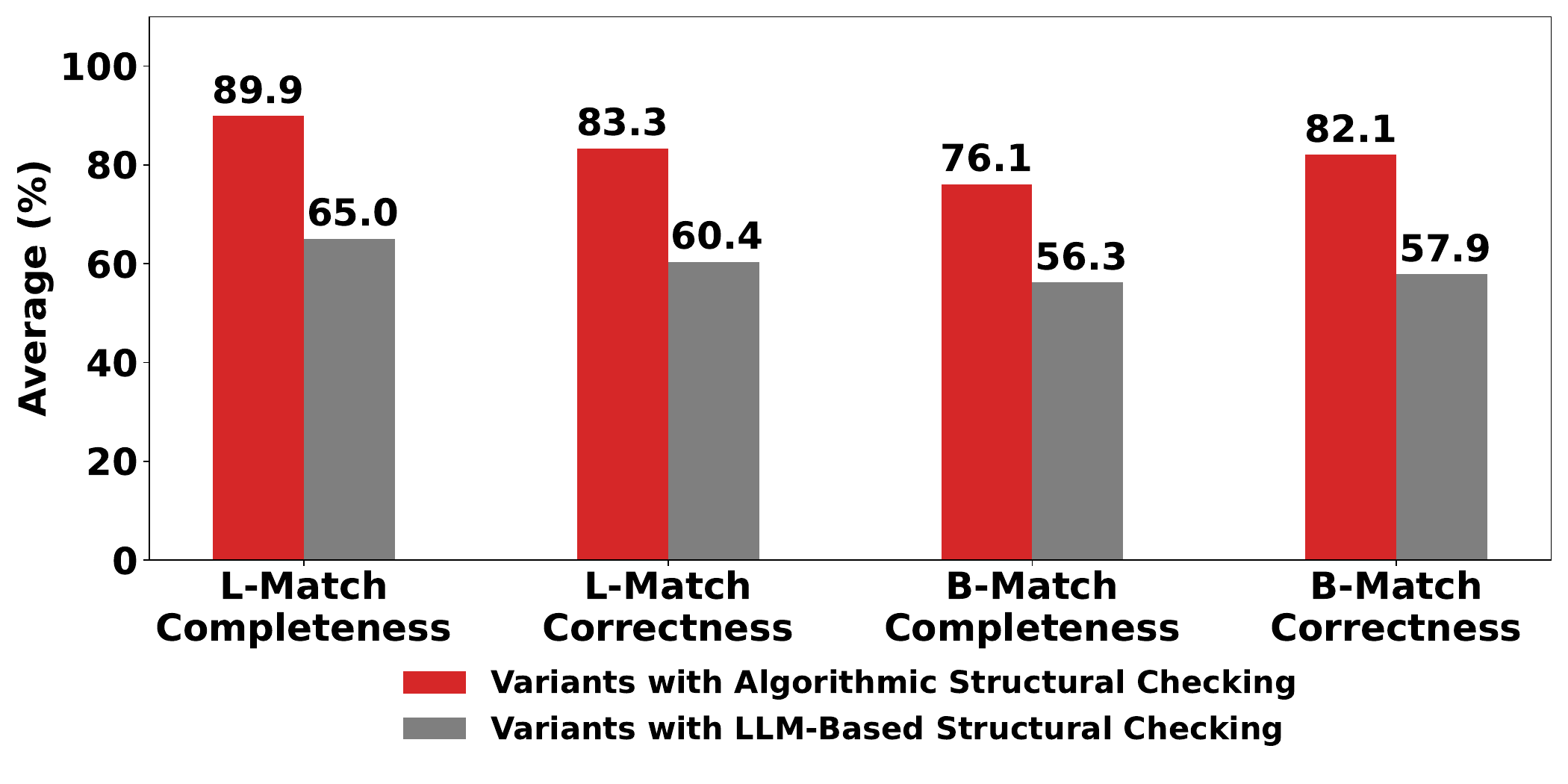}
        \caption{Industry Dataset.}
    \end{subfigure}
    \hfill
    \begin{subfigure}[b]{0.48\linewidth}
        \includegraphics[width=\linewidth]{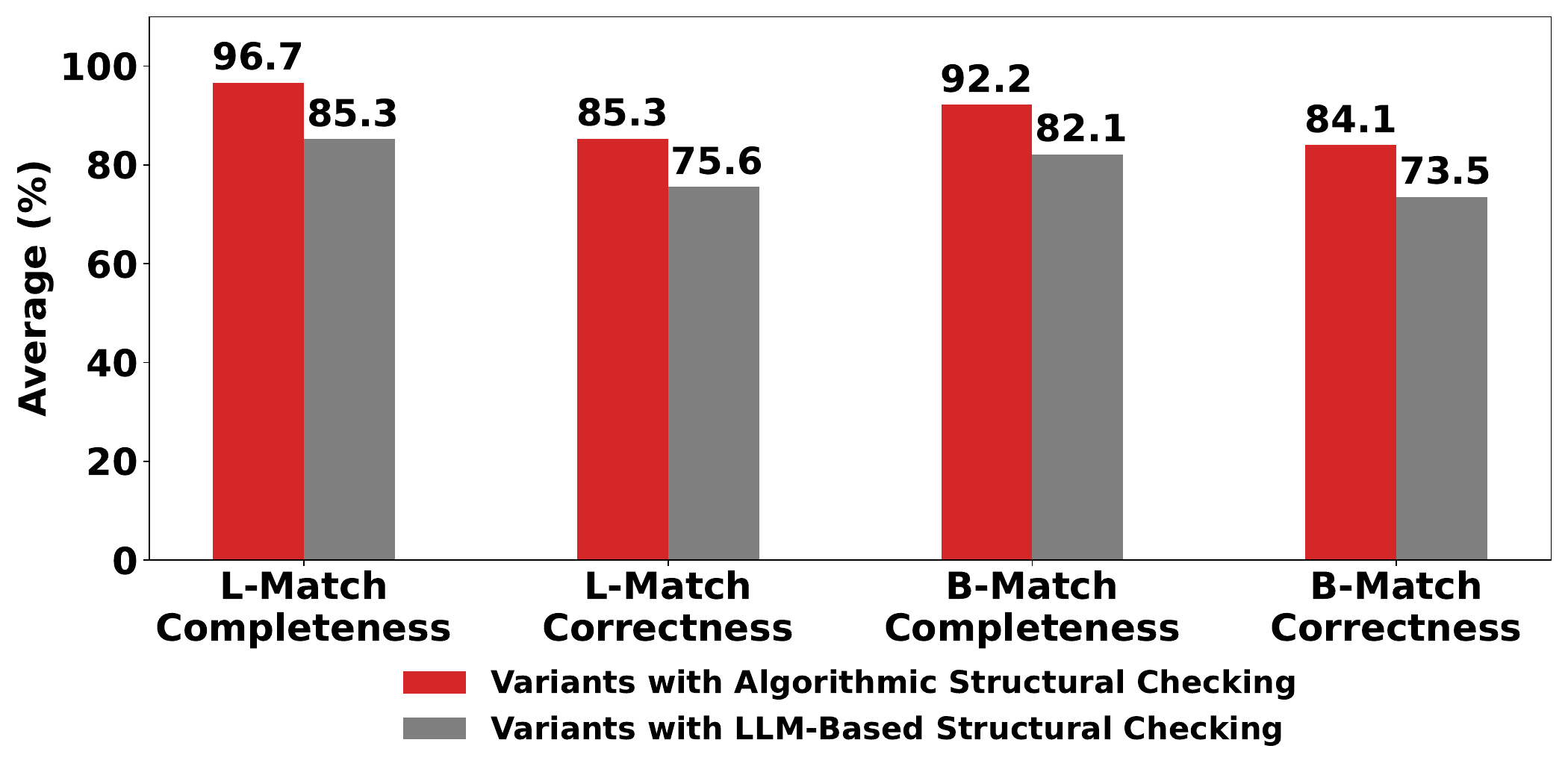}
        \caption{\textsc{Paged} Dataset.}
    \end{subfigure}
    \caption{Comparison of the \approach\ variants using algorithmic versus LLM-based structural checking, showing the average semantic correctness and completeness results obtained from L-Match and the B-Match on the \textsc{Paged} and Industry datasets. All values shown in the bar charts are averaged across all evaluated LLMs.  The red bars represent \approach\ variants with algorithmic structural checking, while the grey bars represent variants with LLM-based structural checking.}
    \label{fig:rq3-compare}
    \vspace*{-.3cm}
\end{figure}

\emph{Correctness and completeness.} Across the 20 comparisons (2 variant pairs $\times$ 5 datasets-LLM pairs $\times$ 2 evaluation methods), the variants of \approach\ that use algorithmic structural checking significantly outperform their LLM-based counterparts in in 13 out of 20 comparisons in terms of \textit{correctness} and 14 out of 20 comparisons in terms \textit{completeness} with large to negligible effect sizes. 
In addition, as shown in Figure~\ref{fig:rq3-compare}, the average \textit{correctness} and \textit{completeness} scores of the algorithmic variants are always higher than those of the LLM-based variants across both datasets and both evaluation methods, i.e., L-Match and B-Match.
On average, the algorithmic variants produce activity diagrams that are up to 23.43\% and 8.77\% more correct, and up to 24.22\% and 9.79\% more complete for the Industry and \textsc{Paged} datasets, respectively, when evaluated with L-Match. For B-Match, the corresponding improvements reach up to 24.63\%, 10.98\%, 17.67\%, and 8.8\%.

\emph{Cost.} \approachStrucutFormalOnly\ and \approachAlignmentLLMStrucutFormal\ require an average of 1.87 and 7.24 LLM calls, respectively, compared to 4.65 and 4.17 calls for \approachStrucutLLMOnly\ and \approachAlignmentLLMStrucutLLM. One might expect that performing structural checks algorithmically would reduce the number of LLM calls by avoiding structural issues misidentified by the LLM, thereby requiring fewer refinement loops. Our results partially confirm this intuition: when alignment checking is absent, algorithmic structural checks indeed reduce the number of calls -- \approachStrucutFormalOnly\ uses, on average, 2.49 times fewer calls than \approachStrucutLLMOnly. However, when alignment checking is present, algorithmic structural checks have the opposite effect, requiring, on average, 1.73 times more calls for \approachAlignmentLLMStrucutFormal\ compared to \approachAlignmentLLMStrucutLLM.  This increase is especially pronounced in the results obtained with the instruction-following LLM, i.e., GPT-4.1 Mini, where \approachAlignmentLLMStrucutFormal\ requires, on average, 3.07 times as many calls as \approachAlignmentLLMStrucutLLM. In contrast, reasoning-based LLMs, i.e., O4 Mini and DeepSeek, show little to no such increase. 

\begin{figure}[t]
\centering
  \includegraphics[width=\linewidth]{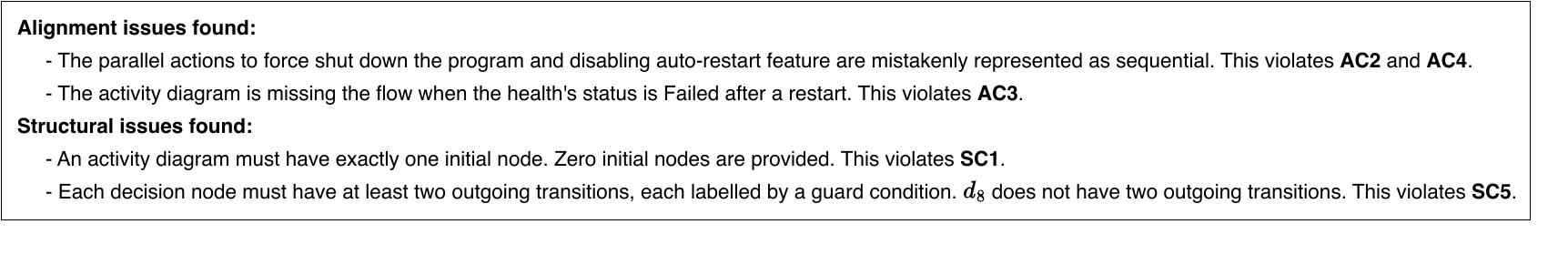}
  \vspace*{-3em}
    \caption{Critique of Figure~\ref{figure:examp}(a) generated by the \approachAlignmentLLMStrucutFormal\ variant.}
    \label{figure:alg-llm-critique-example}
    \vspace*{-.2cm}
\end{figure}

This rather counterintuitive increase in the number of LLM calls arises because when both structural and alignment constraints are evaluated by LLMs, the critique tends to be shorter: the LLM is able to identify flaws that simultaneously trigger both alignment and structural violations. For example, the critique in Figure~\ref{figure:examp}(b), generated by an LLM performing both alignment and structural checking, indicates, in a single item, that a missing flow violates both AC3 and SC5. In contrast, the critique in Figure~\ref{figure:alg-llm-critique-example}, which corresponds to the same example as Figure~\ref{figure:examp} but uses algorithmic structural checking with LLM-based alignment checking, misses the shared root cause, and the missing flow appears twice -- once for AC3 and once for SC5 -- resulting in a longer critique. Based on what we observed in our experiments, longer critiques appear to sometimes cause instruction-following LLMs to overreact by making more changes per refinement, potentially leading to additional refinement rounds. As a result, \approachAlignmentLLMStrucutFormal\ tends to require more LLM calls than \approachAlignmentLLMStrucutLLM, especially when the underlying LLM is instruction-following rather than reasoning-based.
Finally, as detailed in RQ2, because all four variants compared here have a refinement loop, their input, reasoning, and output token usage remains consistent, making the number of LLM calls the primary driver of cost differences.

\begin{tcolorbox}[breakable,colback=gray!10!white, colframe=black!75!black, boxrule=0.5mm, arc=1mm, left=1mm, right=1mm, top=1mm, bottom=1mm, fonttitle=\bfseries]
The answer to \emph{RQ3} is that, in our experiments,  activity diagrams refined based on algorithmic structural checks achieve structural consistency, whereas those refined based on LLM-based checks often still show structural inconsistencies.  Furthermore, across both evaluation methods, algorithmic checking of structural constraints results in an average improvement of 16.95\% in \textit{correctness} and 15.12\% in \textit{completeness} of the generated activity diagrams compared to performing these checks using LLMs.
\end{tcolorbox}

\begin{figure}[t]
    \centering
    \begin{subfigure}[b]{0.48\linewidth}
        \includegraphics[width=\linewidth]{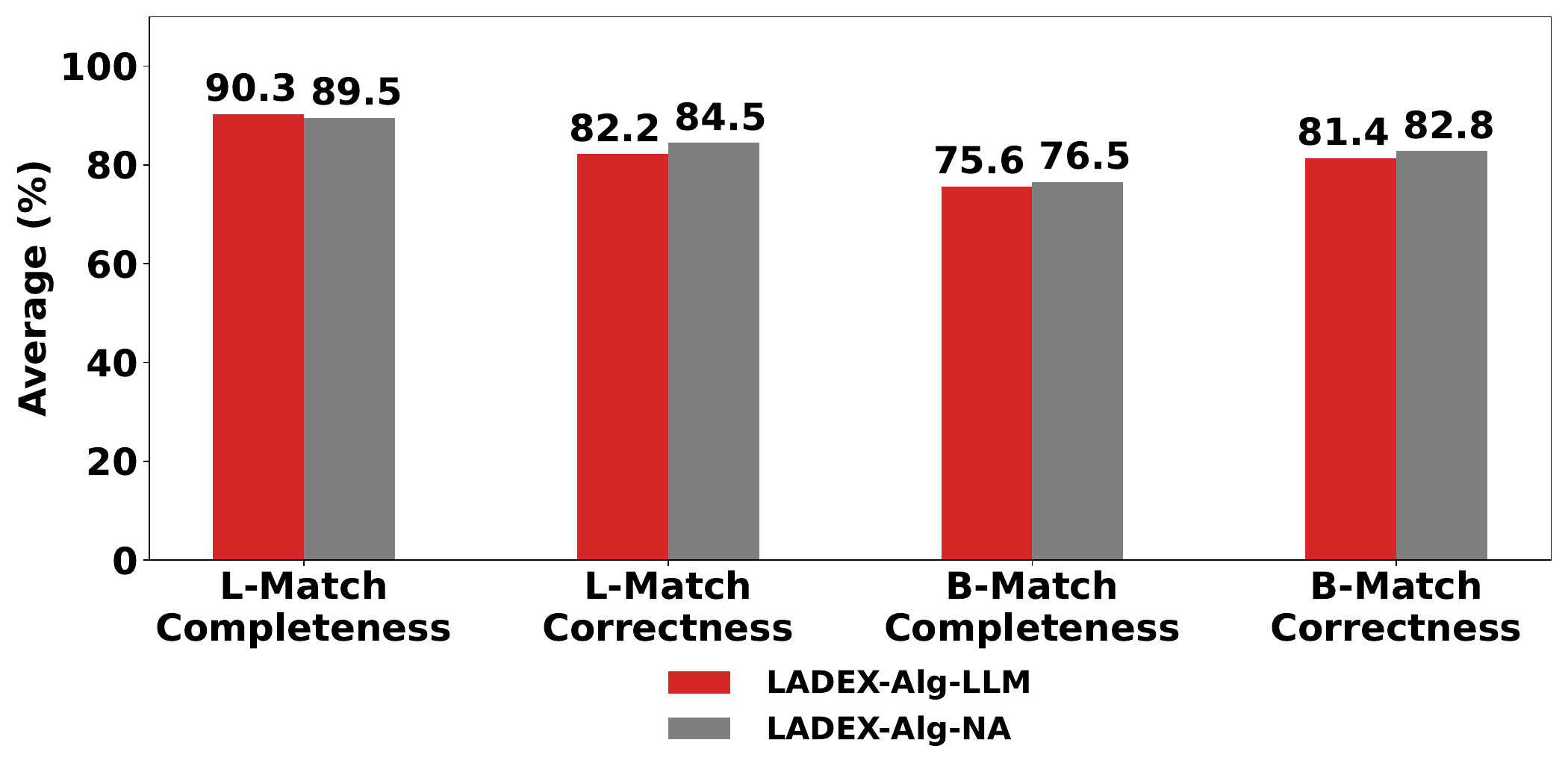}
        \caption{Industry Dataset.}
    \end{subfigure}
    \hfill
    \begin{subfigure}[b]{0.48\linewidth}
        \includegraphics[width=\linewidth]{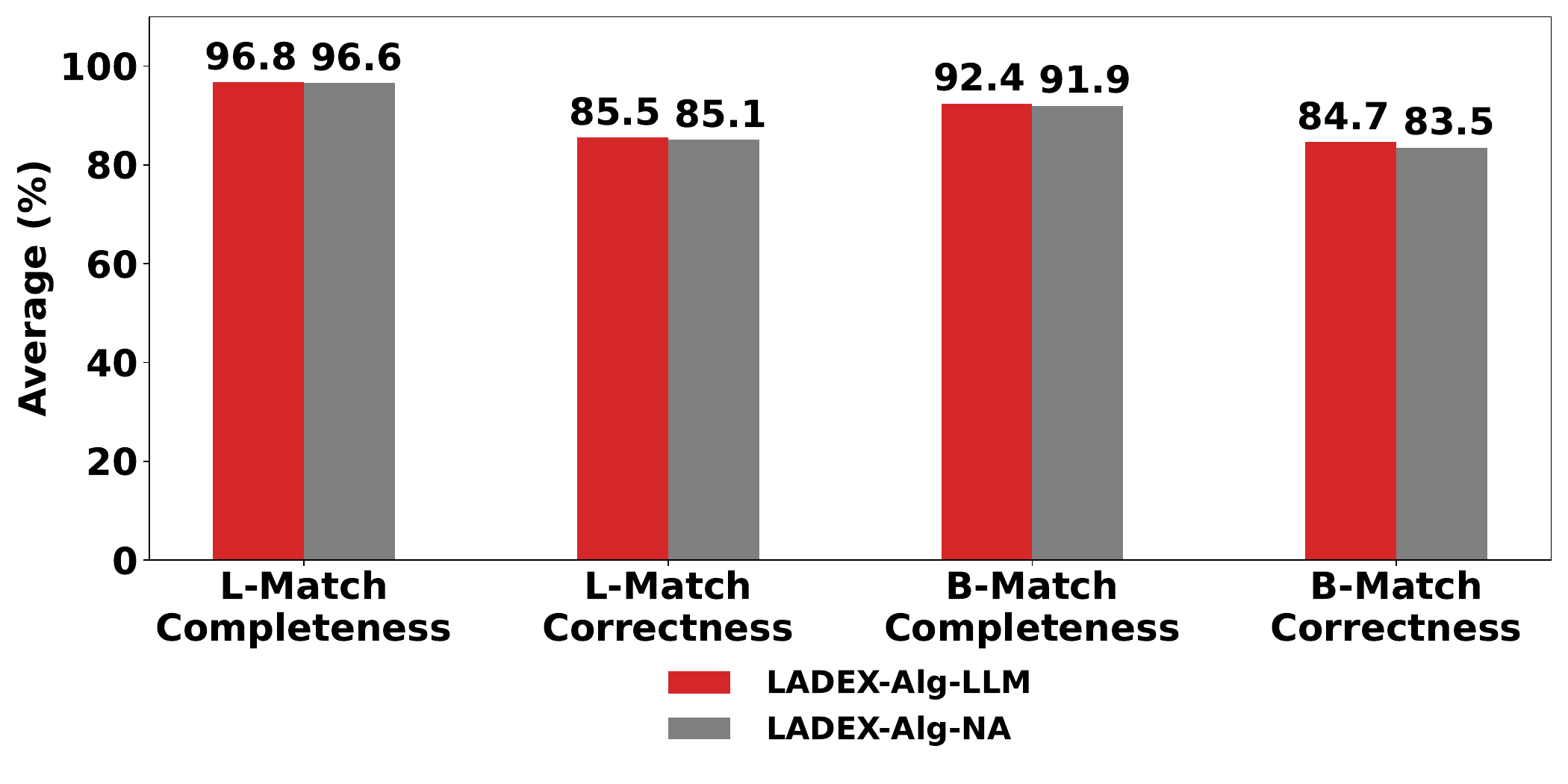}
        \caption{\textsc{Paged} Dataset.}
    \end{subfigure}
    \caption{Comparison of the best-performing variants of \approach, \approachAlignmentLLMStrucutFormal\ and \approachStrucutFormalOnly, showing the average \textit{semantic correctness} and \textit{semantic completeness} scores obtained using L-L-Match and  B-Match on the \textsc{Paged} and Industry datasets. All values shown in the bar charts are averaged across all evaluated LLMs. Red bars represent \approachAlignmentLLMStrucutFormal, and grey bars represent \approachStrucutFormalOnly.}
    \label{fig:rq4-compare}
    \vspace*{-.3cm}
\end{figure}

\textbf{RQ4 (Impact of Alignment Checking).} We compare the best-performing variant of \approach\ with only structural constraint checking (\approachStrucutFormalOnly) with the best-performing variant of \approach\ with both structural constraints and alignment checking (\approachAlignmentLLMStrucutFormal). As discussed earlier, in our experiments, both variants yield structurally sound activity diagrams. Therefore, we compare them in terms of  \textit{correctness}, \textit{completeness}, and \textit{cost}:

\emph{Correctness and completeness.} When evaluated on the Industry dataset using O4 Mini and L-Match, the results indicate a trade-off: \approachAlignmentLLMStrucutFormal\ significantly improves \textit{completeness} over \approachStrucutFormalOnly\ with a negligible effect size, while \approachStrucutFormalOnly\ yields significantly higher \textit{correctness} with a small effect size. This suggests that while alignment checking forces the LLM (specifically O4 Mini) to capture more process details, it introduces a higher risk of semantic errors within complex, proprietary domains. We found no statistically significant difference between the variants when using GPT-4.1 Mini, nor when applying B-Match for either LLM.
For the \textsc{Paged} dataset, the benefits of alignment checking are more consistent. While \textit{completeness} remains comparable across both variants and all LLMs, \approachAlignmentLLMStrucutFormal\ significantly outperforms \approachStrucutFormalOnly\ in \textit{correctness} in three out of six comparisons (3~LLMs~$\times$~2~evaluation methods) with negligible effect sizes.
Overall, Figure~\ref{fig:rq4-compare} shows that the average \textit{semantic correctness} and \textit{completeness} for both variants remain comparable across all datasets, LLMs and evaluation methods.

\emph{Cost.}  As discussed in RQ3, combining algorithmic structural checking with alignment checking increases the number of required LLM calls because \approachAlignmentLLMStrucutFormal\ tends to generate longer critiques with alignment and structural issues presented separately. Overall, \approachAlignmentLLMStrucutFormal\ requires, on average, 5.38 more LLM calls than \approachStrucutFormalOnly. 
Finally, because both \approachAlignmentLLMStrucutFormal\ and \approachStrucutFormalOnly\ have refinement loops, their input, reasoning, and output token usage remains consistent, leaving the number of LLM calls as the main metric that affects cost.

\begin{tcolorbox}[breakable,colback=gray!10!white, colframe=black!75!black, boxrule=0.5mm, arc=1mm, left=1mm, right=1mm, top=1mm, bottom=1mm, fonttitle=\bfseries]
The answer to \emph{RQ4} is that  combining LLM-based alignment checking with algorithmic structural checking improves correctness on the \textsc{Paged} dataset. However, for the  \textsc{Industry} dataset, alignment checking creates a trade-off: it improves completeness but 
degrades correctness compared to the variant with only algorithmic structural checking.
Combining algorithmic structural checking with LLM-based alignment checking, using  O4 Mini, produces structurally sound activity diagrams with an average \textit{correctness} of 86\% and an average \textit{completeness} of 92\% across our two datasets and both evaluation methods, while requiring 4.91 LLM calls on average. As an alternative, we observe in our experiments that applying only algorithmic structural checking -- without alignment checking -- still produces structurally sound activity diagrams, with an average \textit{correctness} of 86\% and an average \textit{completeness} of 90\%, while reducing LLM calls to an average of 1.08, using the same LLM. 
\end{tcolorbox}

\subsection{Threats to Validity} 
\label{sec:threats}
\textbf{Internal Validity.} To enable a fair comparison across \approach\ variants, we ensured that all reported results were produced using identical prompts and identical LLM configurations, run on the same underlying LLMs (GPT-4.1 Mini~\citep{openai2024gpt4ocard}, O4 Mini~\citep{jaech2024openai}, and DeepSeek-R1-Distill-Llama-70B~\citep{guo2025deepseek} for the \textsc{Paged} dataset; GPT-4.1 Mini and O4 Mini for the Industry dataset). 
To mitigate randomness, we repeated each experiment five times. Furthermore, we set the temperature of the instruction-following LLM (GPT-4.1 Mini) to 0.0, as is standard practice for reducing randomness in such LLMs~\citep{guo2025deepseek, openai2024gpt4ocard}. For the reasoning-based LLMs, the O4 Mini LLM does not allow manual adjustment of the temperature, and we executed DeepSeek with its recommended temperature value of 0.6, which is considered optimal for reasoning~\citep{guo2025deepseek, azure_openai_reasoning}. Regarding data leakage, we are confident that our industry dataset, being proprietary, has not been part of any LLM's training data. 
To support further research, we provide an anonymized subset of this dataset online~\citep{replicationPackage}.  
The PAGED dataset is public-domain and may have been included in the training data of LLMs. However, all variants of \approach\ use the same LLMs. Thus, any potential data leakage would affect all variants equally. To evaluate \approach's variants using the LLM matcher, we used the same prompts, the same underlying LLM, and the same LLM configuration. To demonstrate the reliability of the LLM matcher, we show that its outputs correlate strongly with expert judgments and are consistent with the results of our behavioural matching algorithm. Finally, studying how different underlying LLMs impact the evaluation results obtained with the LLM-based matcher is left for future work.
A potential threat arises from the choice of output representation. Direct LLM generation of standard-compliant serializations, such as UML XMI~\citep{Cook2017}, can be prone to syntactic hallucinations, malformed outputs, and metamodel violations because such formats require consistent identifiers, nested elements, and strict conformance to UML metamodel constraints. We mitigated this threat by designing \approach\ to generate a simpler CSV-based intermediate representation that captures the relevant activity-diagram elements and control-flow semantics. As we discuss in Section~\ref{sec:tools}, we provide tool support to transform generated activity diagrams in our CVS-based representation into standard-compliant UML XMI.

\textbf{Conclusion Validity.} We note that when evaluating multiple alternatives -- in our work, different variants of the proposed approach -- some researchers advise initially conducting the Kruskal–Wallis Test to determine if there are significant differences among the alternatives as a whole before proceeding with pairwise  Wilcoxon Rank-Sum Tests. However, because we had only a small number of pairwise comparisons and were primarily focused on direct comparisons between pairs of alternatives, we chose to skip the Kruskal–Wallis Test and move directly to  Wilcoxon Rank-Sum Tests. 
In addition, since running multiple statistical tests can increase the risk of Type I error inflation, we apply the Benjamini-Hochberg (BH) procedure~\citep{benjamini1995controlling} to control the false discovery rate.  All statistical significance tests in our experiments are reported using BH-adjusted p-values.
Finally, we verified the statistical sensitivity of our evaluation. With our  sample sizes ($1000$ for \textsc{Paged} and $100$ for Industry), a power analysis using the Asymptotic Relative Efficiency (ARE) method~\citep{lehmann2006nonparametrics} indicates that we achieve an average observed power of $>0.99$ to detect small effect sizes ($0.56 \leq \hat{A}_{12} < 0.64$ or $0.36 < \hat{A}_{12} \leq 0.44$) on the \textsc{Paged} dataset and $0.55$ on the Industry dataset. Our empirical results confirm this sensitivity: on the \textsc{Paged} dataset, 100\% of the 42 identified small effects reached statistical significance ($p < 0.05$); on the Industry dataset, 20 of the 23 identified small effects (87\%) reached significance. While the smaller sample size of the Industry dataset inherently increases the risk of missing  minor performance variations, our high success rate in identifying significant small effects demonstrates that the evaluation is sufficiently sensitive to capture meaningful differences in both datasets.

\textbf{External Validity.} To improve external validity, our evaluation used two datasets: the proprietary Industry dataset, which contains complex procedural texts for product configuration along with corresponding ground-truth activity diagrams, and the public-domain PAGED dataset, which is made up of textual process descriptions paired with ground-truth activity diagrams from the software-engineering literature~\citep{du-etal-2024-paged}. In the Industry dataset, domain experts created and validated the activity diagrams. 
In the PAGED dataset~\citep{du-etal-2024-paged}, three independent human evaluators validated the ground-truth activity diagrams against the corresponding process descriptions.  While additional benchmarking with a broader range of LLMs would be valuable, the selected LLMs represent state-of-the-art in both instruction-following and reasoning-based LLMs~\citep{chiang2024chatbot}. Moreover, our results show consistent trends across LLMs, reducing the risk of LLM-specific biases.

\textbf{Limitations.} In our work, we treat each action node in an activity diagram as atomic, excluding hierarchical constructs such as swimlanes and composite nodes. These constructs do not appear in our Industry dataset or in the subset of the \textsc{Paged} dataset used in our experiments, and they are  not supported by existing automation-focused activity diagram formalisms~\citep{NejatiS0BM16,maoz,abscon}. According to the UML semantics and constructs documentation~\citep{Cook2017}, swimlanes and composite nodes enhance visual organization but do not increase behavioural expressiveness. Nevertheless, extending \approach\ to support swimlanes and composition could improve the usability of the generated diagrams and is left for future work.

Both our Industry dataset and the \textsc{Paged} dataset provide only a single ground-truth model for each process description. As a result, our evaluation -- consistent with several existing studies on LLM-based generation of structured models from text~\citep{DBLP:conf/models/YangCCMV24, abscon,du-etal-2024-paged} -- relies on assessing the correctness and completeness of the generated activity diagrams against these ground truths. We acknowledge that this approach implicitly assumes a certain determinism in model derivation: in practice, a textual description may legitimately yield multiple valid models, depending on modelling choices such as the level of abstraction or the grouping of activities. Such variations are less likely to occur within the scope of our current setting, which excludes swimlanes and hierarchical structures, as also done in prior work~\cite{abscon}. Nevertheless, extending activity diagrams to include hierarchy and swimlanes, as well as considering alternative evaluation strategies -- such as using multiple expert-created reference models per description -- remains a direction for future investigation.

\section{Tool Support}
\label{sec:tools}
\begin{figure}
    \centering
    \includegraphics[width=0.85\linewidth]{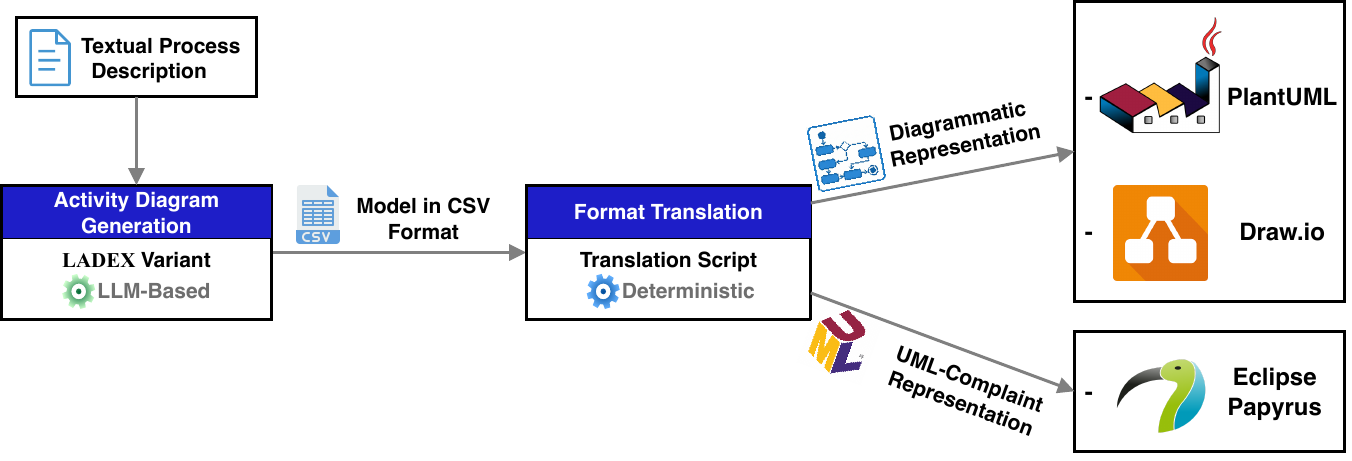}
    \caption{Tool support for transforming \approach-generated CSV representations into UML 2.5 XMI, PlantUML, and Draw.io formats.}
    \label{fig:tools}
\end{figure}

As discussed in Section~\ref{sec:approach}, all variants of \approach\ generate activity diagrams in an intermediate CSV representation. This design choice is to mitigate the risk of syntactic hallucinations that occur when LLMs are tasked with generating complex, strict schemas directly.\\
To allow  activity diagrams in our CSV format to be used in standard MDE workflows, we provide three transformation scripts in our replication package~\citep{replicationPackage}. As shown in Figure~\ref{fig:tools}, these scripts parse the structured intermediate CSV output and transform it into: (1)~a UML 2.5-compliant XMI serialization conforming to the MOF-based UML metamodel~\citep{Cook2017}, supporting interoperability with modeling environments such as Eclipse Papyrus~\citep{Garard2010}; (2)~PlantUML code corresponding to the UML activity model~\citep{roques_plantuml}; and (3)~a Draw.io-compatible graphical representation for lightweight visualization and editing~\citep{drawio}.

Providing these complementary formats enables different stakeholders to use \approach-generated activity diagrams according to their needs. For software engineers, the XMI serialization allows the generated models to be imported into UML-compliant modelling environments for refinement, validation, and execution. At the same time, the PlantUML and Draw.io representations make the models accessible to less technical stakeholders, such as customers and business analysts, thereby supporting efficient review and requirements validation without requiring familiarity with advanced MDE infrastructure.
\section{Related Work}
\label{sec:relwork}
Table~\ref{tab:rel_work} presents a structured comparison with existing research on automated model extraction from natural language descriptions. Below, we describe the comparison criteria and, for each one, discuss how \approach\ relates to existing approaches.

\begin{table}[p]
\caption{Comparison of our approach with existing model-generation methods}
\label{tab:rel_work}
{\fontsize{9.5}{9.5}\selectfont
\begin{center}
\scalebox{0.66}{
\rotatebox{+90}{
\begin{tabular}
{|p{0.1\textheight}|p{0.09\textheight}|p{0.2\textheight}|p{0.2\textheight}|p{0.2\textheight}|p{0.15\textheight}|p{0.1\textheight}|p{0.12\textheight}|p{0.2\textheight}|}
\hline
\textbf{Study} & \textbf{Output Model Type} & \textbf{Methodology} & \textbf{Structural Consistency} & \textbf{Semantic Alignment} & \textbf{Refinement Loop} & \textbf{LLM(s) Used}& \textbf{LLM Temperature} & \textbf{Evaluation Method}  \\ \hline

\citep{DBLP:conf/models/AroraSBZ16} & Domain model & Rule-based NLP with linguistic, syntactic, and semantic parsing & Structural rules defined and enforced during extraction & Ensured using semantic and linguistic rules used for information extraction & \xmark & \xmark & \xmark & Expert assessment on two industrial case studies \\ \hline 

\citep{DBLP:journals/tosem/AroraSNB19} & Domain model & Rule-based NLP with linguistic, syntactic and semantic parsing, plus an active-learning filter & Structural rules defined and enforced during extraction & Ensured using an active-learning relevance classifier and semantic and linguistic rules used for information extraction & Human-in-the-loop active learning for extracted element filtering & \xmark & \xmark & Expert assessment on two industrial case studies \\ \hline 

\citep{DBLP:conf/re/FerrariAA24} & Sequence diagram & Single LLM invocation for model generation & Not addressed & Not addressed & \xmark & GPT-3.5, GPT-4 & Default temperature of public ChatGPT website. Not reported & Expert assessment of completeness, correctness, adherence to standards, terminology consistency, and understandability relative to the requirements text \\ \hline 

\citep{DBLP:conf/refsq/Herwanto24} & Data-flow diagram & Single LLM invocation for model generation & Relies on the LLM’s prior knowledge of data-flow diagrams; structural rules are discussed but not provided to the LLM & Enforced by explicit instructions in the single generation prompt & \xmark & GPT-3.5, GPT-4, Llama 2, Mistral & Default temperature of public ChatGPT website. Not reported & Expert assessment of completeness and correctness against the textual user stories \\ \hline 

\citep{DBLP:conf/models/JahanHGRRRS24} & Sequence diagram & Two methods: (1) rule-based NLP; (2) single zero-shot LLM invocation & (1) Structural rules defined and enforced during extraction; (2) not addressed & (1) Ensured using semantic and linguistic rules used for information extraction; (2) not addressed & \xmark & GPT-3.5 & Not reported & Expert assessment comparing 100 models regarding relevance, object accuracy, message accuracy, and interaction accuracy \\ \hline 

\citep{DBLP:conf/models/YangCCMV24} & Domain model & Iterative element extraction and refinement using LLMs; model elements are extracted through sub-tasks, integrated into partial models, refined iteratively, and combined to form the final domain model & Enforces by iterative extraction and refinement of model elements, followed by post-processing of partial models to produce a syntactically correct final model & Incorporated semantic constraints in LLM prompts for iterative extraction and refinement of model elements and partial models, ensuring alignment with the input text & Refinement loop used for pruning extra elements, adjusting abstractions, refining relationships and partial models & GPT-4 & 0.0 for element extraction tasks; 0.7 for relation and partial model generation tasks & Manual comparison to reference domain models; precision, recall, and F1-score reported by one author \\ \hline 

\citep{abscon} & Graph models & Aggregates multiple candidates into a probabilistic model and produces the final model through concretization & Enforced by an optimization solver using logical structural constraints during the concretization of the final model & Derived from the frequency of elements across multiple candidate models & \xmark & GPT-4o-mini, GPT-4o, Llama 3.1-8B, Llama 3.1-70B & 0.7 & Precision and recall analysis comparing node-level relationships between the generated and the ground-truth models\\ \hline

Our approach \newline (\approach) & Activity diagram & Iterative generation of the entire activity diagram, followed by a critique-refine loop & Formally defined structural rules included in generation and refinement prompts and enforced by an algorithmic critic during refinement & Explicit alignment constraints are defined and applied for activity diagram generation, critique, and refinement steps & Critique-refine loop checks and improves the structural consistency (algorithmically or using an LLM) and semantic alignment (using an LLM, if enabled) of the entire generated  activity diagram & GPT-4.1 Mini, O4 Mini, Deep Seek-R1-Distill-Llama-70B & 0.0 (GPT-4.1 Mini); recommended value, 0.6 (DeepSeek-R1-Distill-Llama-70B); not adjustable (O4 Mini) & Automated evaluation of semantic completeness and correctness of the generated activity diagram against expert-constructed ground-truth activity diagrams using a behavioural matching algorithm and an LLM-based matcher \\ \hline 
\end{tabular}}}
\end{center}}
\end{table}

(1) \emph{Output Model Type} refers to the type of model each approach generates. Most existing techniques focus on extracting structural domain models~\citep{DBLP:conf/models/AroraSBZ16,DBLP:journals/tosem/AroraSNB19,DBLP:conf/models/YangCCMV24}. Three approaches generate behavioural models, i.e., data-flow diagrams and sequence diagrams~\citep{DBLP:conf/re/FerrariAA24,DBLP:conf/refsq/Herwanto24,DBLP:conf/models/JahanHGRRRS24}, and one approach~\citep{abscon} generates graph models such as taxonomies, executable program graphs and flowcharts. Our approach focuses on generating activity diagrams.

(2) \emph{Methodology} refers to the approach used for model generation. Traditional rule-based NLP methods use linguistic analysis, part-of-speech (POS) tagging, and parsing to extract structured knowledge from text, and then translate this knowledge into models~\citep{DBLP:conf/models/AroraSBZ16,DBLP:journals/tosem/AroraSNB19,DBLP:conf/models/JahanHGRRRS24}. Recent work uses LLMs -- which are not restricted to knowledge extracted by hand-crafted linguistic rules -- by prompting LLMs with the textual description, task definitions for model generation, and, optionally, few-shot examples~\citep{DBLP:conf/re/FerrariAA24,DBLP:conf/refsq/Herwanto24,DBLP:conf/models/JahanHGRRRS24}. This prompting strategy is similar to our Baseline variant in Table~\ref{table:approach_variants}. 
Yang et al.~\citep{DBLP:conf/models/YangCCMV24} employ LLMs in a manner inspired by traditional NLP approaches. They instruct an LLM to iteratively extract and refine model elements along with their relationships, integrate them into partial models, further refine these partial models, and ultimately construct a complete domain model.
Chen et al.~\citep{abscon} generate multiple candidate models, aggregate them into a probabilistic partial model, and concretize the result to construct the final model.
Our approach uses LLMs with a critique-refine loop to generate and iteratively refine an entire activity diagram that satisfies structural and semantic alignment constraints derived from the literature~\citep{Cook2017,DBLP:conf/refsq/Herwanto24,DBLP:conf/models/YangCCMV24}.
Unlike Yang et al., who merge partial models from individual elements, \approach\ holistically generates, critiques, and refines the entire model.

(3) \emph{Structural Consistency} refers to how each approach enforces the syntactic correctness and structural well-formedness of the generated models. Rule-based NLP methods~\citep{DBLP:conf/models/AroraSBZ16,DBLP:journals/tosem/AroraSNB19,DBLP:conf/models/JahanHGRRRS24} ensure structural consistency by design, enforcing the construction of syntactically correct models using explicit rules. Most existing LLM-based approaches~\citep{DBLP:conf/re/FerrariAA24,DBLP:conf/refsq/Herwanto24,DBLP:conf/models/JahanHGRRRS24} do not explicitly enforce structural consistency. 
Notable exceptions include Yang et al.~\citep{DBLP:conf/models/YangCCMV24}, who apply a rule-based post-processing step, and Chen et al.~\citep{abscon}, who enforce consistency during concretization using an optimization solver with logical structural constraints.
We ensure structural consistency by checking models against the constraints defined in the UML specification for activity diagrams~\citep{Cook2017}, done either algorithmically or using an LLM. The models are then iteratively refined with the assistance of an LLM, guided by the generated critique,  until no further structural issues are identified.

(4) \emph{Semantic Alignment} refers to the extent to which each approach ensures that the generated models are aligned with the meaning of the input textual descriptions. Traditional NLP methods enforce semantic alignment  using linguistic and semantic extraction and mapping rules~\citep{DBLP:conf/models/AroraSBZ16,DBLP:journals/tosem/AroraSNB19,DBLP:conf/models/JahanHGRRRS24}. LLM-based methods use prompt instructions to guide LLMs in aligning their output with the input textual descriptions~\citep{DBLP:conf/re/FerrariAA24,DBLP:conf/refsq/Herwanto24,DBLP:conf/models/JahanHGRRRS24}. 
Yang et al. include semantic constraints in the prompts used for the iterative extraction and refinement of model elements and partial models to ensure alignment between the generated model and the input text. Chen et al.~\citep{abscon} construct the final model using the most frequent and probable atomic model elements across multiple candidate models to ensure alignment between the generated model and the input text.
In our approach, we include semantic alignment constraints -- gleaned from prior studies~\citep{Cook2017, DBLP:conf/refsq/Herwanto24, DBLP:conf/models/YangCCMV24} -- directly into prompts for model generation and refinement. When alignment checking is included in a given variant of \approach, we also use these prompts with LLMs to iteratively critique the generated candidate activity diagrams and ensure their alignment with the input text until the LLM-based critique identifies no further alignment issues.

(5) \emph{Refinement Loop} refers to whether an approach uses a mechanism to iteratively evaluate and improve the generated models. Some traditional rule-based approaches employ active learning techniques that use human feedback to filter and refine extracted model elements before producing the final model~\citep{DBLP:journals/tosem/AroraSNB19}. Among the LLM-based approaches, Yang et al. introduce self-reflection loops to prune erroneously extracted model elements and refine the remaining elements before generating the final model. \emph{To date, no prior work has systematically assessed the impact of iterative critique-refine loops or investigated effective strategies for implementing them for the purpose of model generation from NL descriptions.} We present the first systematic study addressing this gap. Our findings show that the critique-refine loop improves structural consistency, correctness, and completeness. In addition, we show that implementing structural checking algorithmically helps generate structurally correct models, while incorporating LLM-based alignment checks can improve the \textit{semantic correctness} and \textit{completeness} of the generated models, assuming additional LLM calls are acceptable.

(6) \emph{LLM(s) Used} refers to the specific LLMs applied for model generation. Prior LLM-based work mainly uses instruction-following LLMs -- such as GPT-3.5, GPT-4, or GPT-4o -- in zero- or few-shot settings~\citep{DBLP:conf/re/FerrariAA24, DBLP:conf/refsq/Herwanto24, DBLP:conf/models/JahanHGRRRS24, DBLP:conf/models/YangCCMV24,abscon}. Recent reasoning-based LLMs, such as O4 Mini and DeepSeek-R1-Distill-Llama-70B, provide internal chain-of-thought reasoning and alignment capabilities, though they have not yet been systematically evaluated for model generation. In our evaluation, we compare two families of LLMs: GPT-4.1 Mini~\citep{openai2024gpt4ocard} as the instruction-following baseline; and the reasoning-based LLMs O4 Mini~\citep{jaech2024openai} (on both the Industry and \textsc{Paged} datasets) and DeepSeek-R1-Distill-Llama-70B~\citep{guo2025deepseek} (on \textsc{Paged}). Our evaluation enables a direct comparison between instruction-following and reasoning-based LLMs for the task of NL-to-model transformation.

(7) \emph{LLM Temperature} refers to the value of the temperature parameter used in the underlying LLM of an approach, and whether the approach depends on using a specific value for the temperature.
The temperature controls the balance between creativity and determinism in  LLM's outputs~\citep{openai2024gpt4ocard}. Higher values (e.g., 0.7 or above) increase variability, producing more diverse outputs across different runs. Lower values (e.g., 0.1) make outputs more consistent, reducing randomness. For instruction-following LLMs, the temperature is currently adjustable~\citep{guo2025deepseek}.
However, OpenAI's reasoning-based LLMs -- such as O4 Mini~\citep{jaech2024openai} -- use a fixed internal temperature to ensure deterministic outputs and optimal reasoning~\citep{azure_openai_reasoning}, whereas DeepSeek allows the temperature parameter to be adjusted, though a value of 0.6 is recommended for optimal reasoning performance~\citep{guo2025deepseek}.
Most existing LLM-based approaches do not explicitly discuss temperature. Jahan et al.~\citep{DBLP:conf/models/JahanHGRRRS24} do not report the temperature or how the LLM was used. Ferrari et al.~\citep{DBLP:conf/re/FerrariAA24} and Herwanto et al.~\citep{DBLP:conf/refsq/Herwanto24} use the public OpenAI ChatGPT with its default temperature.
Yang et al.~\citep{DBLP:conf/models/YangCCMV24}, on the other hand, require a low temperature (0.0) when extracting individual elements to ensure deterministic behaviour, while a higher temperature (0.7) is used for generating relationships between elements and partial models.  Similarly, Chen et al.~\citep{abscon} employ a high temperature of 0.7 for candidate models generation, introducing randomness and variation to the candidate models. This makes Yang et al.~\citep{DBLP:conf/models/YangCCMV24}and Chen et al.~\citep{abscon}'s approach dependent on temperature settings, which reasoning-based LLMs such as O4 Mini are designed to avoid~\citep{jaech2024openai,guo2025deepseek}.
Our method, \approach, does not depend on any specific temperature setting.  In our experiments, for the instruction-following LLM (GPT-4.1 Mini~\citep{openai2024gpt4ocard}), we set the temperature to 0.0 to eliminate internal randomness. For DeepSeek, we use the recommended value (0.6), while O4 Mini does not allow temperature adjustment.

(8) \emph{Evaluation Method} refers to the approach used to assess the quality of generated models. Prior work -- both in traditional NLP and in LLM-based model generation -- often relies on experts to manually assess generated models against either the input text or ground-truth models. These assessments typically involve manually rating completeness, correctness, terminology use, or conformity to notation standards~\citep{DBLP:conf/models/AroraSBZ16,DBLP:journals/tosem/AroraSNB19,DBLP:conf/re/FerrariAA24,DBLP:conf/refsq/Herwanto24,DBLP:conf/models/JahanHGRRRS24,DBLP:conf/models/YangCCMV24}. Manual assessment is time-consuming, error-prone, subjective, and difficult to reproduce.
Chen et al.~\citep{abscon} evaluate the generated models using standard recall and precision metrics, comparing the nodes of the generated model with those of a ground-truth model based on the overlap of their labels.
In contrast, our evaluation approaches consider the overall behaviour of the generated activity diagrams. We adopt two automated evaluation methods to compare each generated model with its ground truth. First, we employ an algorithmic approach based on trace-based behavioural model matching techniques~\citep{NejatiSCEZ07,NejatiSCEZ12,maoz}. Second, we leverage an LLM-based matcher, which we show is highly correlated with expert-produced matchings. Both methods produce a similarity mapping between the nodes of the generated and ground-truth models, which we then use to compute \textit{semantic correctness} and \textit{completeness} scores.
\section{Conclusion}
\label{sec:con}
This article introduces \approach, an LLM-based pipeline for generating activity diagrams from natural-language process descriptions via an iterative critique–refine process. We create five variants of \approach\ and systematically evaluate them on  public and industrial datasets, showing that combining algorithmic structural checks with LLM-based alignment yields the highest \textit{semantic correctness} and \textit{completeness}, while algorithmic-only checks remain a strong low-cost option.
\approach\ successfully automates the translation of informal natural-language process descriptions into structurally consistent and semantically aligned activity diagrams. The resulting models are intended to support both technical and non-technical stakeholders, including software engineers, analysts, modelling experts, and clients, by making behavioural workflows explicit in a structured form. These models can be used directly for review and refinement, or translated into standard modelling formats for integration with professional MDE tools when needed. By reducing the manual effort required to construct activity diagrams from text, \approach\ shifts stakeholders' effort from model construction towards the more critical tasks of validating, verifying, and refining the resulting models.
In future work, we plan to examine the use of neuro-symbolic approaches for model generation, in which algorithmic verification acts as a symbolic reasoning layer providing formal checks, while the LLM serves as the neural component that generates models based on these checks~\citep{DBLP:journals/pacmse/IbrahimzadaKPAPSJ25,DBLP:conf/icse/HerboldKRS25}. We will explore how neuro-symbolic methods can be applied systematically throughout different stages of model generation, and how formal symbolic checks can be used to improve the impact of semantic alignment checks during model generation.

\section {Data Availability}
\label{sec:data}
Our \textbf{replication package}, available online~\citep{replicationPackage}, contains our implementation, the \textsc{Paged} dataset used in our evaluation, and an anonymized subset of our Industry dataset.

\section*{Compliance with Ethical Standards}
\sectopic{Conflict of Interest.} The authors declare that they have no conflicts of interest.

\sectopic{Acknowledgments.} We are grateful to the engineering team at Ciena.

\sectopic{Funding.} This research was supported by the Natural Sciences and Engineering Research Council of Canada (NSERC) through the Discovery, Discovery Accelerator, and Alliance programs, by Mathematics of Information Technology and Complex Systems (Mitacs) through the  Accelerate program, and by Ciena.

\sectopic{Ethical Approval.} This research did not involve human participants or animals; therefore, ethical approval was not required.

\sectopic{Informed Consent.} No personal data or identifiable information is reported in this research; informed consent is not applicable.

\sectopic{Author Contributions.} This article is part of the first author's PhD studies, who led the work, including its conceptualization, methodology, implementation, experimentation, analysis, and writing. The second, third, fourth and fifth authors are industry collaborators from Ciena; they contributed to problem identification and provided technical support, as well as access to Ciena’s data and technical resources. The final two authors led and supervised the research, shaped its direction and design, and made substantial contributions to the writing and interpretation of the results.

\section*{Acknowledgment}
We gratefully acknowledge funding from Mitacs Accelerate, Ciena, the Ontario Graduate Scholarship program, and NSERC of Canada under the Discovery  program.

\bibliographystyle{spbasic}
\bibliography{bibliography}

@article{benjamini1995controlling,
  title={Controlling the false discovery rate: a practical and powerful approach to multiple testing},
  author={Benjamini, Yoav and Hochberg, Yosef},
  journal={Journal of the Royal statistical society: series B (Methodological)},
  volume={57},
  number={1},
  pages={289--300},
  year={1995},
  publisher={Wiley Online Library}
}

@article{NejatiSCEZ12,
  author       = {Shiva Nejati and
                  Mehrdad Sabetzadeh and
                  Marsha Chechik and
                  Steve M. Easterbrook and
                  Pamela Zave},
  title        = {Matching and Merging of Variant Feature Specifications},
  journal      = {{IEEE} Transaction on Software Engineering},
  volume       = {38},
  number       = {6},
  pages        = {1355--1375},
  year         = {2012},
  url          = {https://doi.org/10.1109/TSE.2011.112},
  doi          = {10.1109/TSE.2011.112}
}

@inproceedings{NejatiSCEZ07,
  author       = {Shiva Nejati and
                  Mehrdad Sabetzadeh and
                  Marsha Chechik and
                  Steve M. Easterbrook and
                  Pamela Zave},
  title        = {Matching and Merging of Statecharts Specifications},
  booktitle    = {29th International Conference on Software Engineering {(ICSE} 2007),
                  Minneapolis, MN, USA, May 20-26, 2007},
  pages        = {54--64},
  publisher    = {{IEEE} Computer Society},
  year         = {2007},
  url          = {https://doi.org/10.1109/ICSE.2007.50},
  doi          = {10.1109/ICSE.2007.50},
  bibsource    = {dblp computer science bibliography, https://dblp.org}
}

@inproceedings{NejatiS0BM16,
  author       = {Shiva Nejati and
                  Mehrdad Sabetzadeh and
                  Chetan Arora and
                  Lionel C. Briand and
                  Felix Mandoux},
  editor       = {Thomas Zimmermann and
                  Jane Cleland{-}Huang and
                  Zhendong Su},
  title        = {Automated change impact analysis between SysML models of requirements
                  and design},
  booktitle    = {Proceedings of the 24th {ACM} {SIGSOFT} International Symposium on
                  Foundations of Software Engineering, {FSE} 2016, Seattle, WA, USA,
                  November 13-18, 2016},
  pages        = {242--253},
  publisher    = {{ACM}},
  year         = {2016},
  url          = {https://doi.org/10.1145/2950290.2950293},
  doi          = {10.1145/2950290.2950293},
  bibsource    = {dblp computer science bibliography, https://dblp.org}
}

@inproceedings{maoz,
  author       = {Shahar Maoz and
                  Jan Oliver Ringert and
                  Bernhard Rumpe},
  editor       = {Tibor Gyim{\'{o}}thy and
                  Andreas Zeller},
  title        = {ADDiff: semantic differencing for activity diagrams},
  booktitle    = {SIGSOFT/FSE'11 19th {ACM} {SIGSOFT} Symposium on the Foundations of
                  Software Engineering {(FSE-19)} and ESEC'11: 13th European Software
                  Engineering Conference (ESEC-13), Szeged, Hungary, September 5-9,
                  2011},
  pages        = {179--189},
  publisher    = {{ACM}},
  year         = {2011},
  url          = {https://doi.org/10.1145/2025113.2025140},
  doi          = {10.1145/2025113.2025140},
  bibsource    = {dblp computer science bibliography, https://dblp.org}
}

@techreport{Cook2017,
  author = {Steve Cook and 
            Conrad Bock and 
            Pete Rivett and 
            Tom Rutt and 
            Ed Seidewitz and 
            Bran Selic and 
            Doug Tolbert},
  title = {Unified Modeling Language ({UML}) Version 2.5.1},
  institution = {Object Management Group ({OMG})},
  type = {Standard},
  month = Dec,
  year = {2017},
  url = {https://www.omg.org/spec/UML/2.5.1}
}

@inproceedings{DBLP:conf/models/AroraSBZ16,
  author       = {Chetan Arora and
                  Mehrdad Sabetzadeh and
                  Lionel C. Briand and
                  Frank Zimmer},
  editor       = {Benoit Baudry and
                  Beno{\^{\i}}t Combemale},
  title        = {Extracting domain models from natural-language requirements: approach
                  and industrial evaluation},
  booktitle    = {Proceedings of the {ACM/IEEE} 19th International Conference on Model
                  Driven Engineering Languages and Systems, Saint-Malo, France, October
                  2-7, 2016},
  pages        = {250--260},
  publisher    = {{ACM}},
  year         = {2016},
  url          = {http://dl.acm.org/citation.cfm?id=2976769},
  bibsource    = {dblp computer science bibliography, https://dblp.org}
}

@inproceedings{DBLP:conf/re/FerrariAA24,
  author       = {Alessio Ferrari and
                  Sallam Abualhaija and
                  Chetan Arora},
  title        = {Model Generation with LLMs: From Requirements to {UML} Sequence Diagrams},
  booktitle    = {32nd {IEEE} International Requirements Engineering Conference, {RE}
                  2024 - Workshops, Reykjavik, Iceland, June 24-25, 2024},
  pages        = {291--300},
  publisher    = {{IEEE}},
  year         = {2024},
  url          = {https://doi.org/10.1109/REW61692.2024.00044},
  doi          = {10.1109/REW61692.2024.00044},
  bibsource    = {dblp computer science bibliography, https://dblp.org}
}

@inproceedings{DBLP:conf/refsq/Herwanto24,
  author       = {Guntur Budi Herwanto},
  editor       = {Daniel M{\'{e}}ndez et. al.},
  title        = {Automating Data Flow Diagram Generation from User Stories Using Large
                  Language Models},
  booktitle    = {Joint Proceedings of {REFSQ-2024} Workshops, Doctoral Symposium, Posters
                  {\&} Tools Track, and Education and Training Track co-located
                  with the 30th International Conference on Requirements Engineering:
                  Foundation for Software Quality {(REFSQ} 2024), Winterthur, Switzerland,
                  April 8-11, 2024},
  series       = {{CEUR} Workshop Proceedings},
  volume       = {3672},
  publisher    = {CEUR-WS.org},
  year         = {2024},
  url          = {https://ceur-ws.org/Vol-3672/NLP4RE-paper3.pdf},
  bibsource    = {dblp computer science bibliography, https://dblp.org}
}

@inproceedings{DBLP:conf/models/JahanHGRRRS24,
  author       = {Munima Jahan and
                  Mohammad Mahdi Hassan and
                  Reza Golpayegani and
                  Golshid Ranjbaran and
                  Chanchal Roy and
                  Banani Roy and
                  Kevin A. Schneider},
  editor       = {Alexander Egyed and
                  Manuel Wimmer and
                  Marsha Chechik and
                  Beno{\^{\i}}t Combemale},
  title        = {Automated Derivation of {UML} Sequence Diagrams from User Stories:
                  Unleashing the Power of Generative {AI} vs. a Rule-Based Approach},
  booktitle    = {Proceedings of the {ACM/IEEE} 27th International Conference on Model
                  Driven Engineering Languages and Systems, {MODELS} 2024, Linz, Austria,
                  September 22-27, 2024},
  pages        = {138--148},
  publisher    = {{ACM}},
  year         = {2024},
  url          = {https://doi.org/10.1145/3640310.3674081},
  doi          = {10.1145/3640310.3674081},
  bibsource    = {dblp computer science bibliography, https://dblp.org}
}

@inproceedings{DBLP:conf/models/YangCCMV24,
  author       = {Yujing Yang and
                  Boqi Chen and
                  Kua Chen and
                  Gunter Mussbacher and
                  D{\'{a}}niel Varr{\'{o}}},
  editor       = {Manuel Wimmer and
                  Alexander Egyed and
                  Beno{\^{\i}}t Combemale and
                  Marsha Chechik},
  title        = {Multi-step Iterative Automated Domain Modeling with Large Language
                  Models},
  booktitle    = {Proceedings of the {ACM/IEEE} 27th International Conference on Model
                  Driven Engineering Languages and Systems, {MODELS} Companion 2024,
                  Linz, Austria, September 22-27, 2024},
  pages        = {587--595},
  publisher    = {{ACM}},
  year         = {2024},
  url          = {https://doi.org/10.1145/3652620.3687807},
  doi          = {10.1145/3652620.3687807},
  bibsource    = {dblp computer science bibliography, https://dblp.org}
}

@inproceedings{du-etal-2024-paged,
  author       = {Weihong Du and
                  Wenrui Liao and
                  Hongru Liang and
                  Wenqiang Lei},
  editor       = {Lun{-}Wei Ku and
                  Andre Martins and
                  Vivek Srikumar},
  title        = {{PAGED:} {A} Benchmark for Procedural Graphs Extraction from Documents},
  booktitle    = {Proceedings of the 62nd Annual Meeting of the Association for Computational
                  Linguistics (Volume 1: Long Papers), {ACL} 2024, Bangkok, Thailand,
                  August 11-16, 2024},
  pages        = {10829--10846},
  publisher    = {Association for Computational Linguistics},
  year         = {2024},
  url          = {https://doi.org/10.18653/v1/2024.acl-long.583},
  doi          = {10.18653/V1/2024.ACL-LONG.583},
  bibsource    = {dblp computer science bibliography, https://dblp.org}
}

@article{jaech2024openai,
  author       = {Aaron Jaech and
                  Adam Kalai and
                  Adam Lerer and
                  Adam Richardson and
                  Ahmed El{-}Kishky and
                  Aiden Low and
                  Alec Helyar and
                  Aleksander Madry and
                  Alex Beutel and
                  Alex Carney and
                  Others},
  title        = {OpenAI o1 System Card},
  journal      = {CoRR},
  volume       = {abs/2412.16720},
  year         = {2024},
  url          = {https://doi.org/10.48550/arXiv.2412.16720},
  doi          = {10.48550/ARXIV.2412.16720},
  eprinttype    = {arXiv},
  eprint       = {2412.16720},
  bibsource    = {dblp computer science bibliography, https://dblp.org}
}

@article{guo2025deepseek,
  author       = {DeepSeek{-}AI and
                  Daya Guo and
                  Dejian Yang and
                  Haowei Zhang and
                  Junxiao Song and
                  Ruoyu Zhang and
                  Runxin Xu and
                  Qihao Zhu and
                  Shirong Ma and
                  Peiyi Wang and
                  Xiao Bi and
                  Others},
  title        = {DeepSeek-R1: Incentivizing Reasoning Capability in LLMs via Reinforcement
                  Learning},
  journal      = {CoRR},
  volume       = {abs/2501.12948},
  year         = {2025},
  url          = {https://doi.org/10.48550/arXiv.2501.12948},
  doi          = {10.48550/ARXIV.2501.12948},
  eprinttype    = {arXiv},
  eprint       = {2501.12948},
  bibsource    = {dblp computer science bibliography, https://dblp.org}
}

@misc{openai2024gpt4ocard,
  author       = {Aaron Hurst and
                  Adam Lerer and
                  Adam P. Goucher and
                  Adam Perelman and
                  Aditya Ramesh and
                  Aidan Clark and
                  AJ Ostrow and
                  Akila Welihinda and
                  Alan Hayes and
                  Alec Radford and
                  Aleksander Madry and
                  Alex Baker{-}Whitcomb and
                  Alex Beutel and
                  Alex Borzunov and
                  Alex Carney and
                  Others},
  title        = {GPT-4o System Card},
  journal      = {CoRR},
  volume       = {abs/2410.21276},
  year         = {2024},
  url          = {https://doi.org/10.48550/arXiv.2410.21276},
  doi          = {10.48550/ARXIV.2410.21276},
  eprinttype    = {arXiv},
  eprint       = {2410.21276},
  bibsource    = {dblp computer science bibliography, https://dblp.org}
}

@misc{drawio,
  author       = {JGraph},
  title        = {draw.io},
  year         = {2021},
  howpublished = {\url{https://www.draw.io/}},
  note         = {Accessed: 2025-08-21}
}

@Inbook{Garard2010,
author="G{\'e}rard, S{\'e}bastien
and Dumoulin, C{\'e}dric
and Tessier, Patrick
and Selic, Bran",
editor="Giese, Holger
and Karsai, Gabor
and Lee, Edward
and Rumpe, Bernhard
and Sch{\"a}tz, Bernhard",
title="19 Papyrus: A UML2 Tool for Domain-Specific Language Modeling",
bookTitle="Model-Based Engineering of Embedded Real-Time Systems: International Dagstuhl Workshop, Dagstuhl Castle, Germany, November 4-9, 2007. Revised Selected Papers",
year="2010",
publisher="Springer Berlin Heidelberg",
address="Berlin, Heidelberg",
pages="361--368",
isbn="978-3-642-16277-0",
doi="10.1007/978-3-642-16277-0_19",
url="https://doi.org/10.1007/978-3-642-16277-0_19"
}

@misc{roques_plantuml,
  author       = {Arnaud Roques and PlantUML Contributors},
  title        = {PlantUML Software},
  year         = {2009},
  howpublished = {\url{https://github.com/plantuml/plantuml}},
  note         = {Accessed: 2025-08-21}
}

@misc{ollama2023,
  author       = {Ollama},
  title        = {Ollama},
  year         = {2023},
  howpublished = {\url{https://github.com/ollama/ollama}},
  note         = {Accessed: 2025-08-21}
}

@inproceedings{reimers-2019-sentence-bert,
  author       = {Nils Reimers and
                  Iryna Gurevych},
  editor       = {Kentaro Inui and
                  Jing Jiang and
                  Vincent Ng and
                  Xiaojun Wan},
  title        = {Sentence-BERT: Sentence Embeddings using Siamese BERT-Networks},
  booktitle    = {Proceedings of the 2019 Conference on Empirical Methods in Natural
                  Language Processing and the 9th International Joint Conference on
                  Natural Language Processing, {EMNLP-IJCNLP} 2019, Hong Kong, China,
                  November 3-7, 2019},
  pages        = {3980--3990},
  publisher    = {Association for Computational Linguistics},
  year         = {2019},
  url          = {https://doi.org/10.18653/v1/D19-1410},
  doi          = {10.18653/V1/D19-1410},
  bibsource    = {dblp computer science bibliography, https://dblp.org}
}

@misc{zhang2024mgte,
  author       = {Xin Zhang and
                  Yanzhao Zhang and
                  Dingkun Long and
                  Wen Xie and
                  Ziqi Dai and
                  Jialong Tang and
                  Huan Lin and
                  Baosong Yang and
                  Pengjun Xie and
                  Fei Huang and
                  Meishan Zhang and
                  Wenjie Li and
                  Min Zhang},
  editor       = {Franck Dernoncourt and
                  Daniel Preotiuc{-}Pietro and
                  Anastasia Shimorina},
  title        = {mGTE: Generalized Long-Context Text Representation and Reranking Models
                  for Multilingual Text Retrieval},
  booktitle    = {Proceedings of the 2024 Conference on Empirical Methods in Natural
                  Language Processing: {EMNLP} 2024 - Industry Track, Miami, Florida,
                  USA, November 12-16, 2024},
  pages        = {1393--1412},
  publisher    = {Association for Computational Linguistics},
  year         = {2024},
  url          = {https://doi.org/10.18653/v1/2024.emnlp-industry.103},
  doi          = {10.18653/V1/2024.EMNLP-INDUSTRY.103},
  bibsource    = {dblp computer science bibliography, https://dblp.org}
}

@misc{li2023gte,
  author       = {Zehan Li and
                  Xin Zhang and
                  Yanzhao Zhang and
                  Dingkun Long and
                  Pengjun Xie and
                  Meishan Zhang},
  title        = {Towards General Text Embeddings with Multi-stage Contrastive Learning},
  journal      = {CoRR},
  volume       = {abs/2308.03281},
  year         = {2023},
  url          = {https://doi.org/10.48550/arXiv.2308.03281},
  doi          = {10.48550/ARXIV.2308.03281},
  eprinttype    = {arXiv},
  eprint       = {2308.03281},
  bibsource    = {dblp computer science bibliography, https://dblp.org}
}

@standard{omg2011bpmn,
  author       = {OMG},
  title        = {{Business Process Model and Notation (BPMN), Version 2.0}},
  year         = {2011},
  month        = {jan},
  institution  = {Object Management Group},
  organization = {Object Management Group},
  revision     = {2.0},
  url          = {http://www.omg.org/spec/BPMN/2.0},
  note         = {Accessed: 2025-08-21}
}

@inproceedings{10.1007/11691372_28,
  author       = {Oleg Sokolsky and
                  Sampath Kannan and
                  Insup Lee},
  editor       = {Holger Hermanns and
                  Jens Palsberg},
  title        = {Simulation-Based Graph Similarity},
  booktitle    = {Tools and Algorithms for the Construction and Analysis of Systems,
                  12th International Conference, {TACAS} 2006 Held as Part of the Joint
                  European Conferences on Theory and Practice of Software, {ETAPS} 2006,
                  Vienna, Austria, March 25 - April 2, 2006, Proceedings},
  series       = {Lecture Notes in Computer Science},
  volume       = {3920},
  pages        = {426--440},
  publisher    = {Springer},
  year         = {2006},
  url          = {https://doi.org/10.1007/11691372\_28},
  doi          = {10.1007/11691372\_28},
  bibsource    = {dblp computer science bibliography, https://dblp.org}
}

@article{vargha2000critique,
  title={A critique and improvement of the CL common language effect size statistics of McGraw and Wong},
  author={Andr{\'a}s Vargha and 
          Harold D Delaney },
  journal={Journal of Educational and Behavioral Statistics},
  volume={25},
  number={2},
  pages={101--132},
  year={2000},
  publisher={Sage Publications Sage CA: Los Angeles, CA}
}

@incollection{wilcoxon1992individual,
  title={Individual comparisons by ranking methods},
  author={Frank Wilcoxon},
  booktitle={Breakthroughs in statistics: Methodology and distribution},
  pages={196--202},
  year={1992},
  publisher={Springer}
}

@book{lehmann2006nonparametrics,
  title={Nonparametrics: statistical methods based on ranks},
  author={Lehmann, Erich Leo and D'Abrera, Howard JM},
  volume={464},
  year={2006},
  publisher={Springer New York}
}

@article{DBLP:journals/tosem/AroraSNB19,
  author       = {Chetan Arora and
                  Mehrdad Sabetzadeh and
                  Shiva Nejati and
                  Lionel C. Briand},
  title        = {An Active Learning Approach for Improving the Accuracy of Automated
                  Domain Model Extraction},
  journal      = {{ACM} Trans. Softw. Eng. Methodol.},
  volume       = {28},
  number       = {1},
  pages        = {4:1--4:34},
  year         = {2019},
  url          = {https://doi.org/10.1145/3293454},
  doi          = {10.1145/3293454},
  bibsource    = {dblp computer science bibliography, https://dblp.org}
}

@inproceedings{10.1007/978-3-031-29786-1_8,
  author       = {Sarmad Bashir and
                  Muhammad Abbas and
                  Mehrdad Saadatmand and
                  Eduard Paul Enoiu and
                  Markus Bohlin and
                  Pernilla Lindberg},
  editor       = {Alessio Ferrari and
                  Birgit Penzenstadler},
  title        = {Requirement or Not, That is the Question: {A} Case from the Railway
                  Industry},
  booktitle    = {Requirements Engineering: Foundation for Software Quality - 29th International
                  Working Conference, {REFSQ} 2023, Barcelona, Spain, April 17-20, 2023,
                  Proceedings},
  series       = {Lecture Notes in Computer Science},
  volume       = {13975},
  pages        = {105--121},
  publisher    = {Springer},
  year         = {2023},
  url          = {https://doi.org/10.1007/978-3-031-29786-1\_8},
  doi          = {10.1007/978-3-031-29786-1\_8},
  bibsource    = {dblp computer science bibliography, https://dblp.org}
}

@article{DBLP:journals/ese/AbualhaijaASBT20,
  author       = {Sallam Abualhaija and
                  Chetan Arora and
                  Mehrdad Sabetzadeh and
                  Lionel C. Briand and
                  Michael Traynor},
  title        = {Automated demarcation of requirements in textual specifications: a
                  machine learning-based approach},
  journal      = {Empir. Softw. Eng.},
  volume       = {25},
  number       = {6},
  pages        = {5454--5497},
  year         = {2020},
  url          = {https://doi.org/10.1007/s10664-020-09864-1},
  doi          = {10.1007/S10664-020-09864-1},
  bibsource    = {dblp computer science bibliography, https://dblp.org}
}

@misc{azure_openai_reasoning,
  author       = {{Microsoft}},
  title        = {Azure OpenAI reasoning models - GPT-5 series, o3-mini, o1, o1-mini},
  year         = 2025,
  url          = {https://learn.microsoft.com/en-us/azure/ai-foundry/openai/how-to/reasoning},
  note         = {Accessed: 2025-08-21}
}

@misc{chiang2024chatbot,
  author       = {Wei{-}Lin Chiang and
                  Lianmin Zheng and
                  Ying Sheng and
                  Anastasios Nikolas Angelopoulos and
                  Tianle Li and
                  Dacheng Li and
                  Banghua Zhu and
                  Hao Zhang and
                  Michael I. Jordan and
                  Joseph E. Gonzalez and
                  Ion Stoica},
  title        = {Chatbot Arena: An Open Platform for Evaluating LLMs by Human Preference},
  booktitle    = {Forty-first International Conference on Machine Learning, {ICML} 2024,
                  Vienna, Austria, July 21-27, 2024},
  publisher    = {OpenReview.net},
  year         = {2024},
  url          = {https://openreview.net/forum?id=3MW8GKNyzI},
  bibsource    = {dblp computer science bibliography, https://dblp.org}
}

@article{DBLP:journals/pacmse/IbrahimzadaKPAPSJ25,
  author       = {Ali Reza Ibrahimzada and
                  Kaiyao Ke and
                  Mrigank Pawagi and
                  Muhammad Salman Abid and
                  Rangeet Pan and
                  Saurabh Sinha and
                  Reyhaneh Jabbarvand},
  title        = {AlphaTrans: {A} Neuro-Symbolic Compositional Approach for Repository-Level
                  Code Translation and Validation},
  journal      = {Proc. {ACM} Softw. Eng.},
  volume       = {2},
  number       = {{FSE}},
  pages        = {2454--2476},
  year         = {2025},
  url          = {https://doi.org/10.1145/3729379},
  doi          = {10.1145/3729379},
  bibsource    = {dblp computer science bibliography, https://dblp.org}
}

@inproceedings{DBLP:conf/icse/HerboldKRS25,
  author       = {Steffen Herbold and
                  Christoph Knieke and
                  Andreas Rausch and
                  Christian Schindler},
  title        = {Neurosymbolic Architectural Reasoning: Towards Formal Analysis through
                  Neural Software Architecture Inference},
  booktitle    = {1st {IEEE/ACM} International Workshop on Neuro-Symbolic Software Engineering,
                  NSE@ICSE 2025, Ottawa, ON, Canada, May 3, 2025},
  pages        = {5--10},
  publisher    = {{IEEE}},
  year         = {2025},
  url          = {https://doi.org/10.1109/NSE66660.2025.00008},
  doi          = {10.1109/NSE66660.2025.00008},
  bibsource    = {dblp computer science bibliography, https://dblp.org}
}

@misc{replicationPackage,
  title = {{LADEX} Replication Package},
  author={ReplicationPackage},
  howpublished = "\url{https://github.com/parham-box/EMSE-LADEX}",
  year = {2025}, 
  note = {Accessed: 2025-11-24}
}

@inproceedings{DBLP:conf/naacl/PezeshkpourH24,
  author       = {Pouya Pezeshkpour and
                  Estevam Hruschka},
  editor       = {Kevin Duh and
                  Helena G{\'{o}}mez{-}Adorno and
                  Steven Bethard},
  title        = {Large Language Models Sensitivity to The Order of Options in Multiple-Choice
                  Questions},
  booktitle    = {Findings of the Association for Computational Linguistics: {NAACL}
                  2024, Mexico City, Mexico, June 16-21, 2024},
  pages        = {2006--2017},
  publisher    = {Association for Computational Linguistics},
  year         = {2024},
  url          = {https://doi.org/10.18653/v1/2024.findings-naacl.130},
  doi          = {10.18653/V1/2024.FINDINGS-NAACL.130},
  bibsource    = {dblp computer science bibliography, https://dblp.org}
}

@article{spearman,
 ISSN = {00029556},
 url = {https://doi.org/10.2307/1422689},
 author = {Charles Spearman},
 journal = {The American Journal of Psychology},
 number = {3/4},
 pages = {441--471},
 publisher = {University of Illinois Press},
 title = {The Proof and Measurement of Association between Two Things},
 volume = {100},
 year = {1987}
}

@inproceedings{abscon,
  author       = {Boqi Chen and
                  Ou Wei and
                  Bingzhou Zheng and
                  Gunter Mussbacher},
  title        = {Accurate and Consistent Graph Model Generation from Text with Large
                  Language Models},
  booktitle    = {28th {ACM/IEEE} International Conference on Model Driven Engineering
                  Languages and Systems, {MODELS} 2025, Grand Rapids, MI, USA, October
                  5-10, 2025},
  pages        = {130--141},
  publisher    = {{IEEE}},
  year         = {2025},
  url          = {https://doi.org/10.1109/MODELS67397.2025.00018},
  doi          = {10.1109/MODELS67397.2025.00018},
  bibsource    = {dblp computer science bibliography, https://dblp.org}
}

@ARTICLE{6507223,
  author={Whittle, Jon and Hutchinson, John and Rouncefield, Mark},
  journal={IEEE Software}, 
  title={The State of Practice in Model-Driven Engineering}, 
  year={2014},
  volume={31},
  number={3},
  pages={79-85},
  doi={10.1109/MS.2013.65}}

@article{DBLP:journals/ijgi/LiZWJZ24,
  author       = {Diya Li and
                  Yue Zhao and
                  Zhifang Wang and
                  Calvin Jung and
                  Zhe Zhang},
  title        = {Large Language Model-Driven Structured Output: {A} Comprehensive Benchmark
                  and Spatial Data Generation Framework},
  journal      = {{ISPRS} Int. J. Geo Inf.},
  volume       = {13},
  number       = {11},
  pages        = {405},
  year         = {2024},
  url          = {https://doi.org/10.3390/ijgi13110405},
  doi          = {10.3390/IJGI13110405},
  bibsource    = {dblp computer science bibliography, https://dblp.org}
}

@techreport{fUML,
  author = {{Object Management Group}},
  title = {Semantics of a Foundational Subset for Executable {UML} Models ({fUML}) Version 1.5},
  institution = {Object Management Group ({OMG})},
  type = {Standard},
  month = Aug,
  year = {2021},
  url = {https://www.omg.org/spec/FUML/1.5}
}

@article{cohen1960coefficient,
  title={A coefficient of agreement for nominal scales},
  author={Cohen, Jacob},
  journal={Educational and psychological measurement},
  volume={20},
  number={1},
  url={https://doi.org/10.1177/001316446002000104},
  doi={10.1177/001316446002000104}, 
  pages={37--46},
  year={1960},
  publisher={Sage Publications Sage CA: Thousand Oaks, CA}
}

\newpage
\appendix
\section{Formal Definition of Reachability}
\label{ap:path}
Using the notation in Definition~\ref{def:activitydiagram}, we formally define the reachability predicate. The $\mathit{path}(x, y)$ predicate evaluates to true if a sequence of transitions exists from node $\mathit{x}$ to node $y$.  We define a base path of length one as:
$$\mathit{path_1}(x, y) \leftrightarrow ( (x, y) \in \mathit{T} \lor \exists l \in \mathit{TL} : (x, l, y) \in \mathit{T} )$$
For $2 \le m < k$ (where $k = |\mathit{N}|$), a path of length $m$ is defined recursively as: 
$$\mathit{path}_m(x, y) \leftrightarrow \exists z \in \mathit{N} : ( \mathit{path}_{m-1}(x, z) \land \mathit{path_1}(z, y) )$$
We then define the general reachability predicate is the disjunction of all paths up to length $k-1$: 
$$\mathit{path}(x, y) \leftrightarrow \bigvee_{i=1}^{k-1} \mathit{path}_i(x, y)$$

\section{Structural Constraints}
\label{ap:cons}
Table~\ref{table:all_constraints} presents the full list of 17 structural constraints for activity diagrams, derived from the UML 2.5.1 specification~\citep{Cook2017}, indicating whether they are included in the prompts of \approach, along with a justification for their inclusion.
\begin{table}[t]
\centering
\caption{Structural constraints for UML activity diagrams with their inclusion status in \approach\ and justification}
\label{table:all_constraints}
\resizebox{\textwidth}{!}{%
\begin{tabular}{|c|p{6cm}|c|p{6cm}|}
\hline
\centering
 & 
\multicolumn{1}{c|}{\textbf{Constraint}} & 
\multicolumn{1}{c|}{\textbf{Inclusion}} & 
\multicolumn{1}{c|}{\textbf{Reason}} \\ \hline
\textbf{1} & An activity diagram can have zero or more initial nodes & SC1 & Activity diagrams may use more than one initial node when they are hierarchical; otherwise, they have one or none. Our formalization, as detailed in Section \ref{sec:formalization}, does not support hierarchies. In addition, state-of-the-art software modelling practices require every behavioural model to have a starting point so that the model has a clear semantics. Therefore, we require exactly one initial node. \\ \hline
\textbf{2} & An activity diagram can have zero or more end nodes & SC2 & 
State-of-the-art software modelling practices require every behavioural model  to include a termination point. Thus, we require at least one end node to terminate the process.
\\ \hline
\textbf{3} & An initial node must have no incoming edges & SC3 & Included without modification. \\ \hline
\textbf{4} & An end node must have no outgoing edges & SC4 & Included without modification. \\ \hline
\textbf{5} & A decision node must have at least two outgoing edges, each labelled by a guard condition & SC5 & Included without modification. \\ \hline
\textbf{6} & There should be at least one path from the initial node to every other node in the activity diagram & SC6 & Included without modification. \\ \hline
\textbf{7} & An action node can have multiple incoming edges & \xmark & We implicitly support this constraint, as we do not constrain action nodes' incoming edges. \\ \hline
\textbf{8} & An action node can have multiple outgoing edges & \xmark & We implicitly support this constraint, as we do not constrain action nodes' outgoing edges. \\ \hline
\textbf{9} & A merge or join node must have multiple incoming edges & \xmark & We abstract merge and join nodes as action nodes. This simplification eliminates the need for enforcing specific merge and join node constraints. \\ \hline
\textbf{10} & A merge or join node must have exactly one outgoing edge & \xmark & We abstract merge and join nodes as action nodes with a single outgoing flow. This simplification eliminates the need for enforcing specific fork and join constraints. \\ \hline
\textbf{11} & A fork node must have exactly one incoming edge & \xmark & We abstract fork nodes as action nodes. Thus, we do not explicitly enforce this constraint. \\ \hline
\textbf{12} & A fork node must have multiple outgoing edges & \xmark & We implicitly support this constraint by abstracting fork nodes as action nodes, which allows any number of outgoing edges. \\ \hline
\textbf{13} & An object node can have multiple incoming and outgoing edges & \xmark & Since we represent object nodes as action nodes, and action nodes can have multiple incoming and outgoing edges, this constraint is already satisfied. \\ \hline
\textbf{14} & Every fork node that creates parallel flows should have a corresponding join node & \xmark & As we abstract fork and join nodes to action nodes, this constraint does not need explicit enforcement. \\ \hline
\textbf{15} & Object flows must connect compatible object nodes and actions & \xmark & We assume data object compatibility but do not formally constrain it, as we do not explicitly model object nodes. \\ \hline
\textbf{16} & Swimlanes (partitions) must not alter semantics & \xmark & Our formalization, as detailed in Section \ref{sec:formalization}, excludes swimlanes. \\ \hline
\textbf{17} & A node should not have an outgoing transition unless it connects to a target node & \xmark & Our model generation and encoding process inherently enforces this constraint, as it disallows transitions without both a source and a target node. \\ \hline
\end{tabular}%
}
\end{table}

\section{Prompt Outlines}
\label{ap:outlines}
\begin{figure}[t]
	\centering
	\includegraphics[width=\columnwidth]{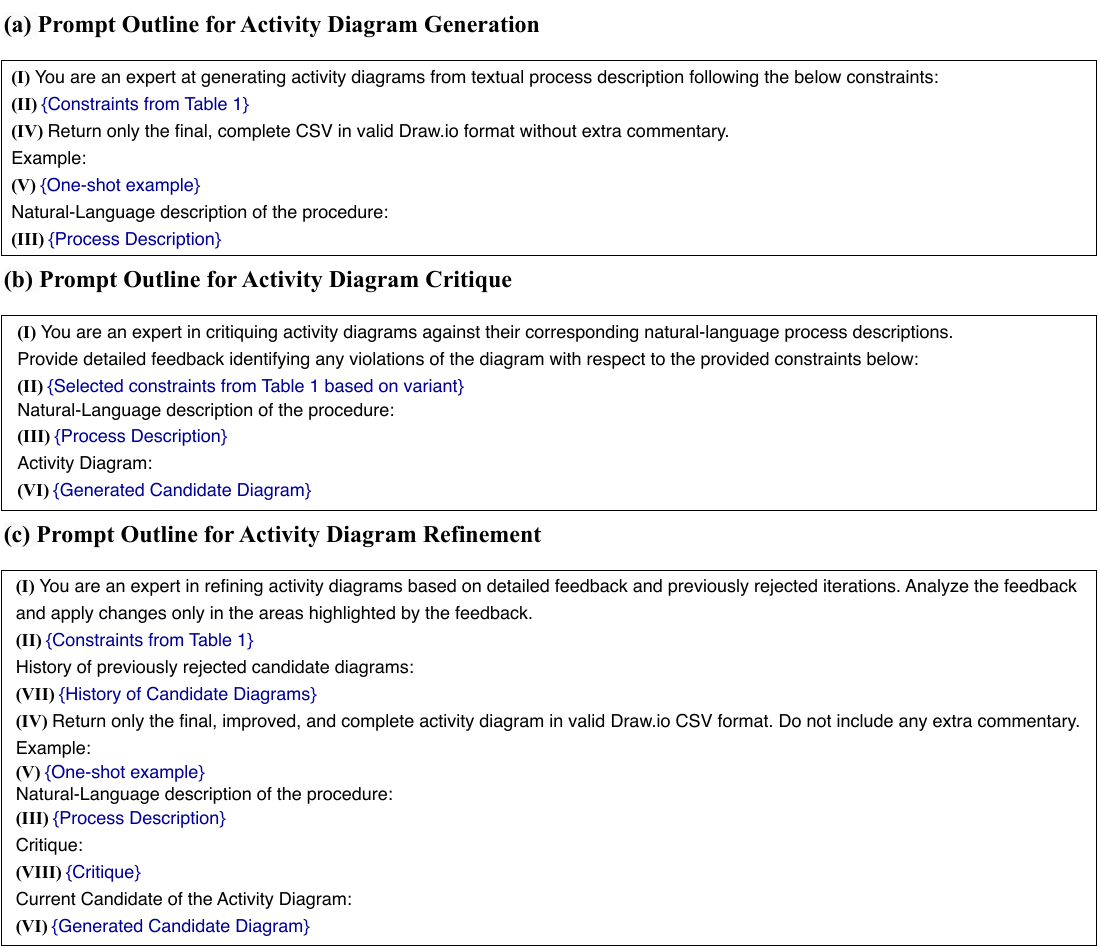} 
	\caption{Outlines of the prompts used at each step of \approach.}
	\label{figure:prompt_outlines}
\end{figure}

Figure~\ref{figure:prompt_outlines} presents the outline of the prompts used at each step of \approach. The numbers in each prompt correspond to the items with the same number in Table~\ref{table:prompt_outline}. The complete prompts are available in our replication package~\citep{replicationPackage}.

\section{Evaluation Results for Exact Matching}
\label{app:exact_match}

Table~\ref{table:semantic_results_e} presents the detailed \textit{semantic correctness} and \textit{semantic completeness} scores obtained using the exact matching algorithm discussed in Section~\ref{sec:compare}. As shown in the Table~\ref{table:semantic_results_e}, the strict string equivalence required by this algorithm results in scores of exactly 0\% on the Industry dataset, and severely degraded average scores of below 1\% on the \textsc{Paged} dataset, across all LLMs and variants. The implementation for this exact matching algorithm is provided in our online replication package~\citep{replicationPackage}.

\begin{table}[t]
    \centering
    \caption{Evaluation results for the traditional exact matching algorithm.}
    \label{table:semantic_results_e}
    \scalebox{0.55}{
    \begin{tabular}{@{}ll l cc cc cc cc cc@{}}
        \toprule
        \multirow{2}{*}{\textbf{Dataset}} & \multirow{2}{*}{\textbf{LLM}} & \multirow{2}{*}{\textbf{Metric}} 
        & \multicolumn{2}{c}{\textbf{Baseline}} 
        & \multicolumn{2}{c}{\textbf{\approachAlignmentLLMStrucutLLM}} 
        & \multicolumn{2}{c}{\textbf{\approachAlignmentLLMStrucutFormal}} 
        & \multicolumn{2}{c}{\textbf{\approachStrucutLLMOnly}} 
        & \multicolumn{2}{c}{\textbf{\approachStrucutFormalOnly}} \\
        \cmidrule(lr){4-5} \cmidrule(lr){6-7} \cmidrule(lr){8-9} \cmidrule(lr){10-11} \cmidrule(lr){12-13}
         & & & \textit{\textbf{Avg (\%)}} & \textit{\textbf{SD}} & \textit{\textbf{Avg (\%)}} & \textit{\textbf{SD}} 
         & \textit{\textbf{Avg (\%)}} & \textit{\textbf{SD}} & \textit{\textbf{Avg (\%)}} & \textit{\textbf{SD}} & \textit{\textbf{Avg (\%)}} & \textit{\textbf{SD}} \\
        \midrule
        
        % ---------------- INDUSTRY ----------------
        \multirow{4}{*}{\textit{\textbf{Industry}}} & \multirow{2}{*}{\textit{\textbf{O4 Mini}}} & \textit{\textbf{Completeness}} & 0 & 0 &  0 & 0 & 0 & 0 & 0 & 0 & 0 & 0 \\
         &  & \textit{\textbf{Correctness}} & 0 & 0 & 0 & 0 & 0 & 0 & 0 & 0 & 0 & 0 \\
        \cmidrule(l){2-13}
         & \multirow{2}{*}{\textit{\textbf{GPT-4.1 Mini}}} & \textit{\textbf{Completeness}} & 0 & 0 & 0 & 0 & 0 & 0 & 0 & 0 & 0 & 0 \\
         &  & \textit{\textbf{Correctness}} & 0 & 0 & 0 & 0 & 0 & 0 & 0 & 0 & 0 & 0 \\
        \midrule
        
        % ---------------- PAGED ----------------
        \multirow{6}{*}{\textit{\textbf{\textsc{Paged}}}} & \multirow{2}{*}{\textit{\textbf{O4 Mini}}} & \textit{\textbf{Completeness}} & 0.5 & 0.16 & 0.53 & 0.21 & 0.57 & 0.24 & 0.39 & 0.21 & 0.52 & 0.16 \\
         &  & \textit{\textbf{Correctness}} & 0.38 & 0.14 & 0.45 & 0.24 & 0.46 & 0.24 & 0.31 & 0.2 & 0.43 & 0.15 \\
        \cmidrule(l){2-13}
         & \multirow{2}{*}{\textit{\textbf{GPT-4.1 Mini}}} & \textit{\textbf{Completeness}} & 0.19 & 0.1 & 0.13 & 0.09 & 0.11 & 0.1 & 0.09 & 0.08 & 0.09 & 0.1 \\
         &  & \textit{\textbf{Correctness}} & 0.15 & 0.1 & 0.1 & 0.07 & 0.09 & 0.09 & 0.05 & 0.05 & 0.09 & 0.1 \\
        \cmidrule(l){2-13}
         & \multirow{2}{*}{\textit{\textbf{DeepSeek-R1}}} & \textit{\textbf{Completeness}} & 0.85 & 0.13 &  0.88 & 0.36 & 0.94 & 0.19 & 0.91 & 0.42 & 1.04 & 0.07 \\
         &  & \textit{\textbf{Correctness}} & 0.79 & 0.11 & 0.73 & 0.29 & 0.76 & 0.16 & 0.8 & 0.37  & 0.89 & 0.07 \\
        \bottomrule
    \end{tabular}
    }
\end{table}

\end{document}